\documentclass[12pt]{article}
\usepackage{epsf,epsfig,here,cite}
\usepackage{amsmath}
\begin{document}
\pagenumbering{arabic}
\pagestyle{empty}
\begin{center}
  \begin{LARGE}
    \textbf{Investigating The Physics Case of Running a B-Factory at the $\mathbf{\Upsilon(5S)}$ Resonance} \\
  \end{LARGE}
\end{center}
\smallskip
\renewcommand{\thefootnote}{\fnsymbol{footnote}}
\begin{center}
\textbf{E.~Baracchini$^{(a)}$, M.~Bona$^{(b)}$, M.~Ciuchini$^{(c)}$, F.~Ferroni$^{(a)}$} \\ 
\textbf{M.~Pierini$^{(d)}$, G.~Piredda$^{(a)}$, F.~Renga$^{(a)}$, L.~Silvestrini$^{(a)}$, A.~Stocchi$^{(e)}$}
\end{center}
\vskip 0.5cm
\begin{center}
  \noindent
    \noindent
    \textbf{$^{(a)}$ Dip. di Fisica, Universit\`a di Roma ``La Sapienza'' and}\\ 
    \textbf{INFN, Sez. di Roma,}\\
    \hspace*{0.5cm}{Piazzale A. Moro 2, 00185 Roma, Italy}\\
    \textbf{$^{(b)}$   Laboratoire d'Annecy-le-Vieux de Physique des Particules,}\\
    \hspace*{0.5cm}{LAPP, IN2P3/CNRS, Universit\'e de Savoie, France}\\
    \textbf{$^{(c)}$ Dip. di Fisica, Universit{\`a} di Roma Tre
      and INFN,  Sez. di Roma Tre,}\\
    \hspace*{0.5cm}{Via della Vasca Navale 84, I-00146 Roma, Italy}\\
    \noindent
    \textbf{$^{(d)}$ Department of Physics, University of Wisconsin,}\\
    \hspace*{0.5cm}{Madison, WI 53706, USA}\\
    \noindent
    \textbf{$^{(e)}$ Laboratoire de l'Acc\'el\'erateur Lin\'eaire,}\\
    \hspace*{0.5cm}{IN2P3-CNRS et Universit\'e de Paris-Sud, BP 34,
      F-91898 Orsay Cedex, France}\\
\end{center}

\smallskip

\begin{center}
\begin{abstract}
We discuss the physics case of a high luminosity $B$-Factory running
at the $\Upsilon(5S)$ resonance.  We show that the coherence of the
$B$ meson pairs is preserved at this resonance, and that $B_s$ can be
well distinguished from $B_d$ and charged $B$ mesons. These facts
allow to cover the physics program of a traditional $B$-Factory and,
at the same time, to perform complementary measurements which are not
accessible at the $\Upsilon(4S)$. In particular we show how, despite
the experimental limitations in performing time-dependent measurements of
$B_s$ decays, the same experimental information can be extracted, in several
cases, from the determination of time-integrated observables.
In addition, a few examples of the potentiality in measuring rare $B_s$ 
decays are given.  Finally, we discuss how the study of $B_s$ meson will 
improve the constraints on New Physics parameters in the $B_s$ sector, in
the context of the generalized Unitarity Triangle analysis.
\end{abstract}
\end{center}

\vfill

\def\bd         {\ensuremath{B_d}}
\def\bdbar      {\ensuremath{\bar{B_d}}}
\def\bs         {\ensuremath{B_s}}
\def\bsbar      {\ensuremath{\bar{B_s}}}
\def\bds        {\ensuremath{B_{d,s}}}
\def\bdsbar     {\ensuremath{\bar{B_{d,s}}}}
\def\bsst       {\ensuremath{B_s^{\ast}}}
\def\bsstbar    {\ensuremath{\bar{B_s^{\ast}}}}
\def\bdst       {\ensuremath{B_d^{\ast}}}
\def\bdstbar    {\ensuremath{\bar{B_d^{\ast}}}}
\def\bdsst      {\ensuremath{B_{d,s}^{\ast}}}
\def\bdsstbar   {\ensuremath{\bar{B_{d,s}^{\ast}}}}
\def\bdpi       {\ensuremath{B_d \bar{B}_d \pi}}
\def\y5s        {\ensuremath{\Upsilon (5S)}}
\def\y4s        {\ensuremath{\Upsilon (4S)}}
\def\de         {\ensuremath{\Delta E}}
\def\mes        {\ensuremath{m_{ES}}}
\def\mmiss      {\ensuremath{m_{miss}}}
\def\mmissd     {\ensuremath{m^{d}_{miss}}}
\def\mmisss     {\ensuremath{m^{s}_{miss}}}
\def\to         {\ensuremath{\rightarrow}}
\def\gev        {\ensuremath{GeV/c^2}}
\def\mev        {\ensuremath{MeV/c^2}}

 \newpage
\renewcommand{\thefootnote}{\arabic{footnote}}
\pagestyle{plain}
 \tableofcontents
\newpage
\section{Introduction}\label{sec:intro}

The study of $B$ mesons at the $B$-Factories has considerably improved
our understanding of flavour physics in the Standard Model (SM),
confirming that CP violation is well described by the
Cabibbo-Kobayashi-Maskawa (CKM) matrix~\cite{CKM}.  In particular, the
agreement between the measured value of $\sin2\beta$ in $b \to c \bar
c s$ decays~\cite{s2b,Paganini:1997cu} and the prediction from the
indirect constraints on the CKM parameters $\bar\rho$ and
$\bar\eta$~\cite{UTs2b} can be considered the first precision test of
the CKM mechanism.  Additional constraints have been added, leading to
a remarkable improvement of the determination of the parameters of the
CKM matrix~\cite{UToggi}.

A large set of these measurements is based on the study of the
coherent $B$--$\bar B$ state, produced by the decay of the
$\Upsilon(4S)$. The possibility of studying simultaneously
time-dependent CP asymmetries and rates in several $B$ decays to
charged and neutral particles pushed the physics program of
BaBar~\cite{babarNIM} and Belle~\cite{belleNIM} well beyond the
initial expectations.  The key ingredients of the success of BaBar and
Belle are the large achieved luminosities and the clean environment in
which $B$ mesons are reconstructed. The two collaborations are
expected to improve the situation by doubling their datasets in the
next two years.

Recently, the study of $B$ physics received an additional boost by the
results from the Tevatron experiments CDF~\cite{CDFNIM} and
D$\O$~\cite{D0NIM}.  In particular, the measurement of the oscillation
frequency of the $B_s$--$\bar B_s$ system, $\Delta m_s$~\cite{dmsexp}
and the comparison to the prediction from the indirect
determination~\cite{Paganini:1997cu} represents an additional test of
the SM, having a value comparable to the measurement of $\sin2\beta$. In
the next future, new measurements will be added, thanks to the
increased dataset of the Tevatron experiments and the start of the
LHCb program~\cite{LHCbNIM}.

Since these measurements will be performed at hadronic machines, where
$B_d$, $B_u$, $B_s$ and $B_c$ mesons are simultaneously produced, it will
be possible to cover a much larger physics sample. In particular, from 
the study of $B_s$ mesons
it is possible to extract some of the fundamental quantities that are
also accessible at the $B$-Factories (as the CKM
phase $\gamma$ or New Physics (NP) parameters) with reduced
theoretical uncertainty with respect to the case of $B_d$ mesons. In
addition, thanks to the different quark content of the initial state,
several $B_s$ decays, which are comparable to interesting decay modes of
the $B_d$ meson, can provide more experimental information. For
instance, the decay $B_s \to K K$ has the same interesting features of
$B_d \to K \pi$, but since it is a CP eigenstate, a study of the
time-dependent CP asymmetry can be performed. Another example is $B_s
\to J/\psi \phi$, similar to $B_d \to J/\psi K^{*0}$, but which can
be reconstructed in a CP eigenstate with all charged particles in the
final state and therefore with a higher efficiency.~\footnote{The
study of $B_d \to J/\psi K^{*0}$ in CP eigenstate requires the
$K^{*0}$ mesons to be reconstructed from $K^0_S \pi^0$ combinations,
with a loss of about a factor of six in efficiency and a large
background, because of the $\pi^0$ in the final state.  Due to the low
reconstruction efficiency, even at the $B$-Factories the time-dependent 
study of this channel~\cite{expJpsiKst} is affected by a 
large statistical error.}

However, we stress that in these latter cases the measurements
will be performed in a very different context
with respect to the clean $e^+e^-$ environment. Because of this, 
several measurements that can be accessed at the $B$-Factories, such as 
those involving neutral particles (i.e. $\pi^0$'s, $\eta'$ decays, 
radiative photons, etc.) can not be carried out with the same high 
accuracy as in an $e^+e^-$ facility. We thus suggest here to investigate the
possibility of performing them at an $e^+e^-$ environment. Such
opportunity might be provided by a next generation $B$-Factory,
running also at the $\Upsilon(5S)$ resonance~\cite{superBcdr}. We
will show how a run at the $\Upsilon(5S)$ could provide, with respect
to already approved experiments, both complementary informations and 
independent determinations of interesting CKM parameters and 
quantities sensitive to NP. 

In the recent past, CLEO and Belle performed short runs at the
$\Upsilon(5S)$, measuring the main features of this resonance with a
few collected data. These first results give just a flavour of the
potentiality of a $B$-Factory running at the $\Upsilon(5S)$, which is
the subject of this paper.

We assume the same detector performances of the existing $B$-Factory
experiments and we use a full detector simulation to study the physics
potential of running at $\Upsilon(5S)$, as a function of the
integrated luminosity.

We show that it is possible to separate $B_s$ mesons from $B_d$ and
charged $B$ mesons, even in presence of several photons in the final
state.  In this way, running at the $\Upsilon(5S)$, it would be
possible to perform the same physics measurements than at the present
$B$-Factories, after imposing kinematic cuts to reject $B_s$
events. 
\begin{boldmath}
\section{The $\Upsilon(5S)$ Production and Decays}\label{sec:upsilon}
\end{boldmath}

\begin{figure}[!tbp]
\begin{center}
\includegraphics[width=9cm]{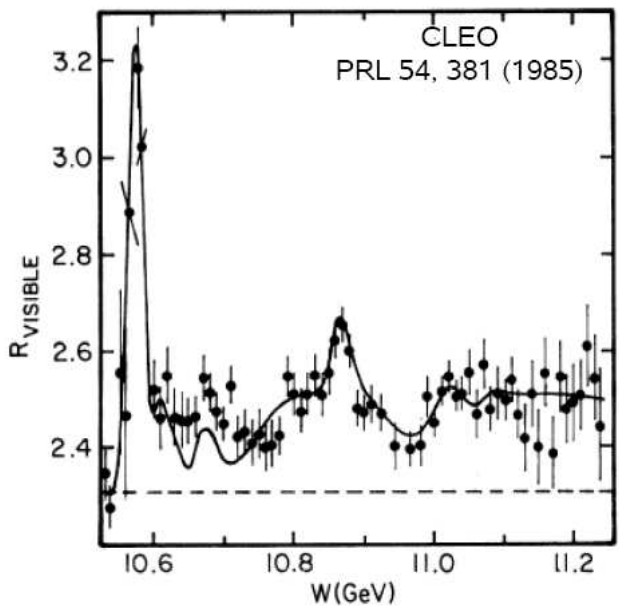}
\caption{\it CLEO measurement of $R = \sigma(e^+e^- \to q \bar
q)/\sigma(e^+e^- \to \mu \mu)$ as a function of the invariant mass $W$, in
the region of $W$ $\sim 11$ GeV~\cite{y5Sdiscovery}.}
\label{fig:CLEOSCAN} 
\end{center}
\end{figure}

The $\Upsilon(5S)$ resonance is a $J^{CP}=1^{--}$ state of a $b \bar
b$ quark pair, having an invariant mass $m_{\Upsilon(5S)} = (10.865
\pm 0.0008)$ GeV~\cite{pdg}.  The cross section of
$\Upsilon(5S)$ production in $e^+e^-$ collisions is $\sigma(e^+e^- \to
\Upsilon(5S)) = 0.301 \pm 0.002 \pm 0.039$ nb~\cite{cleoBR}.  This
resonance was discovered by CLEO at the CESR $e^+e^-$ collider,
measuring the total cross section above the $\Upsilon(4S)$
resonance~\cite{y5Sdiscovery}.  The distribution of $R = \sigma(e^+e^-
\to q \bar q)/\sigma(e^+e^- \to \mu \mu)$ as a function of the
invariant mass is shown in Fig.~\ref{fig:CLEOSCAN}.

The knowledge on the properties of this bound state comes from the
$0.42$ fb$^{-1}$ collected by the CLEO experiment at
CESR~\cite{y5Sdiscovery} and by the $1.86$ fb$^{-1}$ collected by the
Belle detector at KEKB, during the $\Upsilon(5S)$ engineering
run~\cite{belleengineeringrun}.  The Belle collaboration has also
collected a sample of about $20$ fb$^{-1}$ during June 2006.

Unlike the $\Upsilon(4S)$ state, this resonance is heavy enough to
decay in several $B$ meson states. In particular, it can decay to
vector-vector (VV), pseudoscalar-vector (PV), and
pseudoscalar-pseudoscalar (PP) combinations of $B^{(*)\pm}$,
$B_d^{(*)}$ and $B_s^{(*)}$ mesons.  The $\Upsilon(5S)$ resonance can
also decay into $B^{(\ast)}\bar B^{(\ast)} \pi$.

In the case of PP decays, the two $B$ mesons are produced in a
$1^{--}$ state, as in the case of the $\Upsilon(4S)$ resonance. When
the produced mesons are neutral, they exhibit flavour oscillations,
because of the coherence of the initial state, the oscillation
frequency being determined by the mass differences $\Delta m_{d,s}$.

The case of PV and VV decay is more complicated. The vector states
$B_q^\ast$ ($q=u,d,s$) decay in $B_q^\ast \to B_q \gamma$ through an
electromagnetic interaction which, up to a very good level of
approximation, can be considered as an instantaneous process. Under
this assumption, it can be shown (see Sec.~\ref{sec:timeevolution})
that the time evolution of VV states is similar to that of the PP
state. For events generated by PV decays, a difference comes from the
opposite eigenvalue of the charge-conjugation operator.

Quark models~\cite{quarkmodelsY} predict the production to be mainly
in the excited final states $\bdsst \bar{\bdsst}$ with $\bdsst \to
\bds \gamma$.  Even if relative branching ratios (BR) of
$\Upsilon(5S)$ are not precisely known yet, the current measurements
confirm this picture. We summarize in Tab.~\ref{tab:Y5Sinput} the
corresponding experimental results, as reported by CLEO \cite{cleoBR}
and BELLE \cite{belleengineeringrun}, along with the values used in
this paper.  For unmeasured values, we made educated guesses based on
the predicted (and observed) VV dominance.  
\\
\begin{table}[!tbp]
\begin{center}
\begin{tabular}{cccc}
\hline
\hline
$\Upsilon(5S)$ Decay Modes & CLEO & BELLE & This Paper \\
\hline
$B_s^{(*)} \bar B_s^{(*)}$ $\left[\%\right]$ & $26^{+7}_{-4}$ & $21^{+6}_{-3}$ & 26 \\
$(B_s^* \bar B_s^*)/(B_s^{(*)}\bar B_s^{(*)})$ & $ - $ & $0.94^{+0.06}_{-0.09}$ & 0.94 \\
$(B_s^* \bar B_s + B_s \bar B_s^*)/(B_s^{(*)}\bar B_s^{(*)})$ & $ - $ & $ - $ & 0.06 \\
$(B_s \bar B_s)/(B_s^{(*)}\bar B_s^{(*)})$ & $ - $ & $ - $ & $ - $ \\
\hline
$B_{d,u}^* \bar B_{d,u}^*$ $\left[\%\right]$ & $ 43.6 \pm 8.3 \pm 7.2 $ & $ - $ & 44 \\
$B_{d,u} \bar B_{d,u}^* + B_{d,u}^* \bar B_{d,u} $ $\left[\%\right]$ & $ 14.3 \pm 5.3 \pm 2.7 $ & $ - $ &  14 \\
$B_{d,u} \bar B_{d,u}$ $\left[\%\right]$ & $ <13.8 $ & $ - $ & - \\
\hline
$B_{d,u} \bar B_{d,u}^{(*)} \pi + B_{d,u}^{(*)} \bar B_{d,u} \pi$ $\left[\%\right]$ & $ <19.7 $ & $ - $ & 16 \\
$B_{d,u} \bar B_{d,u} \pi \pi$ $\left[\%\right]$ & $< 8.9$ & $ - $ & - \\
\hline
\hline
\end{tabular}
\caption{\it \label{tab:Y5Sinput} BR of $\Upsilon(5S)$ decays, measured by
CLEO~\cite{cleoBR} and BELLE~\cite{belleengineeringrun}
collaborations. The last column shows the values used throughout this
paper.}
\end{center}
\end{table}

The numbers quoted in the table, together with the different values of
the cross sections, show that $B_d$ and charged $B$ mesons can be
produced at the $\Upsilon(5S)$, with a rate about six times smaller
than at the $\Upsilon(4S)$. This fact should be kept in mind: a reduction 
in the statistics of $B_d$ and charged $B$ mesons is the price to pay in 
order to study $B_s$ particles at a $B$-Factory.
\\
\begin{boldmath}
\section{Reconstruction of $B$--$\bar B$ pairs}\label{sec:kinematic}
\end{boldmath}

The reconstruction of $B$--$\bar B$ pairs at the $\Upsilon(5S)$
proceeds like at the traditional $B$-Factories. The only relevant
difference is the fact that several final states can be
produced, with different momenta.

In the case of $B_s$ mesons, the largest fraction ($\sim 94\%$ of the
events) is produced in a VV mode. One can then tune the event
reconstruction and selection on these events, considering the PV and
the PP modes as a source of background. To avoid decreasing the
reconstruction efficiency, one should not look for either the photons
produced by the $B_{s,d}^*$ decay or the one or more pions produced in
the continuum $b \bar b$ production.

At a traditional $B$-Factory, $B \bar B$ events are distinguished from
$q \bar q$ background ($q=u,d,s,c$) using a set of variables related
to the distribution of the decay products in the center-of-mass (CM)
system of the $\Upsilon$ resonance. We consider here the two quantities
$L_0$ and $L_2$, defined as~\cite{badfrancesi}
\begin{equation}
L_j\equiv\sum_i |{\bf p}^*_i| |\cos \theta^*_i|^j ~(j=0,2),
\end{equation}
where ${\bf p}^*_i$ is the momentum of particle $i$ in the $e^+e^-$
rest frame, $\theta^*_i$ is the angle between ${\bf p}^*_i$ and the
thrust axis of the $B$ candidate and the sum runs over all
reconstructed particles except for the $B$-candidate daughters. In
data analyses, $L_0$ and $L_2$ are typically combined into a Fisher
discriminant ${\mathcal F}$~\cite{Fisher}. Alternatively, the ratio
$L_2/L_0$ is used~\cite{kspi0paper}.  The discriminating power against
the $q \bar q$ background for these two variables is very close, so
that we use one or the other in the subsequent sections.

\begin{figure}[!tbp]
\begin{center}
\includegraphics[width=9cm]{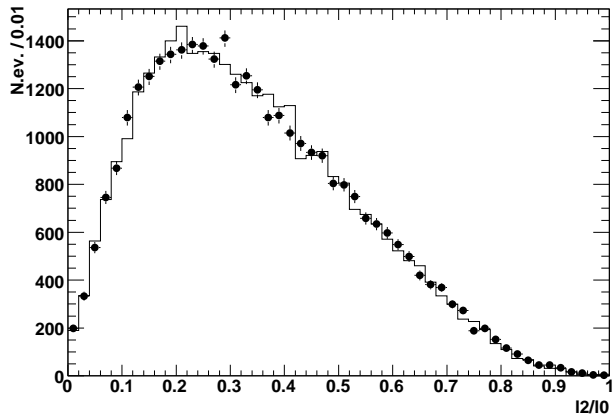}
\caption{\it Distribution of $L_2/L_0$ for $B_s \to J/\psi \phi$ events
simulated at the $\Upsilon(5S)$ resonance in a VV (dots) and PP
(black line) configurations.}
\label{fig:norml12} 
\end{center}
\end{figure}

At the $\Upsilon(5S)$, $L_0$ and $L_2$ preserve all their features,
even in presence of additional unreconstructed particles (photons and
pions) produced with the $B \bar B$ pair.  This is shown in
Fig.~\ref{fig:norml12}, where the distribution of the ratio $L_2/L_0$
for $B_s \to J/\psi \phi$ generated events is given for PP(continuous
line) and VV (dots) events. We do not distinguish between VV, PV, and
PP in the following, when dealing with topological variables.

Kinematic variables are also used to reject background. At the
$\Upsilon(4S)$, a $B \bar B$ event is usually characterized by the two
following variables:
\begin{itemize}
\item $\de = E_{B}^{\ast}-\sqrt{s}/2$, the energy difference between
the reconstructed $B$ candidate in the center of mass (CM) frame
$E_{B}^{\ast}$ and one half the total CM energy $\sqrt{s}$.
\item the beam-energy substituted mass $\mes=\sqrt{(s/2+{\mathbf
{p}}_i\cdot {\mathbf {p_B}})^2/E_i^2- {\mathbf {p_B}}^2}$, where the
$B$ momentum ${\mathbf {p_B}}$ and the four-momentum $(E_i, {\mathbf
{p_i}})$ of the $e^+ e^-$ initial state are defined in the laboratory
frame.
\end{itemize}

These two variables are found to be Gaussian distributed and almost
uncorrelated for those $B$ decays having only charged tracks in the
final state.  For $B$ mesons coming from $\Upsilon(4S)$ decays,
$m_{ES}$ ($\Delta E$) peaks at the value of the $B_d$ mass (at zero).
The presence of photons in the final state (coming from decays of
intermediate particles, radiative $B$ decay or {\it bremsstrahlung})
introduces a correlation (typically less than $20\%$), induced by
the missing energy. In these cases, it was found~\cite{kspi0paper}
that a better signal identification is achieved replacing $m_{ES}$
with $m_{miss}$, defined as $m_{miss} = |p_{e^+e^-} - \tilde{p}_B|$,
where $p_{e^+e^-}$ is the four-momentum of the $e^+e^-$ initial state
and $\tilde{p}_B$ is the four-momentum of the $B$ candidate after
applying a mass constraint on it.
 
\begin{figure}[tb!]
\begin{center}
\includegraphics[width=9cm]{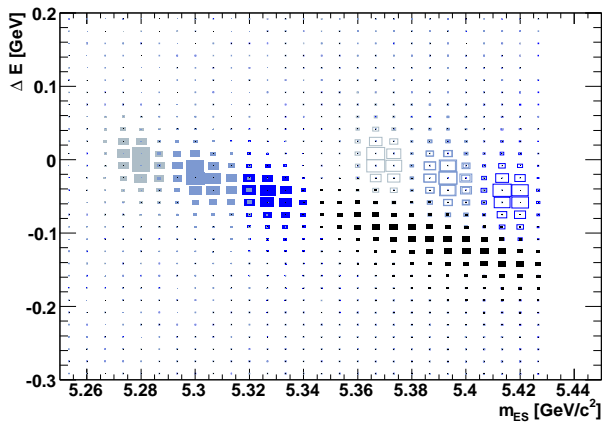}
\caption{\it Distribution of $\Delta E$ vs. $m_{ES}$ for a sample of
simulated $B_{d,s} \to J/\psi \phi$ decays at the $\Upsilon(5S)$
resonance.  Events coming from $B_q^{(*)} \bar B_q^{(*)}$ ($q=d,s$)
are all generated with the same relative rate. We use full boxes for
$q=d$ and empty boxes for $q=s$. The color scale identifies VV, VP and
PP events (from the darker to the lighter).  Events from $B_d \bar B_d
\pi$ events are also shown (full dark boxes on the bottom-right).}
\label{fig:mesvsde_Y5S_jpsiphi}
\end{center}
\end{figure}

To illustrate the kinematic properties of $B$ mesons coming from
$\Upsilon(5S)$ decays, we consider the case of $\Upsilon(5S) \to B_s^*
B_s^*$ events.  After the two $B_s^* \to B_s \gamma$ decays, the final
state is represented by the decay products of the two $B_s$ mesons and
by the two additional photons which are not reconstructed.  When
calculating $\Delta E$ from the observed decay products,
$E_{B}^{\ast}$ assumes the value of the energy of the $B_s$ meson,
while the value to use for $\sqrt{s}/2$ corresponds to half the energy
of the $B_s^* B_s^* $ system.  This difference generates a shift of
the $\Delta E$ distribution of $-47$ MeV, corresponding to the mass
difference $m_{B_{s}}-m_{B^*_{s}}$. In addition, the $B_s$ momentum
distribution is smeared by the spread of the photon energy in the CM
frame, since the photon is monochromatic only in the $B_s^*$ rest
frame, while the laboratory frame is boosted~\footnote{In order to be
able to study also $B_d$ physics, $\Upsilon(5S)$ decays should be
studied at an asymmetric $B$-Factory, where also time-dependent
measurements of $B_d$ decays can be performed.}  and the energy of the
photon becomes a function of the polar angle in the CM frame.  At the
same time, $m_{ES}$ values are shifted by the same quantity to larger
values. This is shown in Fig.~\ref{fig:mesvsde_Y5S_jpsiphi}, where the
distribution of $B_{d,s} \to J/\psi \phi$ events in the $\Delta E$ vs
$m_{ES}$ plane is given. For $B_s$ decays the three bumps
(corresponding to PP, PV and VV decays) are aligned along a straight
line, identified by the relation: $\Delta E =
-(m_{ES}-m_{B_s})$. Similar considerations are valid for $B_d$ and
charged $B$ mesons.

For $m_{miss}$ the situation is more complicated, since the
calculation of this variable requires a mass hypothesis for the
reconstructed $B$ ($B_d$ or $B_s$ meson).  One is forced to define two
variables, $m_{miss}^d$ and $m_{miss}^s$, and to use one or the other
according to the case under study.

In the following, we present three examples of $B$ decays at the
$\Upsilon(5S)$ resonance.  We start from the simple case of a final
state without neutral particles ($B_{d,s} \to J/\psi \phi$).  After
showing the good performances in separating VV events from PV and PP
events and $B_s$ from $B_d$, we consider the case of $B$ final states
with two photons ($B_{d,s} \to \pi^0 K^0_S$) and three photons
($B_{d,s} \to K^{*0} \gamma$ with $K^{*0} \to \pi^0 K^0_S$). 
In principle, one would expect these decays (and
in general all the decays to final states with photons) to be more
problematic, since only a part of the energy of the photons is detected in
the calorimeter (introducing asymmetries in the shapes of the
kinematic variables).  As we show, the use of $m_{miss}$ instead of
$m_{ES}$ can provide a good separation even in this case. Other choices 
adopted in literature (such the use of $m_B$ rather than $\Delta E$) do 
not give advantages and are not considered in the following.

\begin{figure}[tb!]
\begin{center}
\includegraphics[width=7cm]{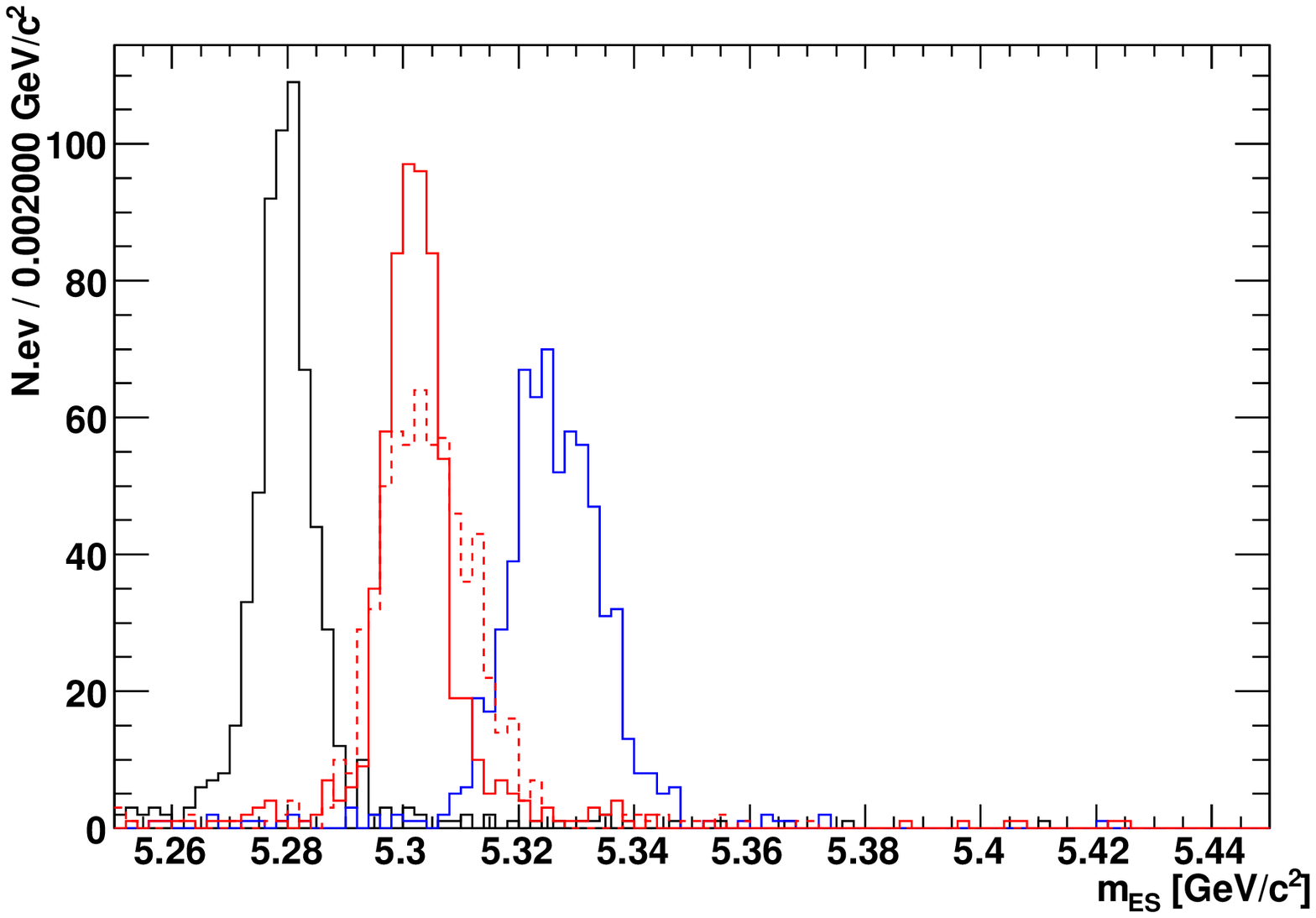}
\includegraphics[width=7cm]{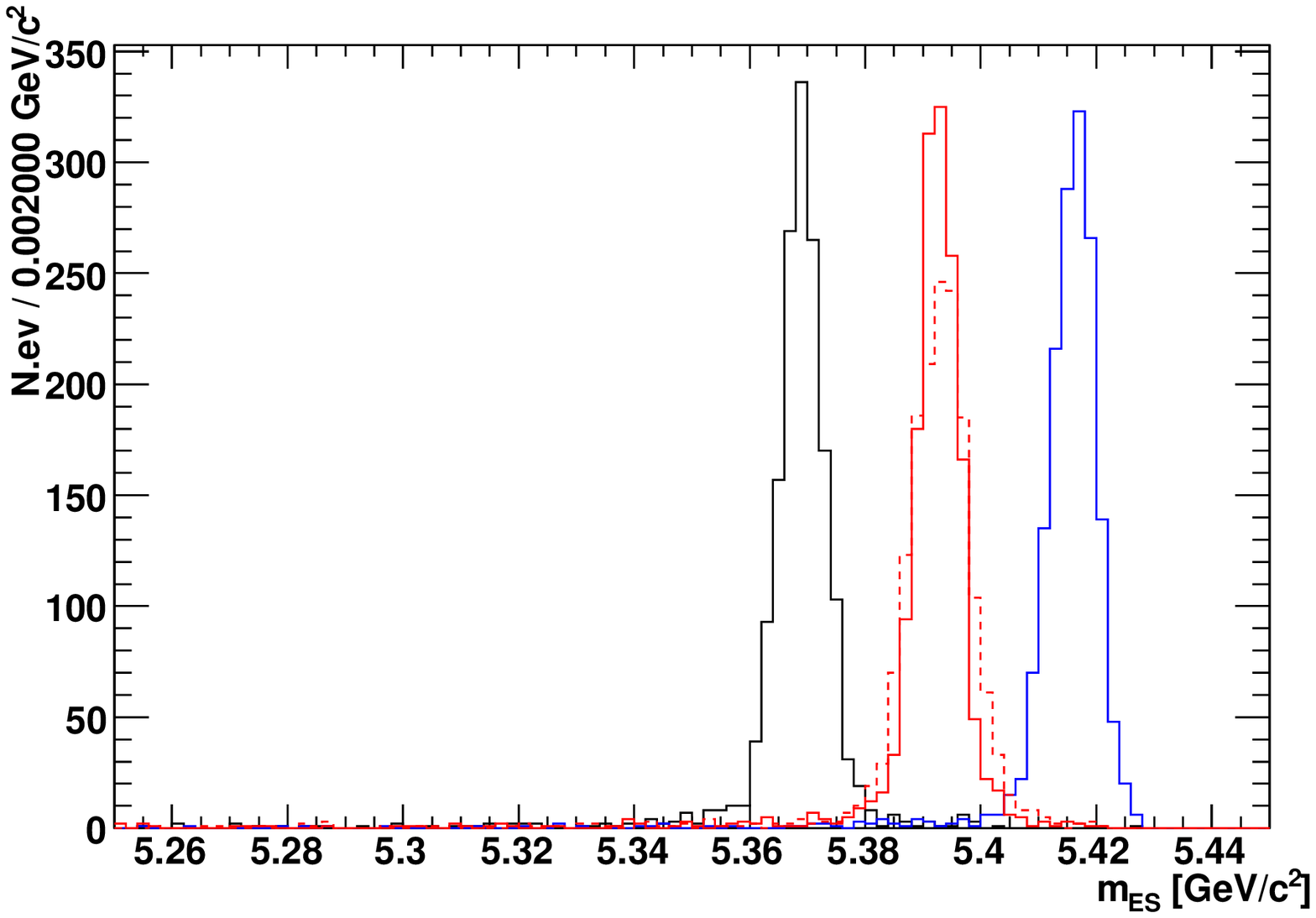} \\
\includegraphics[width=7cm]{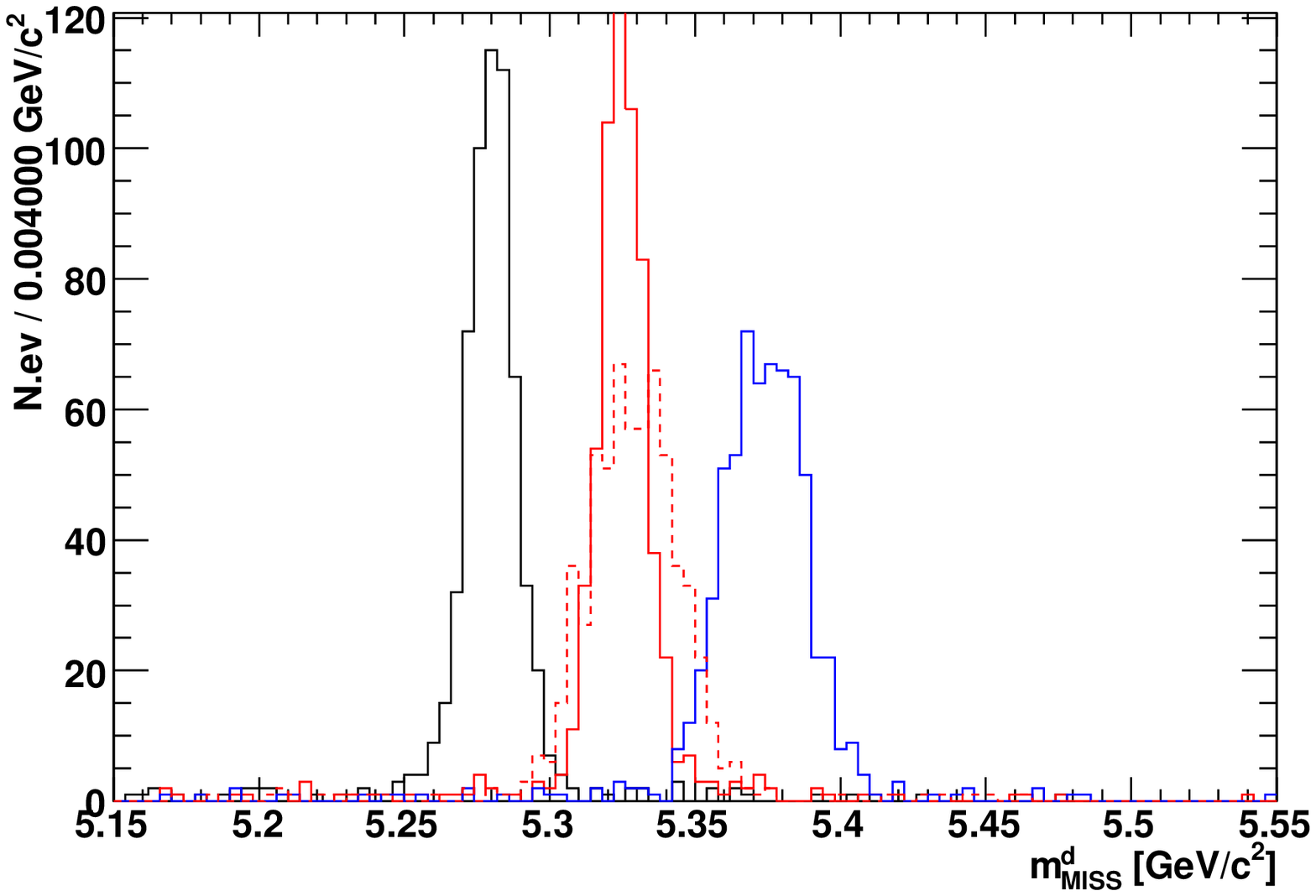}
\includegraphics[width=7cm]{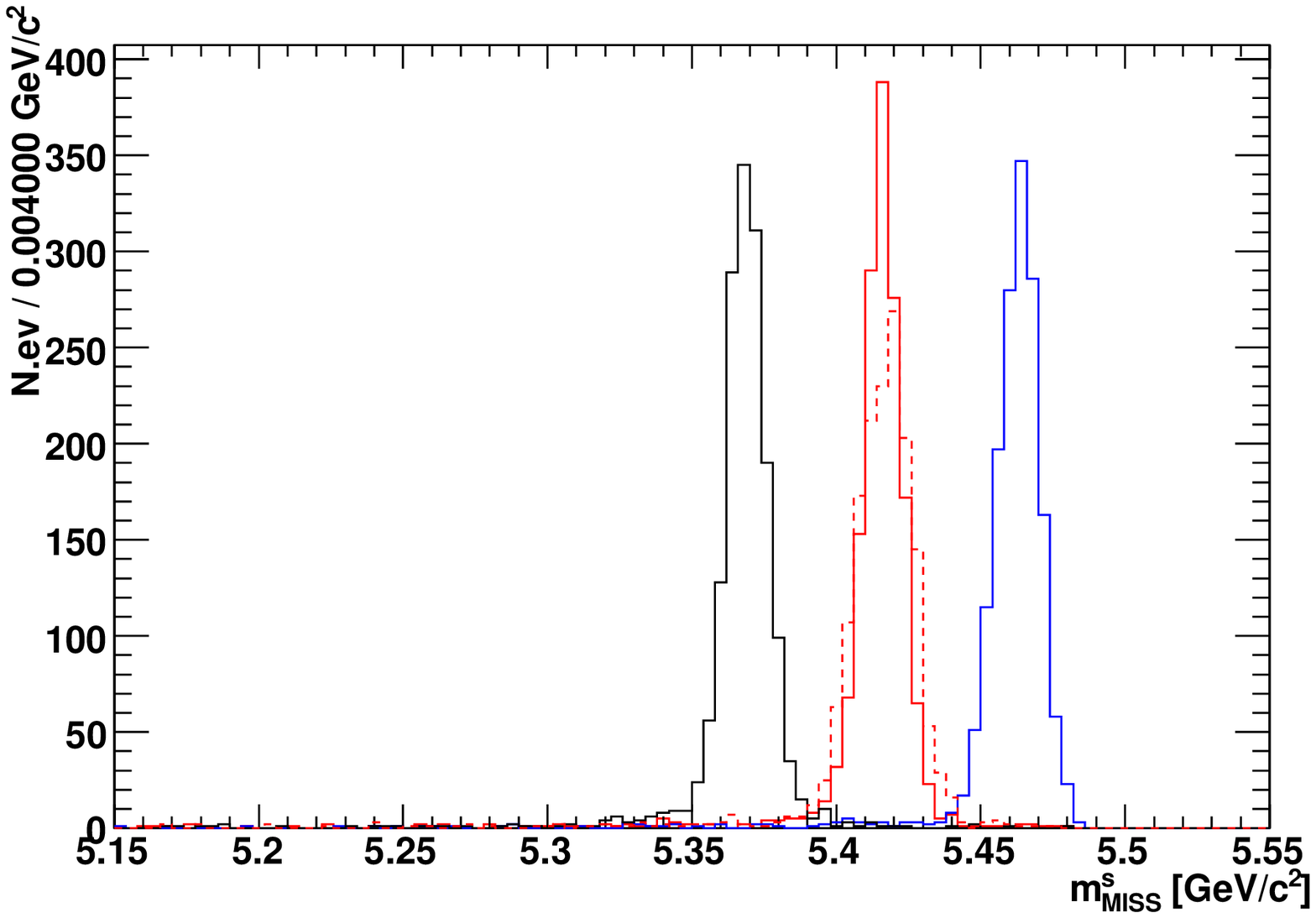}
\caption{\it Top row : distribution of $m_{ES}$ for a sample of simulated
$B_{d}$ (left) and $B_s$ (right) decaying to $J/\psi \phi$ at the
$\Upsilon(5S)$ resonance.  Bottom row : distribution of $m^d_{miss}$
for a sample of simulated $B_{d}$ (left) and $m^s_{miss}$ for a sample
of $B_s$ (right) decaying to $J/\psi \phi$ at the $\Upsilon(5S)$
resonance. From the left to right, we show events coming from $B_q
B_q$, $B_q B^*_q$, $B^*_q B_q$ (dashed) and $B^*_q B^*_q$ ($q=d,s$),
all generated with the same relative rate.  In writing $B^{(*)}_q
B^{(*)}_q$ the first $B^{(*)}_q$ is intended as the reconstructed
one.}
\label{fig:mes_jpsiphi}
\end{center}
\end{figure}

\begin{figure}[tb!]
\begin{center}
\includegraphics[width=7cm]{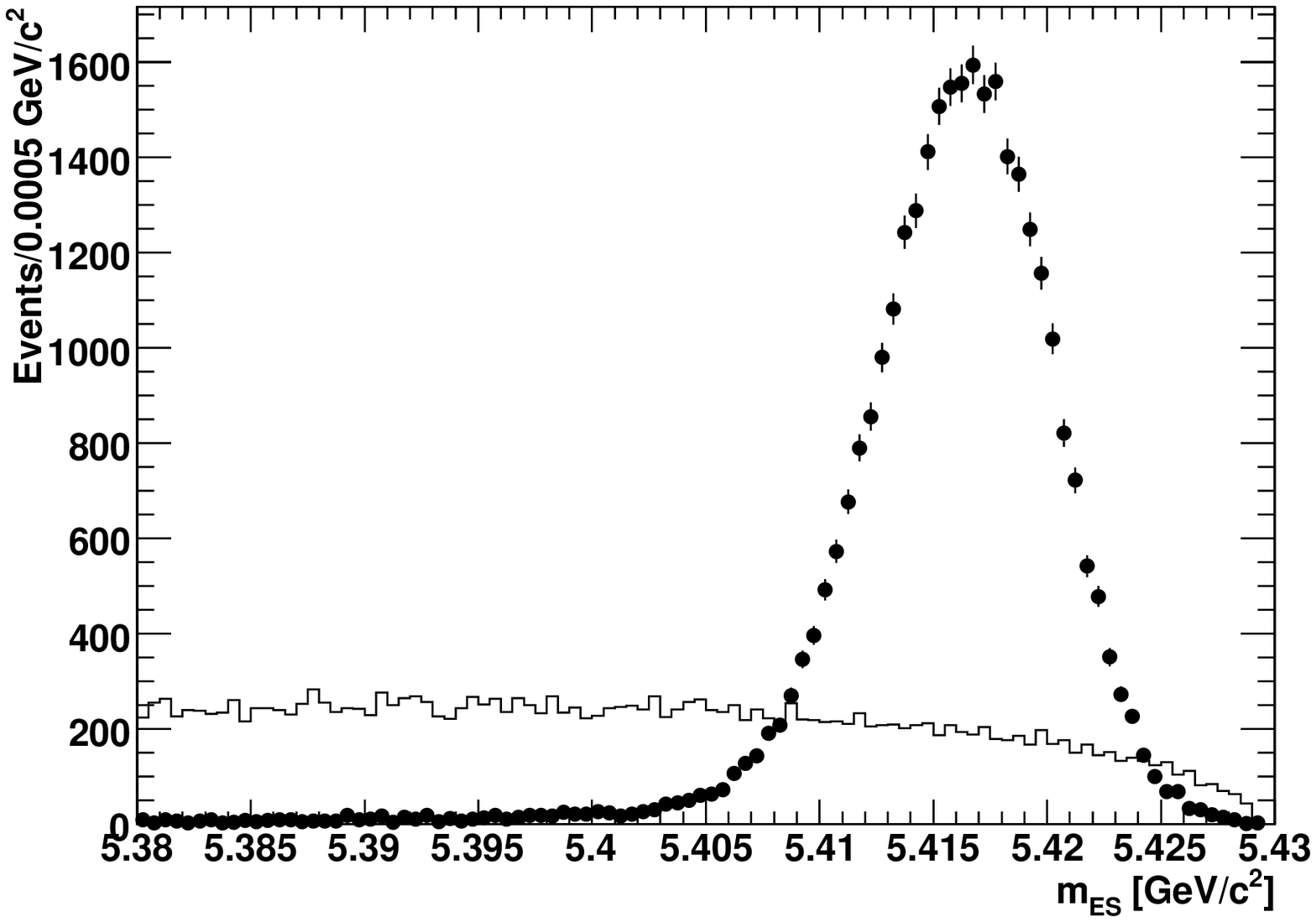}
\includegraphics[width=7cm]{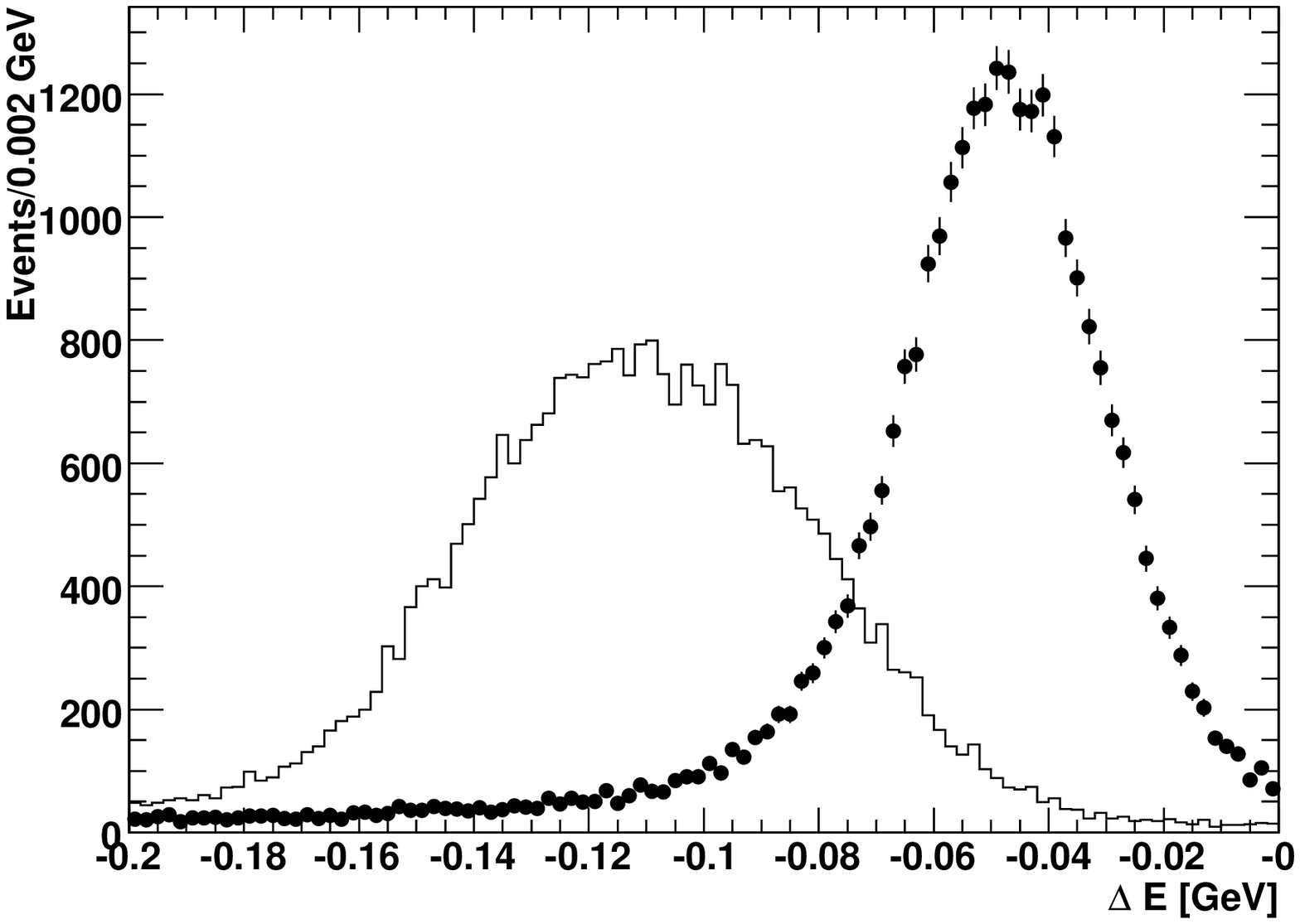}
\caption{\it Distribution of $m_{ES}$ (left) and $\Delta E$ (right) for
$B_d \to J/\psi \phi$ candidates, from a sample of simulated $\Upsilon
(5S) \to B_d \bar B_d \pi$ events (line).  $B_s \to J/\psi \phi$
events, generated in $\Upsilon (5S) \to B_s^* \bar B_s^*$ events
(dots), are also shown for comparison.
\label{fig:mesdeBBpi}}
\end{center}
\end{figure}

\begin{figure}[tb!]
\begin{center}
\includegraphics[width=7cm]{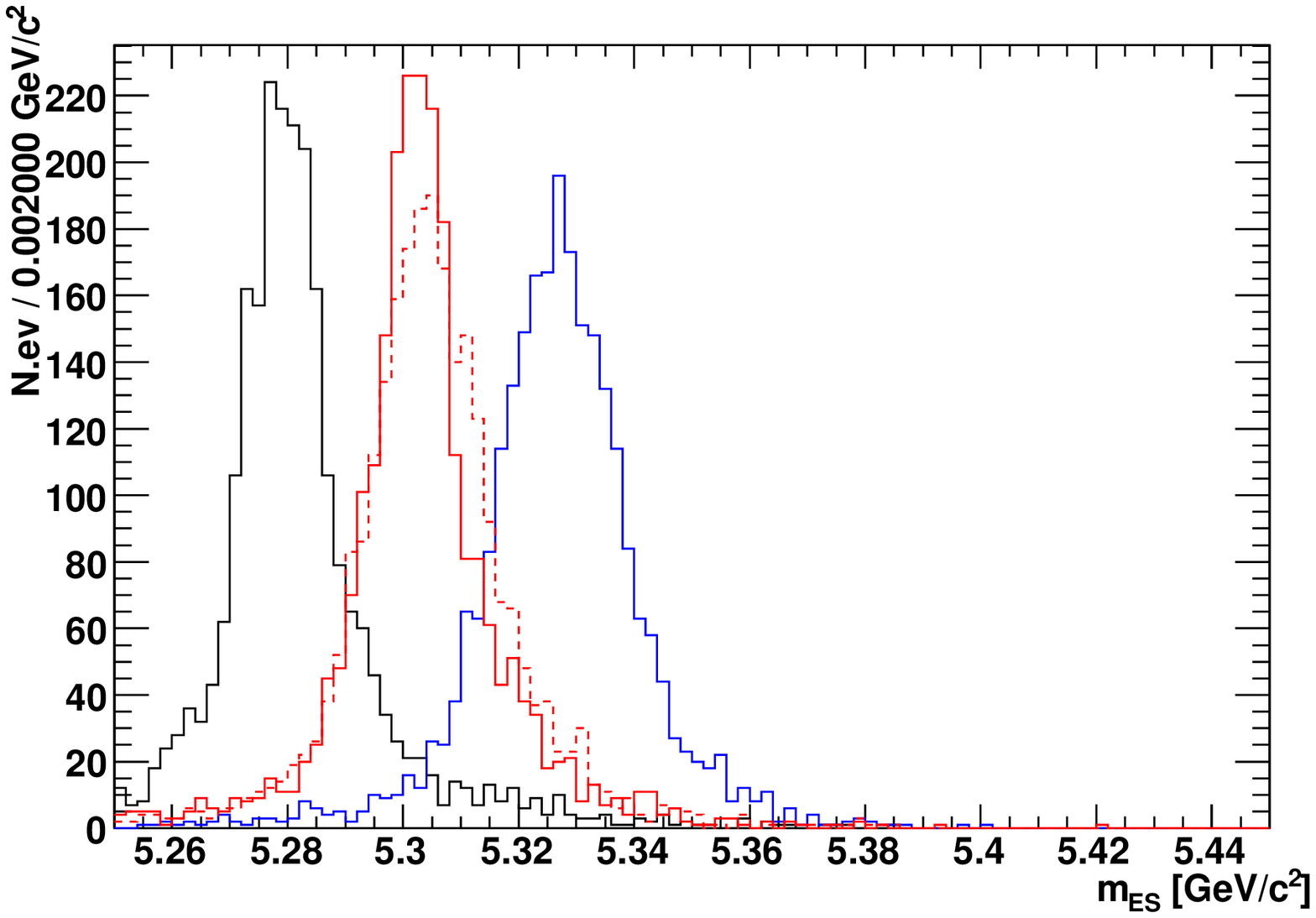}
\includegraphics[width=7cm]{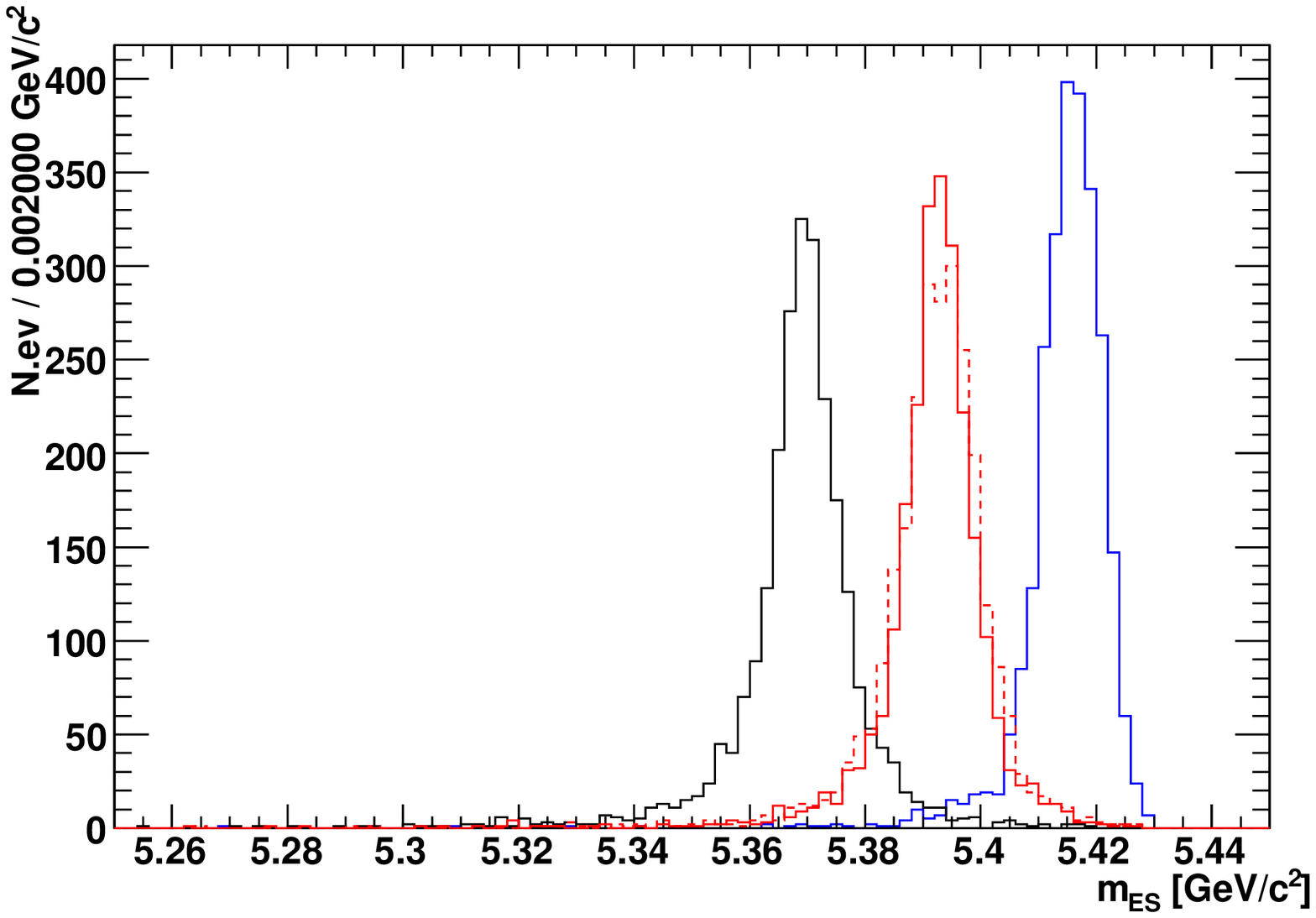} \\
\includegraphics[width=7cm]{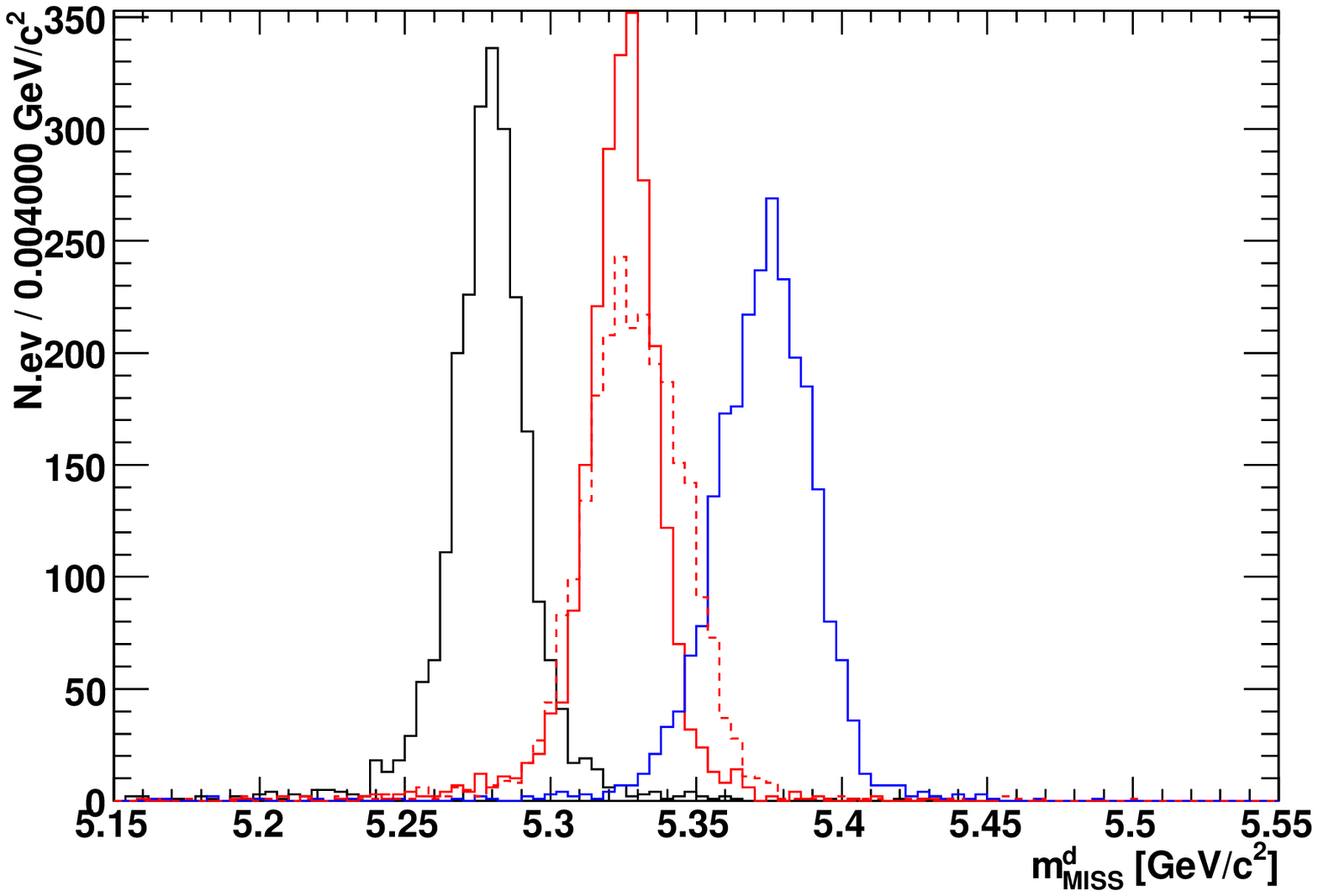}
\includegraphics[width=7cm]{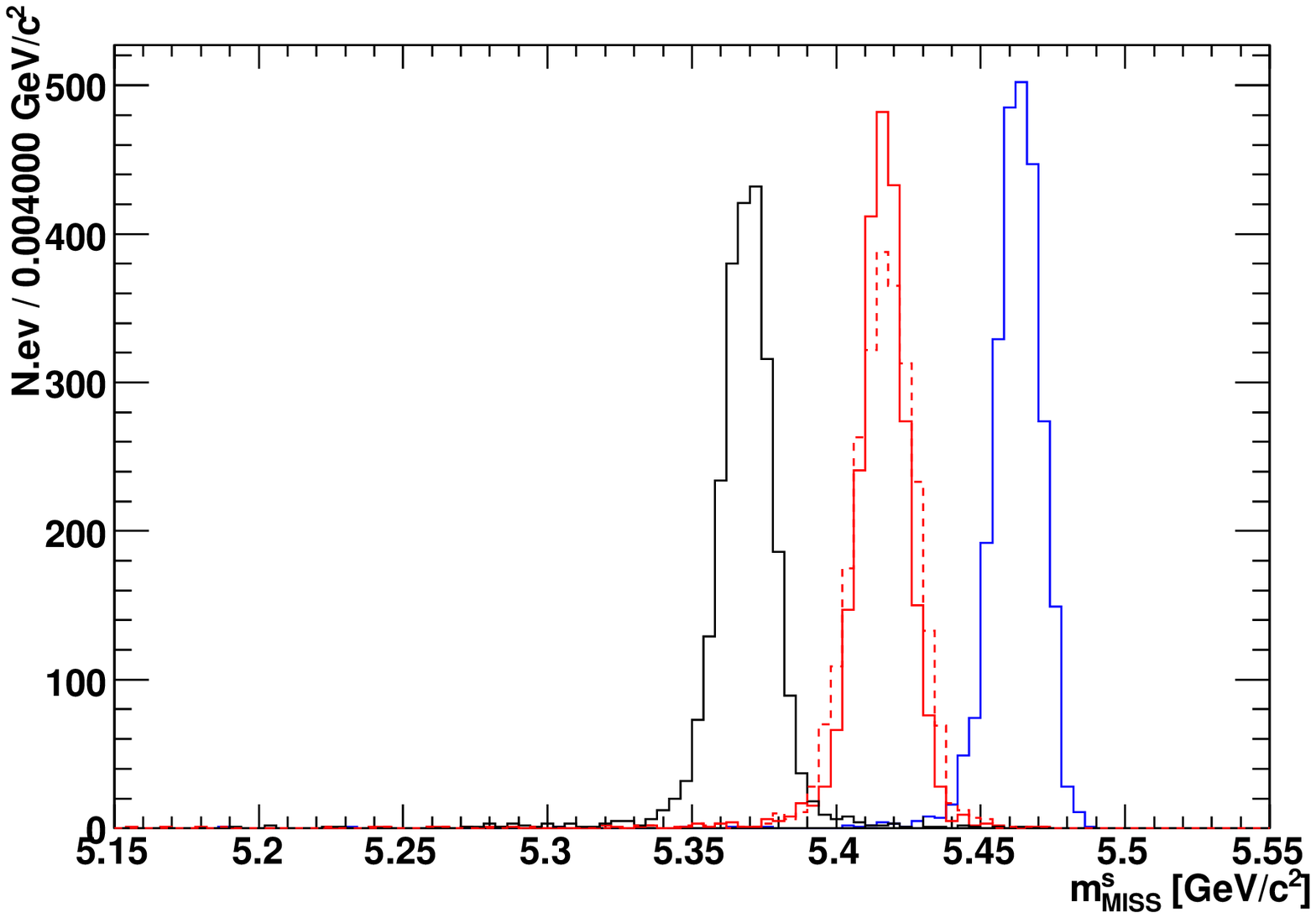}
\caption{\it Top row : distribution of $m_{ES}$ for a sample of simulated
$B_{d}$ (left) and $B_s$ (right) decaying to $K^0_S \pi^0$ at the
$\Upsilon(5S)$ resonance.  Bottom row : distribution of $m^d_{miss}$
for a sample of simulated $B_{d}$ (left) and $m^s_{miss}$ for a sample
of $B_s$ (right) decaying to $K^0_S \pi^0$ at the $\Upsilon(5S)$
resonance.  From the left to right, we show events coming from $B_q
B_q$, $B_q B^*_q$, $B^*_q B_q$ (dashed) and $B^*_q B^*_q$ ($q=d,s$),
all generated with the same relative rate.  In writing $B^{(*)}_q
B^{(*)}_q$ the first $B^{(*)}_q$ is intended as the reconstructed
one.}
\label{fig:mes_kspi0}
\end{center}
\end{figure}

\begin{boldmath}
\subsection{Selection of $B_{d,s} \to J/\psi \phi$}
\end{boldmath}

Fig.~\ref{fig:mesvsde_Y5S_jpsiphi} shows the distribution of signal
Monte Carlo events for $B_{d,s} \to J/\psi \phi$ decays, obtained
assuming all the $b \bar b$ decays of the $\Upsilon(5S)$ to have the
same production rate. The values of the two variables are calculated
from the daughters of the reconstructed $B$ mesons, without searching
for additional particles (photons or pions) produced in the
$\Upsilon(5S)$ decay tree.

It can be noted that the PP, PV and VV configurations are well
separated in the ($m_{ES}$,$\Delta E$) plane. $\bds \bds$ events are
centered at $\de \sim 0$ while $\bds \bdsst$ and $\bdsst \bdsst$ are
shifted to negative values of $\de$ and higher values of $\mes$ due to
the missing energy of the photons originating from $B_s^* \to B_s \gamma$ 
decays. In a similar way, $B\bar B \pi$ events are further shifted.  
Three important features should be noted:
\begin{itemize}
\item The resolution of both variables is worse in the case of $\bd$
mesons than for $\bs$ mesons, since $B_d$ mesons receive a larger
momentum than $B_s$ mesons in the CM frame of the $\Upsilon(5S)$,
resulting in a wider distribution of the $B$ momentum in the
laboratory frame.
\item The resolution for VV events is worse than for PP ones, since
the presence of the photon introduces an additional source of energy
spread.
\item In the case of PV events, the distribution is the sum of two
components: a broad one, due to reconstructed $B$ mesons originating
from the $B^*$ decay; and a narrow one, due to reconstructed $B$
mesons originating directly from the $\Upsilon(5S)$.  This is shown in
Fig~\ref{fig:mes_jpsiphi}, where the two components are independently
shown in the $m_{ES}$ and $m_{miss}$ projections.
\end{itemize}

We consider only VV events and we study the cross-feed generated by
the tails of similar events (in this case $B_{d,s} \to J/\psi \phi$)
coming from the other configurations (PP and PV).  We have chosen
specific selection criteria for $m_{ES}$ or $m_{miss}$ to separate the
different components.  We find that in general, and mainly for the
channels with photons in the final state as shown in the following
sections, the $m_{miss}$ variables provide better efficiency and
purity.  In Tab.~\ref{tab:cuts} we summarize the cuts applied on
$m_{miss}$, the efficiency and we quantify the cross-feed among $B_s$
and $B_d$ samples.

The situation is different in the case of $B \bar B \pi$ continuum
production. These events are overlapped to $B_s$ mesons from VV
events, in such a way that a simple cut in the $\Delta E$ vs. $m_{ES}$
plane cannot separate the two components without affecting the
distribution of $q \bar q$ background.

The separation is indeed possible since these events are characterized
by a continuum-like distribution in $m_{ES}$ (see left plot of
Fig.~\ref{fig:mesdeBBpi}), while they exhibit a peaking structure in
$\Delta E$ (see right plot of Fig.~\ref{fig:mesdeBBpi}), shifted to
negative values with respect to $B_{s}$ mesons, because of the
undetected energy of the pion.  At present, concerning the relative
amount of this component with respect to $B_s$ mesons, we can rely
only the upper limit (UL) quoted by CLEO~\cite{cleoBR} (see
Tab.~\ref{tab:Y5Sinput}), so that we are not allowed to neglect this
kind of background.  Anyhow, the differences in the kinematic
variables allow to isolate these events from the signal, adding a
component in the Maximum Likelihood (ML) fit as it is currently done
with $B \bar B$ background in many charmless analyses in
BaBar~\cite{charmlessBB}.

\begin{table}[tb!]
\begin{center}
\begin{tabular}{ccc} 
\hline\hline
& $B_d$ & $B_s$  \\
\hline
 & \multicolumn{2}{c}{$B \to J/\psi \phi$} \\
\hline
Cut & $5.355<m^d_{miss}\,[GeV]<5.410$ & $5.440<m^s_{miss}\,[GeV]<5.490$ \\ 
Efficiency & 81\% & 91\% \\
Contamination & 1.1\% & 0.0\%                                     \\
              &[PV=0.7\%, PP=0.3\%]       &       \\
\hline
& \multicolumn{2}{c}{$B \to K^0_S \pi^0$}\\
\hline
Cut & $ 5.350<m^d_{miss}\,[GeV] <5.410 $ & $5.440<m^s_{miss}\,[GeV]<5.490$ \\ 
Efficiency & 89\% & 98\% \\
Contamination & 1.6\% & 0.0\%                                     \\
              &[PV=1.5\%, PP=0.1\%]       &       \\
\hline
& \multicolumn{2}{c}{$B \to K^*(K^0_S \pi^0) \gamma$}   \\
\hline
Cut &$5.355<m^d_{miss}\,[GeV]<5.410$ & $5.440<m^s_{miss}\,[GeV]<5.490$\\ 
Efficiency & 75\% & 84\% \\
Contamination & 4.4\% & 0.5\%                                      \\
              &[PV=3.3\%, PP=1.0\%]       &[PV=0.4\%, PP=0.1\%]        \\
\hline\hline
\end{tabular}
\end{center}
\caption{\it \label{tab:cuts} Cuts, efficiencies and contaminations in the
channels $B \to J/\psi \phi$ (no photons), $B \to K^0_S \pi^0$ (2
photons), $B \to K^* \gamma$ (3 photons).  The quoted values take into
account the different production rates of the $\Upsilon(5S)$
resonance. The contamination of $B_d$ to $B_s$ ($B_s$ to $B_d$) is
negligible (less than $0.1\%$).}
\end{table}   

\begin{boldmath}
\subsection{Selection of $B_{d,s} \to K^0_S \pi^0$}
\end{boldmath}

As mentioned in Sec.~\ref{sec:kinematic}, the presence of neutral particles in
the final state smears the distribution of the interesting kinematic
variables. The ability of the cuts to separate the different channels
is therefore reduced by this effect, so that the result of the
previous study is not guaranteed to hold in this case. To investigate
this point, we consider a sample of $B_{d,s} \to K^0_S \pi^0$ decays,
reconstructing signal candidates from $K^0_S \to \pi^+ \pi^-$ and
$\pi^0 \to \gamma \gamma$ decays.

Fig.~\ref{fig:mes_kspi0} shows the $m_{ES}$, $m^{d}_{miss}$ and
$m^{s}_{miss}$ distributions in this channel. Although the width of
the distribution increases in a significant way, it is still possible
to separate the different contributions.  In particular, the use of
$m^{d}_{miss}$ rather than $m_{ES}$ provides a separation almost as
good as in the case of $J/\psi \phi$ final states.  \\
\begin{boldmath}
\subsection{Selection of $B_{d,s} \to K^* \gamma$}
\end{boldmath}

As a further step, we repeat the previous study using $B_{d,s} \to
K^*(K^0_S \pi^0) \gamma$ signal events, reconstructed from $K^0_S \to
\pi^+ \pi^-$ and $\pi^0 \to \gamma \gamma$.  In this case, there are
three photons in the final state.  As in the previous section, an
improvement in the separation between different channels can be
achieved using $m_{miss}$ instead of $m_{ES}$.  In
Tab.~\ref{tab:cuts}, we summarize the cuts applied on $m_{miss}$ in
the three channels, with efficiencies and the details on the
contaminations.  \\
\begin{boldmath}
\subsection{A Remark on $B$ reconstruction}
\end{boldmath}

We would like to highlight a point that is relevant for the rest of
this paper. Since it is possible to isolate $B_s$ mesons from $B_d$
and charged $B$ mesons, it is clear that it is possible to continue
the $\Upsilon(4S)$ physics program when running at the
$\Upsilon(5S)$. This is particularly crucial for measurements that are
difficult to perform at other facilities, such as inclusive
measurements ($b \to s \gamma$, $b \to s l l$, and semileptonic
decays) as well as some exclusive measurements with open kinematic ($B
\to \tau \nu$, exclusive semileptonic decays). All these measurements
represent important milestones of the physics program of the current
$B$-Factories and it is hard to imagine a precision test of the
flavour sector of NP without the possibility of reducing the errors
on these observables with future facilities. At the same time, as
already stressed in Sec.~\ref{sec:upsilon}, one should keep in mind
that there is a price to pay in terms of statistics, corresponding to
a factor about three (six) for inclusive (exclusive) measurements.  \\

\begin{boldmath}
\section{Accessing CP asymmetries in $B^0_d$--$\bar B^0_d$ events}
\end{boldmath}
\label{sec:BdBdbarphase}

In order to cover the physics program of a $B$-Factory running at the
$\Upsilon(4S)$ resonance, not only one should be able to identify
$B_d$ and charged $B$ mesons, but also to perform time-dependent
measurements of CP asymmetries in $B_d$ decays, which are based on the
fact that the initial $B \bar B$ pair is produced in a coherent state.
Moving to the $\Upsilon(5S)$ does not bring problems in terms of
vertexing and tagging, since the experimental environment is similar
to that of BaBar and Belle. In addition, as for the $\Upsilon(4S)$,
the $B_d \bar B_d$ pair is in a coherent state when the $\Upsilon(5S)$
decays in a PP final state. We show below that this is also true
for the dominant contribution, namely VV decays. We also discuss
how, using time-integrated measurements, PV decays could be used to
access the value of ${\mathcal Im}\lambda^f_{CP}$, with $\lambda^f_{CP} =
\frac{q}{p}\frac{\bar A_f}{A_f}$.

This last feature represent a novelty with respect to the study of
$\Upsilon(4S)$ decays and it allows to obtain an information
equivalent to the measurement of the coefficient $S$ of the
time-dependent CP asymmetry even for those decays for which a
measurement of the $B$ decay vertex, needed for time-dependent
measurements, cannot be performed. This is for instance the case of
$B_d \to \pi^0 \pi^0$ and $B_d \to \gamma \gamma$ decays, for which
otherwise only $\gamma$ conversion processes would allow to try a
time-dependent measurement, paying a big price in terms of
reconstruction efficiency.
\\
\begin{boldmath}
\subsection{The coherent time evolution of the $B^* \bar B^*$ mesons}
\label{sec:timeevolution}
\end{boldmath}

As explained in Sec.~\ref{sec:upsilon}, at the $\Upsilon(5S)$
resonance $B^* \bar B^*$ meson pairs are produced in a $J^{CP}=1^{--}$
state.  The question is whether the $B_q \bar B_q$ pairs resulting
from the $B_{q}^* \rightarrow B_{q} \gamma$ and $\bar B_{q}^*
\rightarrow \bar B_{q} \gamma$ decays exhibit the same properties of
quantum coherence of the initial state. Since the $B_{q}^* \rightarrow
B_{q} \gamma$ decay (and its CP conjugate) is an electromagnetic decay
and can be considered instantaneous, one is allowed to regard the $B_q
\bar B_q \gamma \gamma$ final state as a direct product of the
$\Upsilon(5S)$ resonance. This implies that the state constituted by
these four final particles has to preserve the initial quantum
numbers. Since the two photons have to satisfy the Bose-Einstein
statistic, their sum has to be a symmetric state. This request forces
the $B_q \bar B_q$ pairs to be in an antisymmetric state. This implies
that $B_q \bar B_q$ pairs produced in a VV event of a $\Upsilon(5S)$
decay have the same quantum coherence properties of $B^0 \bar B^0$
pairs generated at the $\Upsilon(4S)$ resonance. The full expression
for the decay amplitude is:
\begin{flalign}
A^{\mu} &= B_{s}(p_{1}+k_{1})\cdot B_{s}(p_{2}+k_{2})\cdot G_{3}\cdot \nonumber\\
&\cdot\Big(\frac{\epsilon(q,p_1,k_1,\epsilon_{1})\epsilon(q,p_2,k_2,\epsilon_{2})
  p_{1}^{\mu}}{2M_{B_{s}^{\ast}}^2}-\frac{\epsilon(q,p_1,k_1,\epsilon_{1})
  \epsilon(q,p_2,k_2,\epsilon_{2})p_{2}^{\mu}}{2M_{B_{s}^{\ast}}^2}
+\nonumber\\
&+\frac{\epsilon(q,p_1,k_1,\epsilon_{1})\epsilon(q,p_2,k_2,\epsilon_{2})k_{1}^{\mu}}
{2M_{B_{s}^{\ast}}^2}-\frac{\epsilon(q,p_1,k_1,\epsilon_{1})
  \epsilon(q,p_2,k_2,\epsilon_{2})k_{2}^{\mu}}{2M_{B_{s}^{\ast}}^2}
\Big) +\nonumber\\ 
&+ B_{s}(p_{1}+k_{1})\cdot B_{s}(p_{2}+k_{2})\cdot G_{2}\cdot \nonumber\\
&\cdot\Big(-\epsilon(q,p_1,k_1,\epsilon_{1})\epsilon(p_2,k_2,\epsilon_{2},\mu)
+\epsilon(q,p_2,k_2,\epsilon_{2})\epsilon(p_1,k_1,\epsilon_{1},\mu)+\nonumber\\
&+2\epsilon(p_1,k_1,\epsilon_{1},\sigma)\epsilon(p_2,k_2,\epsilon_{2},\sigma)
p_{1}^{\mu}-2\epsilon(p_1,k_1,\epsilon_{1},\sigma)\epsilon(p_2,k_2,
\epsilon_{2},\sigma)p_{2}^{\mu}+\nonumber
\end{flalign}
\begin{flalign}
&+2\epsilon(p_1,k_1,\epsilon_{1},\sigma)\epsilon(p_2,k_2,\epsilon_{2},
\sigma)k_{1}^{\mu}-2\epsilon(p_1,k_1,\epsilon_{1},\sigma)
\epsilon(p_2,k_2,\epsilon_{2},\sigma)k_{2}^{\mu}\Big)+\nonumber\\
&+ B_{s}(p_{1}+k_{2})\cdot B_{s}(p_{2}+k_{1})\cdot G_{3}\cdot \nonumber \\
&\cdot\Big(\frac{\epsilon(q,p_1,k_2,\epsilon_{2})\epsilon(q,p_2,k_1,
\epsilon_{1})p_{1}^{\mu}}{2M_{B_{s}^{\ast}}^2}-\frac{\epsilon(q,p_1,k_2,\epsilon_{2})
\epsilon(q,p_2,k_1,\epsilon_{1})p_{2}^{\mu}}{2M_{B_{s}^{\ast}}^2}
-\nonumber\\
&-\frac{\epsilon(q,p_1,k_2,\epsilon_{2})\epsilon(q,p_2,k_1,\epsilon_{1})k_{1}^{\mu}}
{2M_{B_{s}^{\ast}}^2}+\frac{\epsilon(q,p_1,k_2,\epsilon_{2})\epsilon(q,p_2,k_1,
\epsilon_{1})k_{2}^{\mu}}{2M_{B_{s}^{\ast}}^2}
\Big) +\nonumber\\ 
&+ B_{s}(p_{1}+k_{2})\cdot B_{s}(p_{2}+k_{1})\cdot G_{2}\cdot \nonumber\\
&\cdot\Big(-\epsilon(q,p_1,k_2,\epsilon_{2})\epsilon(p_2,k_1,\epsilon_{1},\mu)+
\epsilon(q,p_2,k_1,\epsilon_{1})\epsilon(p_1,k_2,\epsilon_{2},\mu)+\nonumber\\
&+2\epsilon(p_1,k_2,\epsilon_{2},\sigma)\epsilon(p_2,k_1,\epsilon_{1},\sigma)
p_{1}^{\mu}-2\epsilon(p_1,k_2,\epsilon_{2},\sigma)\epsilon(p_2,k_1,
\epsilon_{1},\sigma)p_{2}^{\mu}-\nonumber\\
&-2\epsilon(p_1,k_2,\epsilon_{2},\sigma)\epsilon(p_2,k_1,\epsilon_{1},\sigma)
k_{1}^{\mu}+2\epsilon(p_1,k_2,\epsilon_{2},\sigma)
\epsilon(p_2,k_1,\epsilon_{1},\sigma)k_{2}^{\mu}\Big),
\label{eq:timeev1}
\end{flalign}
where
\begin{eqnarray}
\nonumber && B_{s}(P) = \frac{1}{P^2-M_{B_{s}^{\ast}}^2+iM_{B_{s}^{\ast}}\Gamma_{B_{s}^{\ast}}} \\
\nonumber && \epsilon(a,b,c,d) = \epsilon_{\alpha \beta \gamma
\delta}a^{\alpha}b^{\beta}c^{\gamma}d^{\delta}=\epsilon^{\alpha \beta
\gamma \delta}a_{\alpha}b_{\beta}c_{\gamma}d_{\delta} \\
&&\epsilon(a,b,c,\sigma)\epsilon(d,e,f,\sigma) = \epsilon_{\alpha
\beta \gamma \sigma}\epsilon^{\delta \lambda \nu
\sigma}a^{\alpha}b^{\beta}c^{\gamma}d_{\delta}e_{\lambda}f_{\nu},
\label{eq:timeev2}
\end{eqnarray}
and $\mu$ is the index contracted by the $\Upsilon(5S)$ polarization
vector, $q$ is the $\Upsilon(5S)$ momentum, $p_1 (p_2)$ is the $B_{s}
(\bar B_{s})$ momentum, $k_1$ and $k_2$ ($\epsilon_1$ and
$\epsilon_2$) are the photons momenta (polarization vectors), and
$G_2$ and $G_3$ are two hadronic form factors ~\cite{brodsky}
\footnote{The $G_i$ form factors are defined from Eq.~(3.2)
of Ref.~\cite{brodsky} after imposing the relation $G_1$ = -2 $G_2$, 
which is valid for neutral mesons in the limit where the $Q^2$ dependence
in the charge form factor can be neglected.}.  
Among the variables entering
Eq.~(\ref{eq:timeev1}) and Eq.~(\ref{eq:timeev2}), $G_2$ and $G_3$ are
the only whose knowledge is not precise.  Different estimates exist in
literature~\cite{brodsky,ff}, but at present the theoretical
expectation is controversial. Anyhow, since the ratio of the two
determines the angular distribution of the $B^* \bar B^*$ mesons (and
so also of the $B_q \bar B_q$ pairs) in the $\Upsilon(5S)$ rest
frame, the study of the angular distribution of the final states in
$\Upsilon(5S)$ decays allows to experimentally ascertain them. In
any case, whatever the value of the two form factors is, the coherence
of the $B_q \bar B_q$ is guaranteed by Eq.~(\ref{eq:timeev1}).
\\
\begin{boldmath}
\subsection{Time-integrated CP asymmetries in  $B^* B$ events}
\end{boldmath}
\label{sec:bstarb}

When the $\Upsilon(5S)$ decays to $B^* \bar B$ events, the final $B
\bar B$ system is in a C$=+1$ state, after the decay $B^* \to B \gamma$.
This difference implies that, unlike the case of $\Upsilon(4S)$, for
each value of the time $t$, before one of the two $B$ mesons decays,
the two mesons are both $B_H$ or $B_L$ CP eigenstates. Integrating the
time out of the wave function of the $B \bar B$ pair, the
time-integrated CP asymmetry becomes~\cite{branco}:
\begin{equation}
A^f_{CP} = \left (\frac{1-y^2}{1+x^2}\right)^2
\frac{(1-x^2)(1-|\lambda^f_{CP}|^2)+4x {\mathcal Im}(\lambda^f_{CP})}
     {(1+y^2)(1+|\lambda^f_{CP}|^2)-4y {\mathcal Re}(\lambda^f_{CP})}
\label{eq:acpdirectBstB}
\end{equation}
where $x = \Delta m/\Gamma$, $y = \Delta \Gamma/2\Gamma$,
$\lambda^f_{CP} = \frac{q}{p}\frac{\bar A_f}{A_f}$, $q/p$ is the mixing
parameter of the $B$--$\bar B$ mixing, $A_f$ ($\bar A_f$) is the
amplitude for $B \to f$ ($\bar B \to f$) decays, and
$\eta_f$ is the CP eigenvalue of the final state $f$.

\begin{figure}[!tbp]
\begin{center}
\includegraphics[width=8cm]{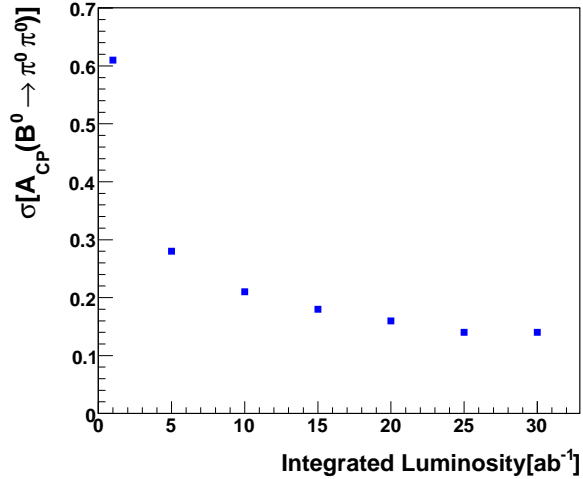}
\caption{\it Statistical error on the measurement of $A_{CP}(B_d \to \pi^0
\pi^0)$ for PV decays collected at the $\Upsilon(5S)$, as a function
of the integrated luminosity.\label{fig:aCP_PI0PI0}}
\end{center} 
\end{figure}

While this might not be relevant for those channels for which
$\lambda^f_{CP}$ can be determined at the $\Upsilon(4S)$, it opens new
perspectives in the study of CP asymmetries for $B_d$ decays to
neutral particles, among which two remarkable examples are $B_d \to
\gamma \gamma$ and $B_d \to \pi^0 \pi^0$.  $B_q \to \gamma \gamma$
decays are sensitive to NP effects in $b \to q$ transitions and they
are complementary to the measurement of $b \to q \gamma$ processes
(see Sec.~\ref{sec:bsgammagamma}). We do not discuss this specific
case, while we concentrate our attention to the impact of
measuring the direct CP asymmetry of $B_d \to \pi^0 \pi^0$ decays in
PV events.

Measurements of rate and asymmetry of $B_d \to \pi^0 \pi^0$ are
currently used in the isospin analysis of $B \to \pi \pi$ decays 
for extracting $\alpha$.  The isospin analysis provides eight
solutions, corresponding to the trigonometric ambiguities of the
isospin construction~\cite{glisospin}. Recently, it was shown that the
use of basic QCD properties allows to exclude some of the
solutions~\cite{utfitalpha}, still leaving several ambiguities in the
determination of $\alpha$.  The measurement of ${\mathcal Im}
(\lambda^f_{CP})$ from Eq.~(\ref{eq:acpdirectBstB}) allows to remove
some of these ambiguities, providing an additional information with
respect to BR's and CP asymmetries of the other channels.

\begin{figure}[tb!]
\begin{center}
\includegraphics[width=5cm]{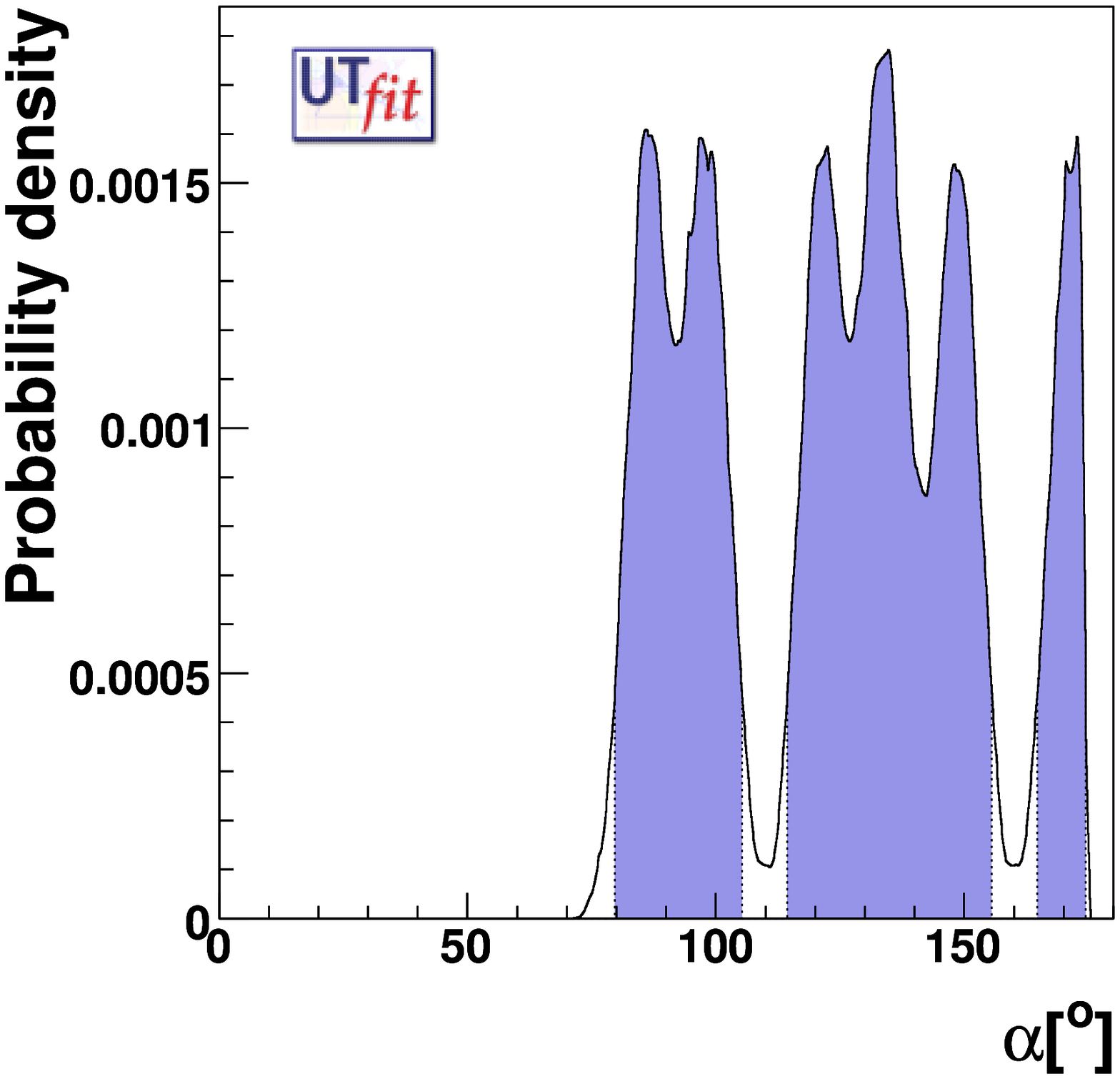}
\includegraphics[width=5cm]{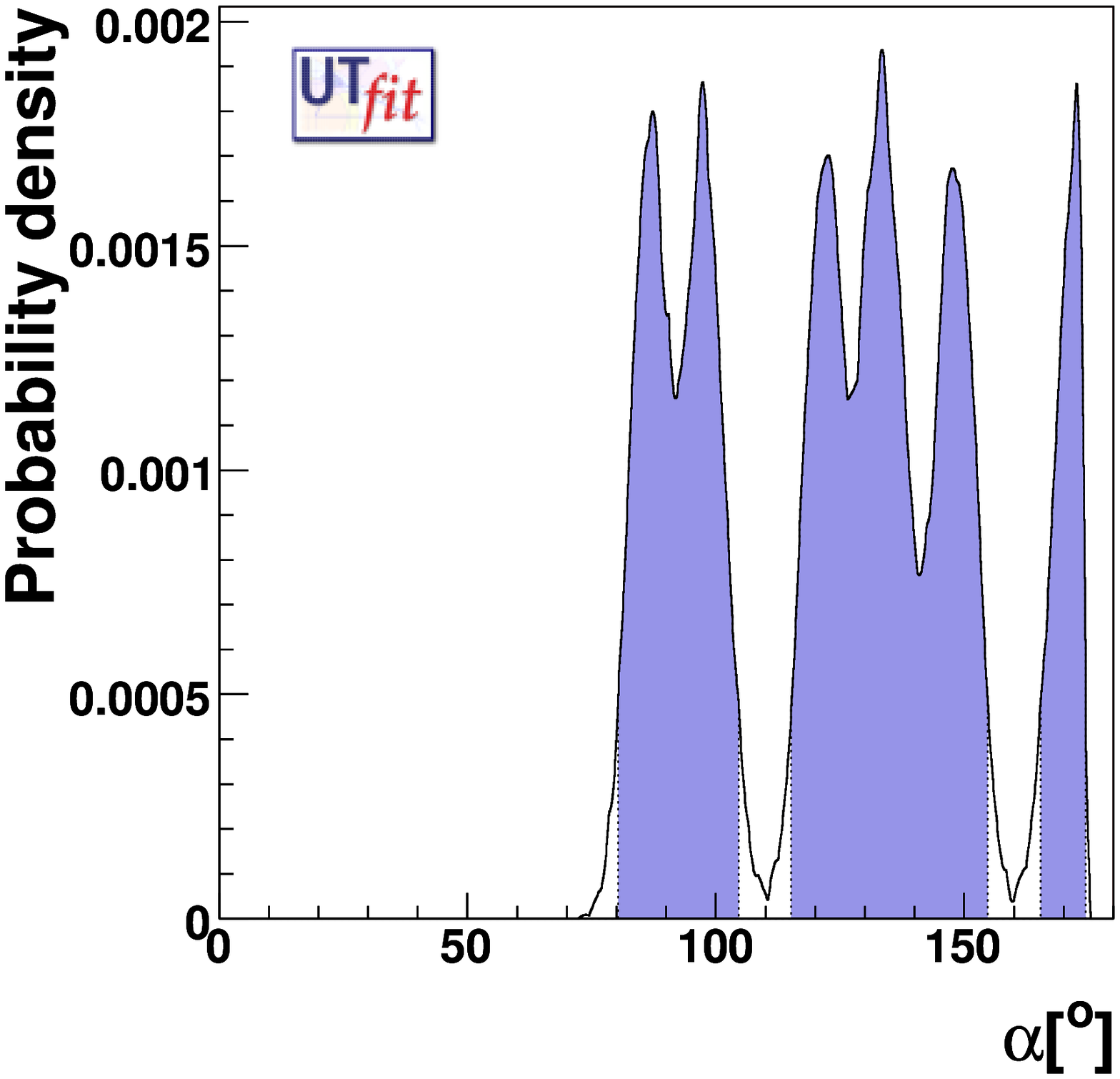}
\includegraphics[width=5cm]{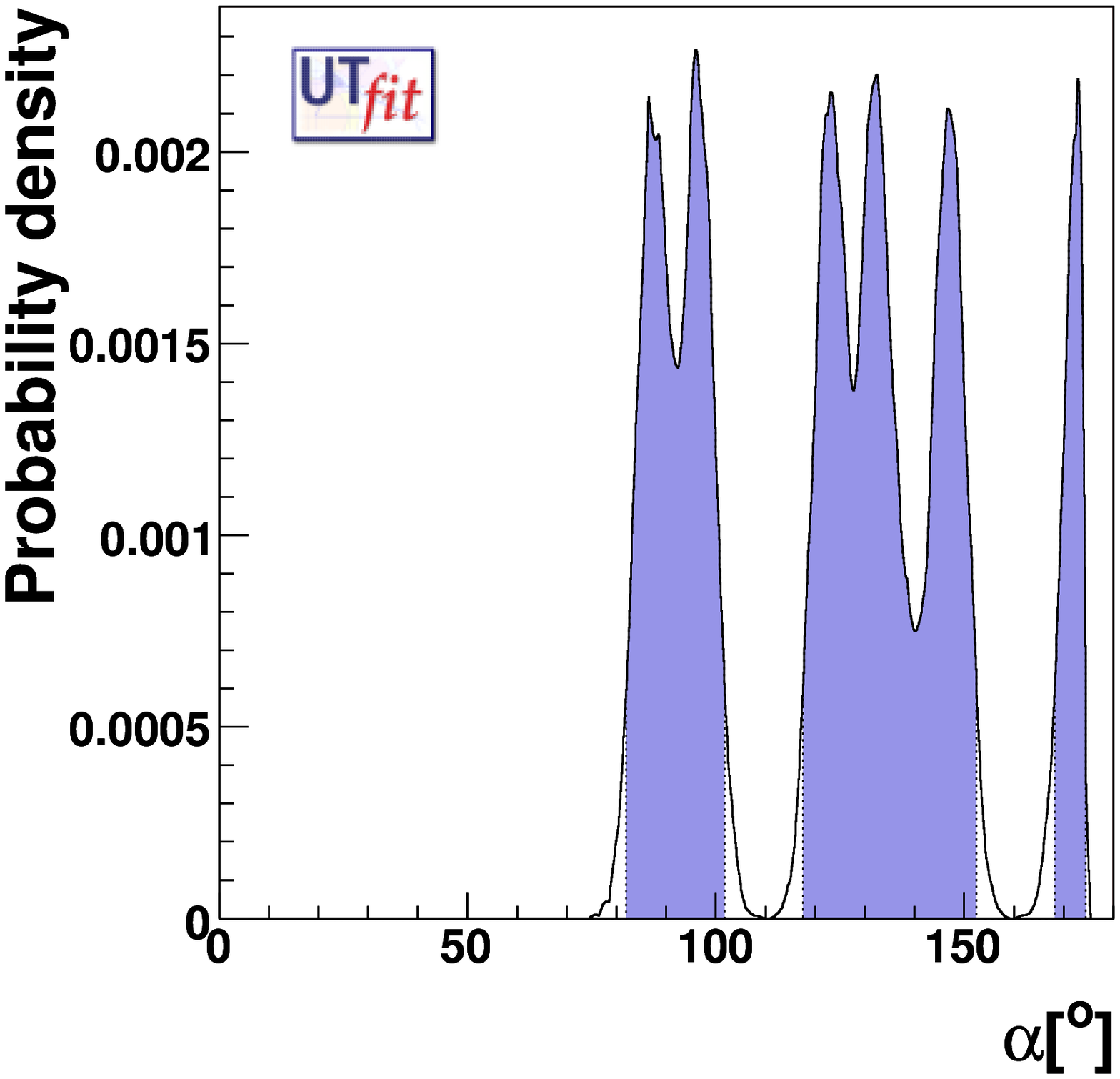}\\
\includegraphics[width=5cm]{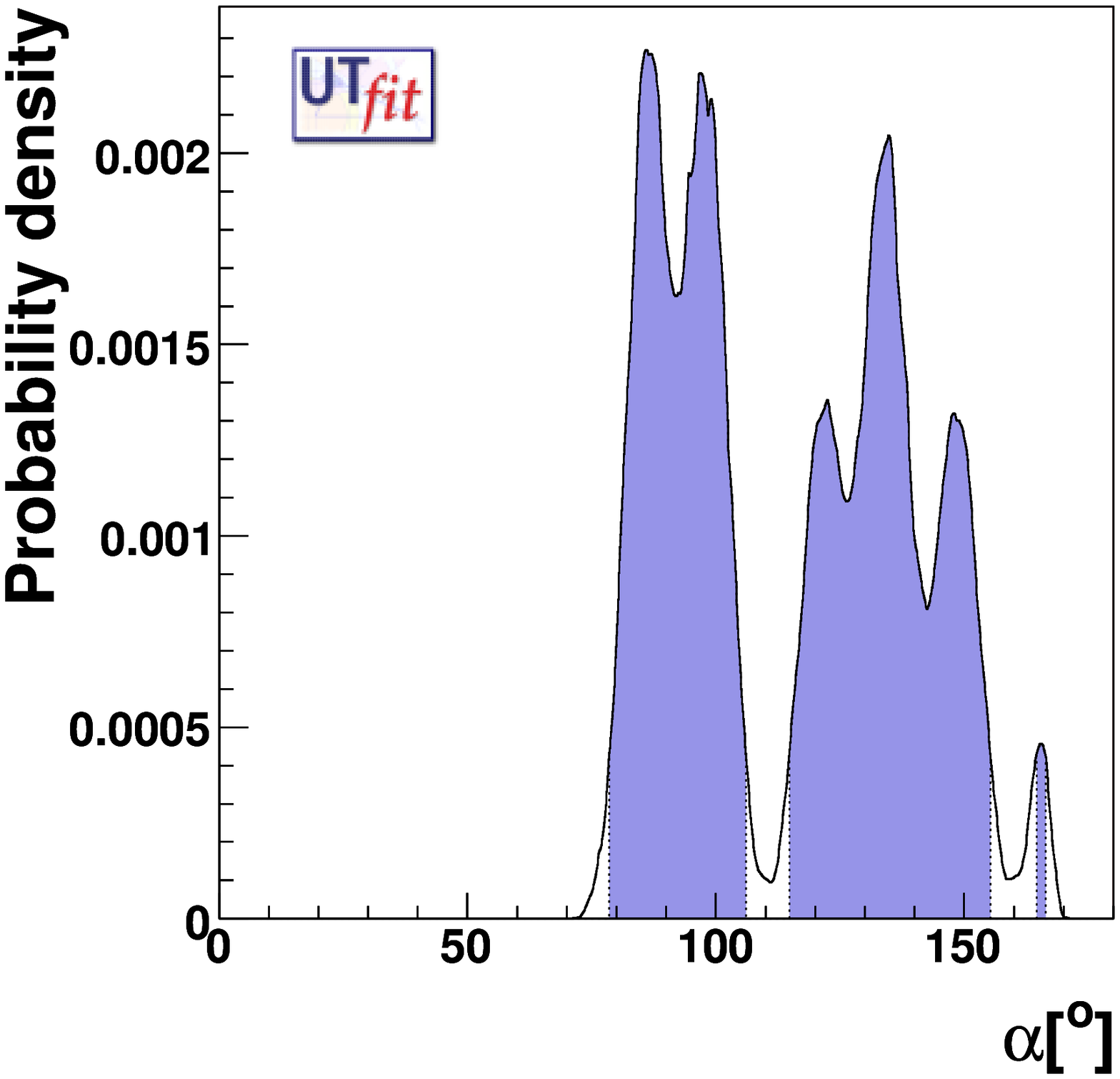}
\includegraphics[width=5cm]{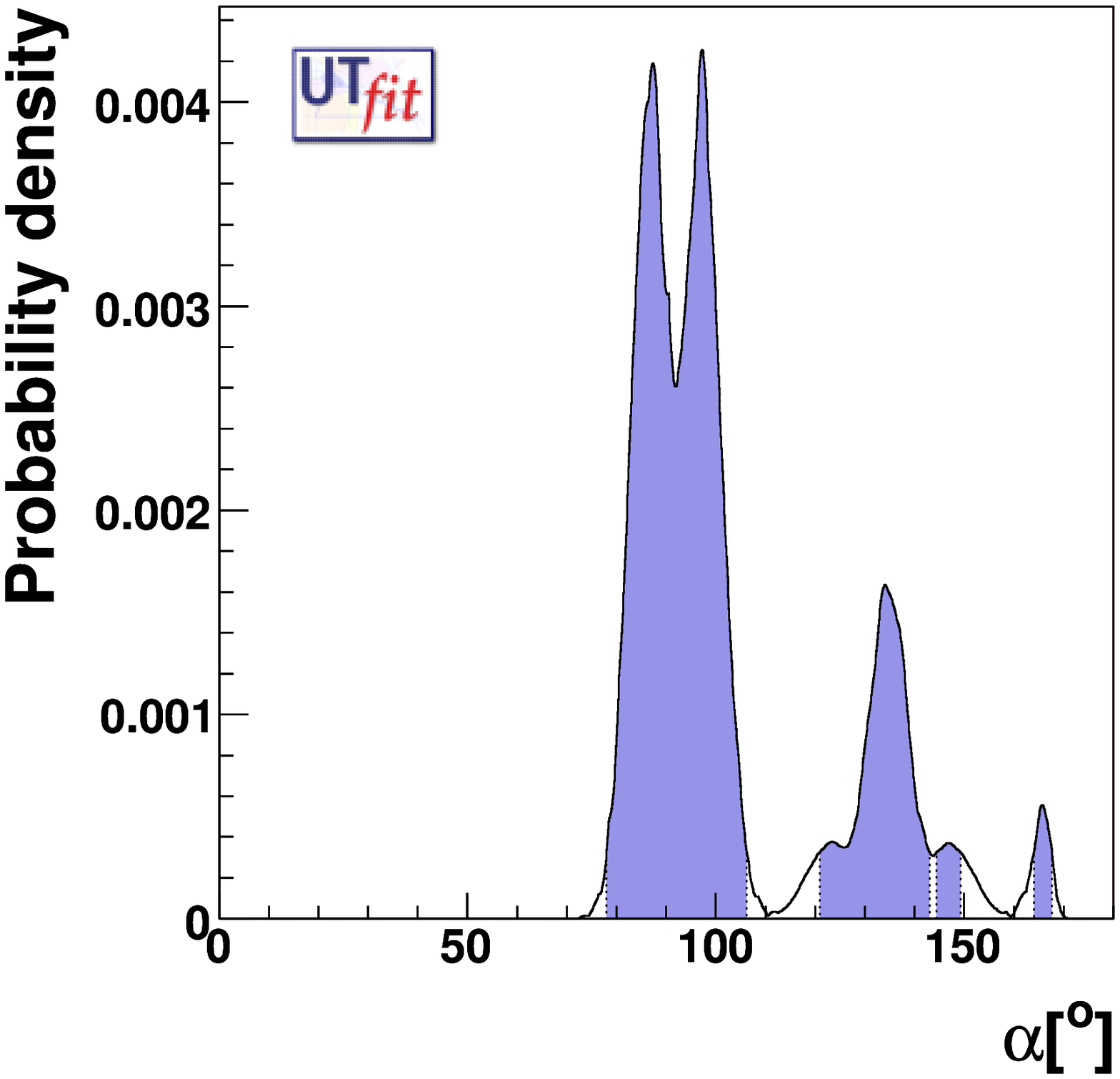}
\includegraphics[width=5cm]{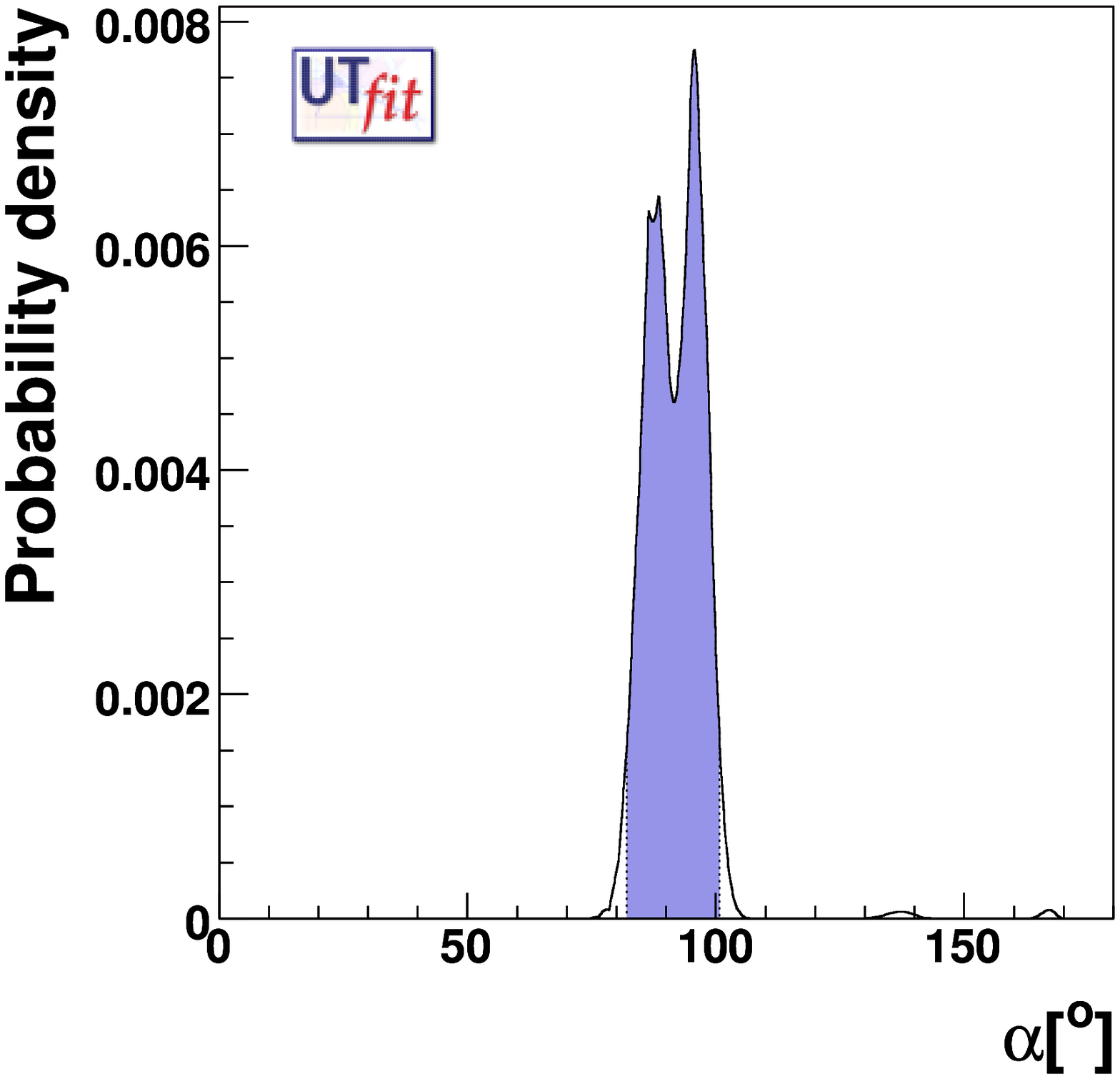}
\caption{\it Determination of the angle $\alpha$ without (top) or with
(bottom) the constrain from direct CP asymmetries of $B_d \to
\pi\pi$ decays in PV events, combining 1 ab$^{-1}$(left), 5
ab$^{-1}$(center) and 30 ab$^{-1}$(right) integrated at the
$\Upsilon(5S)$ for 2 ab$^{-1}$ collected at the $\Upsilon(4S)$ (end of
BaBar and Belle data taking).\label{fig:alpha}}
\end{center} 
\end{figure}

From an experimental point of view, the determination of the direct CP
asymmetry in PV decays is equivalent to what is done at the
$\Upsilon(4S)$ for the same decay channel~\cite{exppi0pi0}. Using
kinematic cuts and requirements on the energy deposits in the
electromagnetic calorimeter, signal events are isolated from continuum
background in a ML fit that uses topological and kinematic
variables. Using the information obtained applying the tagging
algorithm on the rest of the event, the flavour of the decaying $B$
meson is determined on a statistical basis, taking into account the
misidentification probability of the tagging algorithm. Scaling the
currently available BaBar result to the statistics available at a
$B$-Factory machine running at the $\Upsilon(5S)$ and assuming the same
tagging performances, we obtain the experimental error as a function
of the luminosity shown in Fig.~\ref{fig:aCP_PI0PI0}.

To illustrate the impact of this measurement on the isospin analysis,
we implemented this new measurement in both $B_d \to \pi^0 \pi^0$ and
$B_d \to \pi^+ \pi^-$ modes in the analysis by ${\bf
UT}fit$~\cite{utfitalpha}. The measurements currently available will
be improved using VV (PV and VV) events for CP asymmetries (decay
rates). We assume also that the results of the $\Upsilon(5S)$ runs
will be combined to the outcome of 2 ab$^{-1}$ collected by BaBar and
Belle at the end of their data taking.  At the same time, the
measurement of direct CP asymmetry for neutral $B$ decays in PV events
will provide the experimental determination of the observable in
Eq.~(\ref{eq:acpdirectBstB}).  While $A_{CP}(\pi^0\pi^0)$ provides a
new information, with respect to what is currently used in the isospin
analysis, $A_{CP}(\pi^+\pi^-)$ contributes to reduce the error on
${\mathcal Im}(\lambda_{CP})(\pi^+\pi^-)$, already constrained by
$S(\pi^+\pi^-)$.  For all the measurements, we use the current central values
and the errors are obtained scaling the statistical errors to the
assumed luminosity, without reducing the systematic error. The error
on the measurements of direct CP asymmetries in PV decays are obtained
scaling those of PP decays to the ratio of relative production
fractions (see Tab.~\ref{tab:Y5Sinput}).

The result in terms of the determination of the angle $\alpha$ is
shown in Fig.~\ref{fig:alpha} using the current isospin analysis (top)
and adding the information from PV decays (bottom), for three
different values of integrated luminosities. The plots clearly show
that the ambiguity is substantially reduced already with 5 ab$^{-1}$
of integrated data.
\\

\begin{boldmath}
\section{Accessing the $B^0_s$--$\bar B^0_s$ mixing phase}
\end{boldmath}
\label{sec:BsBsbarphase}

One of the main ingredients of the success of the current $B$-Factories
is the possibility of using the coherence of the initial $B$--$\bar B$
state to exploit the interference between CP violation in the decay
and in the mixing. Thanks to this technique the measurement of
$\sin2\beta$ has been possible with a high level of accuracy.

In the same way, one would like to exploit the coherence of the
$B^{*}_s$--$\bar B^{*}_s$ final state at $\Upsilon(5S)$. 
The paradigmatic physics goal
in this field is the determination of the weak phase $\beta_s$ using
$B_s \to J/\psi \phi$ decays.

As described in the following, this does not seem achievable with the
precision of the current vertexing detectors and the design parameters
of the next-generation $B$-Factories, currently under
discussion~\cite{superBcdr}. The main problem comes from the large value
of the ratio $\Delta m_s/\Delta m_d = 35.0 \pm 0.4$, which requires
a presently unreachable vertexing resolution to be sensitive to $B_s$
oscillations.

Nevertheless, information on the mixing phases can still be obtained
using time-integrated measurements. In general, comparing the tagged
decay rates for $B_s$ decays into CP eigenstates $f$ for positive and
negative values of $\Delta t$, it is possible to determine the value
of ${\mathcal Re} \lambda^f_{CP}$ and ${\mathcal Im} \lambda^f_{CP}$.
Considering different decay modes, several combinations
of weak phases can be extracted. In particular, we consider
the case of the determination of $\beta_s$ from tree-level $B^0 \to
J/\psi \phi$ decays and from penguin dominated $B_s \to K^0 \bar K^0$
decays. In addition, complementary information on $\beta_s$ can be
obtained from the charge asymmetry in flavour specific final states or in
dimuon events, or from the angular analysis of $B^0 \to J/\psi \phi$
decays.
\\
\subsection{Time-dependent CP asymmetry}\label{sec:timedepcpasymmetry}

The study of time-dependent CP asymmetries is one of the milestones of the 
physics program for currently running $B$-Factories. It is based on the
simple idea that, provided that the laboratory frame is
sufficiently boosted with respect to the CM frame of the $\Upsilon$
resonance, it is possible to access the $B_q$ oscillation even though the
vertex resolution is smaller than $\Delta m_q$, thanks to the $\beta
\gamma$ Lorentz factor~\cite{Oddone}. 
In the case of $B_d$--$\bar B_d$ oscillation,
a boost of $\beta\gamma \sim 0.5$ was large enough to allow the
measurement of $\sin 2 \beta$ with the vertex precision available when
BaBar and Belle were designed. With the currently available technology
on silicon vertex detectors, the boost could be in principle reduced
by a factor of two or the resolution improved by the same factor using the
current boost. Nevertheless, this is not enough for $B_s$--$\bar
B_s$ oscillations, which are too much faster than in the case of $B_d$
mesons.

We used Monte Carlo simulated experiments (\emph{toy Monte Carlo experiments}) 
to identify the minimal vertexing resolution needed to detect 
$B_s$--$\bar B_s$ oscillations at a
$B$-Factory. We quantify the minimal resolution as the largest value
of the $\Delta t$ resolution that allows to extract the $S$ and $C$
parameters of the time-dependent CP asymmetry of $B_s \to J/\psi \phi$
decays with an accuracy comparable to the precision on $\sin 2 \beta$.
In this way, it is possible to provide an answer to the problem
without necessarily relying on a particular set of machine and
detector parameters. The result obtained can then be translated into a
requirement on a given machine, once the energies of the beams (that
fix the Lorentz boost) are defined. In fact, the precision on
$\sigma(\Delta t)$ can be written in terms of the resolution of the
vertex detector $\sigma(\Delta z)$ and the Lorentz boost $\beta\gamma$
of the CM frame with respect to the laboratory frame, by using
the relation $\sigma(\Delta z) \sim \sigma(\Delta t)\beta\gamma c$.

We determine the shape of signal and background using fully simulated
samples, selected requiring~\cite{jpisphi_babar}:
\begin{itemize}
\item $3.06$ GeV $<m_{J/\psi}<3.14$ GeV for $J/\psi \to \mu^+
\mu^-$
\item $2.95$ GeV $<m_{J/\psi}<3.14$ GeV for $J/\psi \to e^+ e^-$
\item $1.004$ GeV $<m_{\phi}<1.034$ GeV
\item particle identification (PID) requirements based on Cherenkov
      angle and $dE/dx$, both on $J/\psi$ and $\phi$ daughters.
\item a cut on the polar angle of the thrust axis:
$|\cos(\theta_{Thr})|<0.8$
\end{itemize}
in order to reduce contamination from $q \bar q$ background and from
tracks combinatoric.  We also apply the following kinematic
requirements:
\begin{itemize}
\item $5.440$ GeV $< m^s_{miss} < 5.490$ GeV 
\item $-0.2$ GeV $<\Delta E< 0.1$ GeV
\end{itemize}
to select an almost pure sample of $\Upsilon(5S) \to B_s^*B_s^*$
events.  This selection has a global efficiency of $\sim 14.7 \%$ on
signal events. We generate and fit samples containing signal and
background events (scaled to the considered luminosity),
according to the following likelihood function:
\begin{eqnarray}
  {\mathcal L} = \frac{e^{-(N_{sig}+N_{bkg})/N}}
  {^N\sqrt{(N_{sig}+N_{bkg})!}}  &\cdot& \prod_{i=1}^N \{ N_{sig}\cdot
  P_{sig}(m_{miss})_{i} \cdot P_{sig}(\Delta E)_{i} \cdot
  P_{sig}({\cal F})_{i} \cdot P_{sig}(\Delta t)_{i} +\nonumber \\
  &&N_{bkg}\cdot P_{bkg}(m_{miss})_{i} \cdot P_{bkg}(\Delta E)_{i}
  \cdot P_{bkg}({\cal F})_{i} \cdot P_{bkg}(\Delta t)_{i} \},
\label{eq:jpsiphilik}
\end{eqnarray}
where $P_{sig}(x)$ ($P_{bkg}(x)$) is the probability density function
(PDF) of the variable $x$ for the signal ($q \bar q$ background)
component. Background events from other $B \bar B$ decays are
negligible after the selection cuts.  $P_{sig}(\Delta t)$ takes into
account the knowledge of the proper time of $B$ mesons, coming from
the vertex reconstruction.  The full expression uses the convolution
of the angular distribution function with the resolution function
associated to the vertex reconstruction (usually, the sum of three Gaussians,
with mean and width scaled according to the per-event error
$\sigma_{\Delta t}$). We assumed the models adopted for BaBar time-dependent
measurements~\cite{babarRF}, for both the shape of $\Delta_t$
and the performances of the tagging algorithm.  

All the other functions are used to separate signal from background.
Their shapes are obtained from unbinned maximum-likelihood fits to
signal and background samples, coming from full Monte Carlo
simulations.

The signal distributions for $m_{miss}^s$, $\Delta E$, and the Fisher
discriminant ${\cal F}$ are parameterized as:
\begin{eqnarray}
f(x) = \exp
\Big[-\frac{(x-m)^2}{2\sigma^2_{\pm}+\alpha_{\pm}(x-m)^2}\Big]
\label{eq:Cruijff}
\end{eqnarray}
where $m$ is the maximum of the distribution, while $\sigma_\pm$ and
$\alpha_\pm$ quantify the width and the tail, the positive
(negative) sign corresponding to positive (negative) values of
$x-m$. The invariant masses of the two resonances, $m_\phi$ and
$m_{J/\psi}$ are described by relativistic Breit-Wigner
functions.

In order to describe the $q \bar q$ background, we use a second-order
polynomial for $\Delta E$, while ${\cal F}$ is given by the function
of Eq.~(\ref{eq:Cruijff}) and $m_{miss}^s$ is parameterized by a phase-space
threshold function~\cite{ARGUS}:
\begin{eqnarray}
f(x) = x\sqrt{1-x^2} \exp \Big[- \xi \cdot(1-x^2)\Big]~~~~~;~~~~~~x =
m_{miss}^s/m_T
\label{eq:argus}
\end{eqnarray}
where $m_T$ is the value of the threshold and $\xi$ is the parameter
determining the shape. For $\Delta t$ we use a Gaussian resolution
function.

The $B_d$ background is parameterized using the function of
Eq.~(\ref{eq:Cruijff}) for $\Delta E$, and the phase-space threshold
function (as in Eq.~(\ref{eq:argus})) for $m_{miss}^{s}$. All the other
PDF's are similar to signal ones.

\begin{figure}[!tbp]
\begin{center}
\includegraphics[width=9cm]{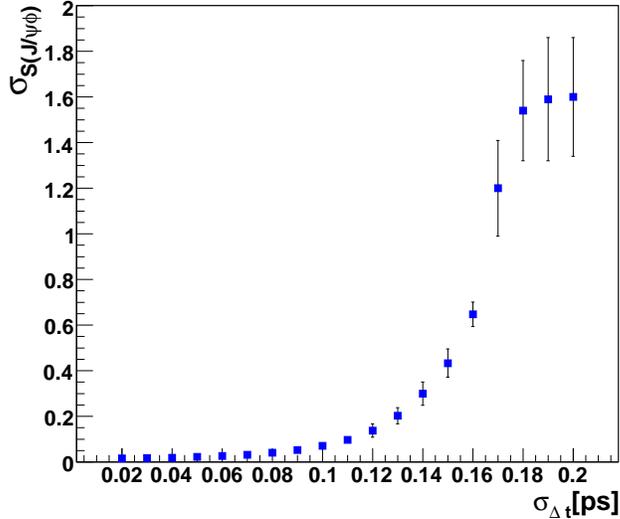}
\caption{\it Error on $S(J/\psi \phi)$ as a function of the $\Delta t$
resolution $\sigma_{\Delta t}$.  Error bars represent the RMS of the
error distribution for a given set of toys.\label{fig:SCjpsiphi}}
\end{center}
\end{figure}

For simplicity, we assume a fully polarized final state, which allows
us to discard from the likelihood the PDF describing the angular
distribution, and we parameterize the resolution function with a single Gaussian
function, centered around $-0.2$~ps$^{-1}$ (to include the effect of the
charm bias on the tag side). Here, we do not use the per-event error but an
average value $\sigma_{\Delta t}$ as the RMS of the resolution function, varied
in the range [0.02,0.20] ps$^{-1}$. We assume an integrated luminosity
of $30~ab^{-1}$ and we generate and fit a set of toy Monte Carlo
samples for each value of $\sigma_{\Delta t}$.  The result of this
study is reported in Fig.~\ref{fig:SCjpsiphi} for the error on $S$.
The error on $C$ is not shown, since the values of the errors on $S$
and $C$ are the same for all the chosen values of $\sigma_{\Delta t}$.

It is clear from the plot that, in order to achieve an acceptable
precision on the value of the CP parameters, values of $\sigma(\Delta
t) \leq 0.11 $~ps are needed. We also found that for values larger
than $0.11 $~ps not only the average error increases, but also a bias
is introduced in the fit, generating a larger spread in the error
distribution (the error bars in the plot).  For these values of
$\sigma(\Delta t)$ the $B_s$ oscillation is too fast to be
detected. The bad news is that the values of resolution required are
beyond the possibilities of the current designs. The existing
$B$-Factories are outside the range of the plot, having $\sigma(\Delta
t) \sim 0.65 $~ps. The improvements coming from new technology,
together with the possibility of adding to the vertex detector a {\it layer
0}~\cite{CDFNIM} on top of the beam pipe, allows to push $\sigma(\Delta t)$ down
to $\sim 0.4$~ps for a boost of $\beta\gamma \sim 0.28$~\cite{talkneri},
corresponding to $\sim 0.2 $~ps for a boost of $\beta\gamma \sim 0.6$.

We stress the fact that statistics alone does not
allow to reduce the errors in a significative way, since the limiting
factor is in the vertexing resolution and does not have a statistical
nature (as for instance in the case of decay modes with large
backgrounds).  \\
\begin{boldmath}
\subsection{Tagged rates at positive and negative $\Delta t$}
\end{boldmath}

Even assuming that the needed time resolution for the study of
$B_s$--$\bar B_s$ oscillations is outside the capability of the
next-generation $B$-Factories, it remains possible to measure weak phases in
$B_s$ decays measuring the sign of $\Delta t$ for $B_s$--$\bar B_s$
meson pairs. In fact, in this case one can measure decay rates as a
function of $B_s$ flavour and $\Delta t$ sign, which provide four
experimental observables, depending on the weak phase of the chosen
final state. Using this technique, it is possible to access the
$B_s$--$\bar B_s$ mixing phase $\beta_s$ in $B_s \to J/\psi \phi$
decays, as well as combinations of this phase and the phase $\gamma$
of the CKM matrix (as for the case of $B_s \to D_s K$, which we do not
discuss here).  In this section, we introduce the basic formalism and
we provide two examples of such measurements.  \\
\subsubsection{Theoretical Formalism and Experimental Details}\label{sec:formalism}

Let us consider a $B_s$ pair produced at the $\Upsilon(5S)$ resonance,
through a $B_s^* \bar B_s^*$ state. If one of the two $B_s$ decays into
a CP eigenstate with eigenvalue $\eta_f$ and the other one in a flavour
tagging final state, the PDF of the proper time difference $\Delta t$
can be written as:
\begin{equation}
P(\Delta t) \propto e^\frac{-|\Delta t|}{\tau}\left[ \kappa_1
\cosh\left (\frac{\Delta \Gamma_s \Delta t}{2}\right ) + \kappa_2
\cos\left (\Delta m_s \Delta t\right ) + \kappa_3 \sinh\left
(\frac{\Delta \Gamma_s \Delta t}{2}\right ) +\kappa_4 \sin\left (\Delta
m_s \Delta t\right ) \right],\label{eq:prob_dt}
\end{equation}
where the $\kappa_i$ coefficients are defined as:
\begin{eqnarray}
\kappa_1 = \frac{1}{2}(1+|\lambda^{f}_{CP}|^2) ~~~~~&;&~~~~~ 
\kappa_2 = -q_{tag}\frac{1}{2}(1-|\lambda^{f}_{CP}|^2) \nonumber \\
\kappa_3 = -{\mathcal Re}\lambda^{f}_{CP} ~~~~~&;&~~~~~ 
\kappa_4 = -q_{tag}{\mathcal Im}\lambda^{f}_{CP}.
\end{eqnarray}
These four quantities are effective parameters that depend on the tag
sign $q_{tag}$ ($+1$ ($-1$) when the tag meson is a $B_s$ ($\bar
B_s$)) and on the CP parameter $\lambda^f_{CP}$.  We use, as a good
approximation, $q/p = e^{-2i\beta_s}$ in the following.

In the case of the current $B$-Factories, it is usually imposed that
$\Delta \Gamma \sim 0$, which holds with a very good accuracy for
$B_d$ mesons. This assumption reduces the number of parameters to two
(the usual $S$ and $C$). In the case of time-integrated measurements,
the tagged rates are measured without any requirement on $\Delta t$,
which is equivalent to integrating Eq.~(\ref{eq:prob_dt}) between
$-\infty$ and $\infty$. In the case of time-dependent measurements,
the convolution between Eq.~(\ref{eq:prob_dt}) and the resolution
function of the vertex detector is used in the fit, which allows to
extract the $S$ and $C$ coefficients.

As discussed in the previous section, it is hard to imagine that
time-dependent measurements of $B_s$ decays will be performed at a
$B$-Factory. On the other hand, the presence of $\Delta t$-odd terms
introduces an asymmetry between the numbers of events with $\Delta t >
0$ and $\Delta t < 0$, depending on $\lambda^f_{CP}$ and $q_{tag}$. In
addition, the sign of $\Delta t$ can be determined with good precision
even without improving the vertex resolution, since $\tau_{B_s} \sim
\tau_{B_d}$.  It is then possible to extract the four parameters
$\kappa_i$ from the tagged decay rates for positive and negative
values of $\Delta t$, the dependence on the $\kappa_i$ being provided
by the integral of Eq.~(\ref{eq:prob_dt}) and the resolution function
of the vertex detector in the range $[-\infty,0]$ and $[0,\infty]$.
As an alternative, one can measure a flavour integrated rate and quote
flavour asymmetries (as currently done by the $B$-Factories), which
allows to cancel some of the systematics.

Exploiting the technique described above, it is possible to measure
the parameter $\lambda^{f}_{CP}$ for all CP eigenstates of the $B_s$,
without any need of improving the vertex resolution.  Although this
procedure is not as powerful as the time-dependent measurements
performed by experiments at hadron collider, it can be applied to all
the channels for which the $\Delta t$ sign can be detected. Given the
very clean environment of $e^+$--$e^-$ machines, the set of accessible
channels is larger than in the case of LHCb. In particular, since the
position of the $B$ decay vertex can be measured even using $K^0_S \to
\pi^+ \pi^-$ rather than prompt tracks~\cite{kspi0paper,k0pi0primo},
the possibility of accessing $\lambda^{f}_{CP}$ even in these cases
opens up interesting possibilities.

Covering all the possibilities goes beyond the purpose of this
paper. Nevertheless, we give below two examples of the potentiality of
this new technique.
\\
\begin{boldmath}
\subsubsection{Determination of $\beta_s$ from tree-level processes}
\end{boldmath}
\label{sec:betas_dtsign}

The study of $B_s$ decays at hadron colliders allows to determine
the absolute value and phase of the $B_s$--$\bar B_s$ mixing
amplitude. The comparison of the experimental measurements to the
theoretical expectations allows to test the presence of NP in $b \to
s$ transitions. The recent measurement of $\Delta m_s$~\cite{dmsexp}
already provided the first milestone of this physics program.  A first
measurement of the mixing phase is available from
D$\O$~\cite{D0phase}, obtained from a time-dependent angular analysis
of $B_s \to J/\psi \phi$ decays.  

A large improvement will come from LHCb~\cite{lhcbtdr}. In fact, this 
experiment is expected to measure, with the same technique, both the 
width difference and the phase up to 
$\sigma(\Delta \Gamma_s/\Gamma_s) = 0.0092$ and 
$\sigma(\beta_s) = 0.023$~\cite{schune_paris}
in just one year of nominal data taking. With the full expected B physics
dataset (10 $fb^{-1}$) the error on $\beta_s$ can be reduced to
 $\sigma(\beta_s) \simeq 0.01$. It is clear that such a precision
cannot be achieved with the Lorentz boost available at present $e^+e^-$
facilities and the resolution of actual vertexing detectors
(see Sec.~\ref{sec:timedepcpasymmetry}). Nonetheless, we will show how 
a $B$-Factory running at $\Upsilon(5S)$ could provide alternative and
independent informations on the same quantities, even if with not the
same accuracy.


In fact, one can use the approach described
in Sec.~\ref{sec:formalism} to measure the four $\kappa_i$ parameters
in the case of $B_s \to J/\psi \phi$ and then extract from them a
value for $\beta_s$. As for the measurement of $\beta$ from $B_d \to
J/\psi K^0_S$, one can neglect CKM suppressed subleading 
contributions~\footnote{As it is done
for $B_d \to J/\psi K^0_S$~\cite{CPS_beta}, the impact of the CKM
suppressed subleading contributions can be obtained from the
measurements of $\lambda^{f}_{CP}$ for similar channels, such $B_s \to J/\psi
K^{*0}$} and assume that $\lambda^{f}_{CP} = \eta_f e^{2i\beta_s}$.

\begin{figure}[!tbp]
\begin{center}
\includegraphics[width=7cm]{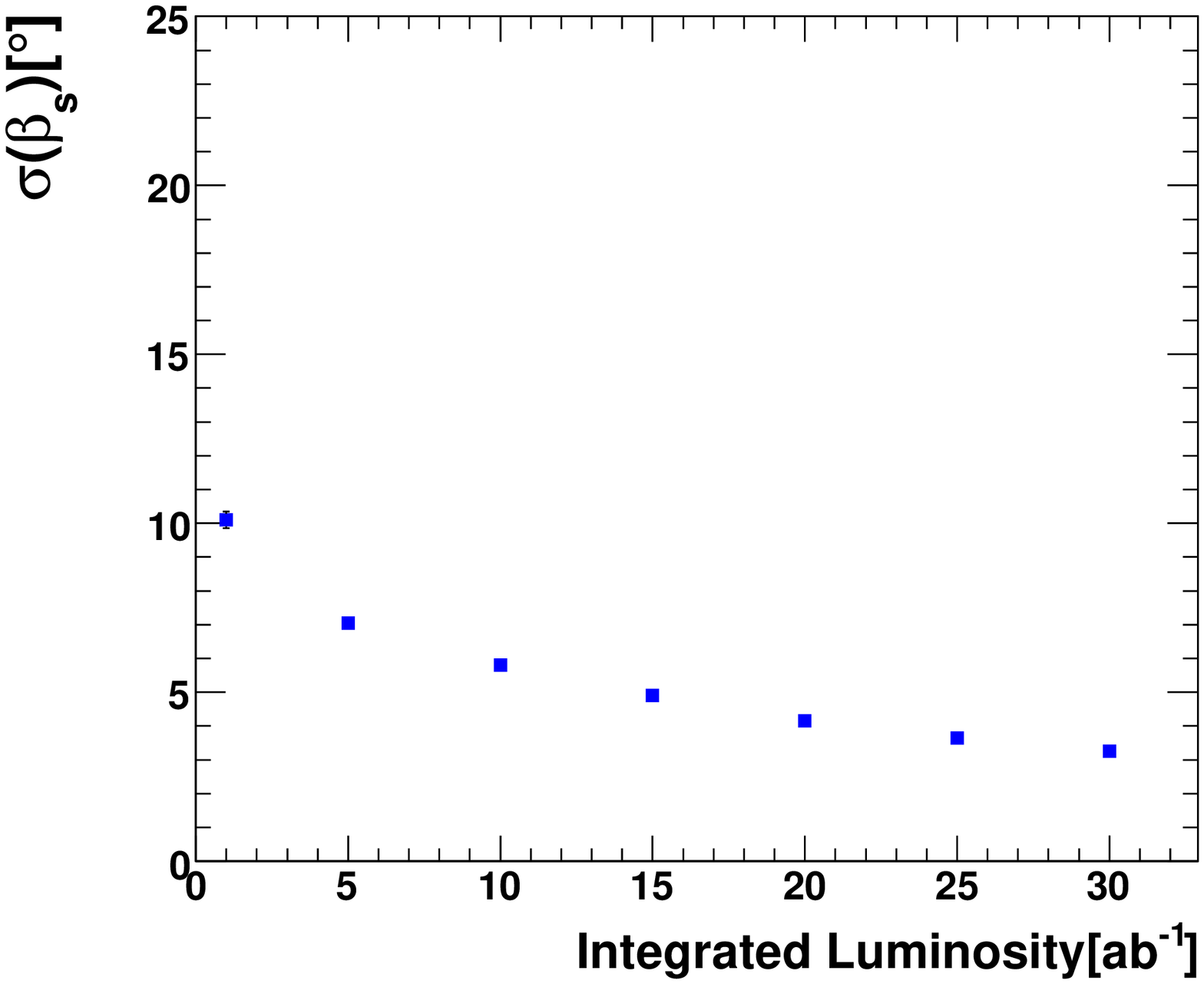}
\includegraphics[width=7cm]{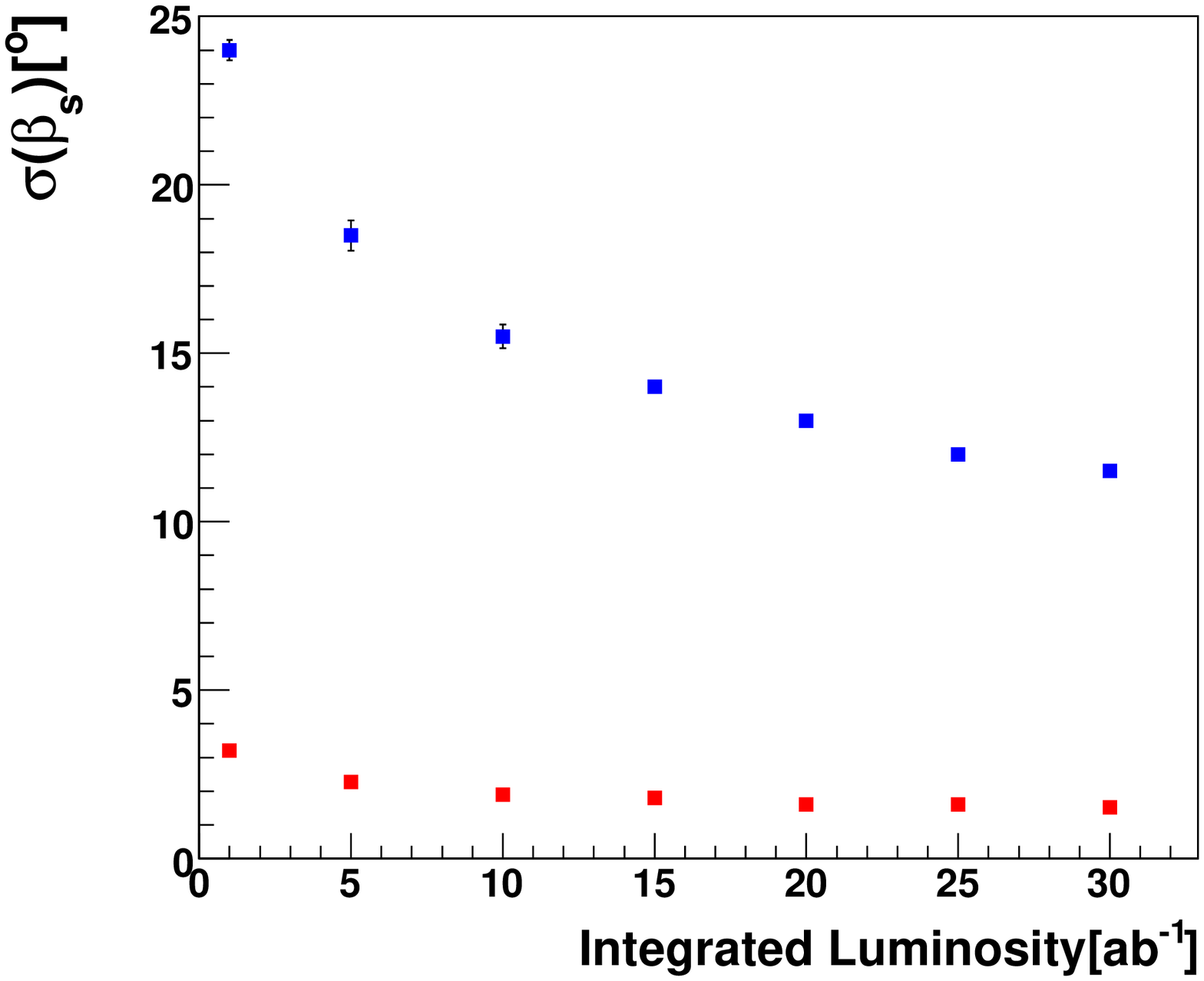}
\caption{\it Distribution of the error on $\beta_s$ from $B_s \to J/\psi
\phi$ (left) and $B_s \to K^0 \bar K^0$ (right) decays, as a function
of the integrated luminosity, obtained from the measurement of tagged
rates with $\Delta t>0$ and $\Delta t<0$. For the right plot,
the estimate of the theoretical error is also given (see text for
details).\label{fig:betas_tdsign}}
\end{center}
\end{figure}

In order to evaluate the sensitivity to $\beta_s$, we used the same
framework described in Sec.~\ref{sec:timedepcpasymmetry} to perform a
set of toy Monte Carlo experiments, as function of the integrated
luminosity.  The statistical error as a function of the luminosity,
shown in the left plot of Fig.~\ref{fig:betas_tdsign}, clearly proves
the potentiality of this technique and the possibility to measure
$\beta_s$ with a relatively good precision. We found a 
two-fold ambiguity between $\beta_s$ and $-\beta_s$. When the value of .
$\beta_s$ is close to zero (as it should be in the SM), two different 
and partially superimposed peaks appear, which produce a (almost) two-times 
larger resolution in the total pdf. 
Anyway, it is important to stress that a similar approach can be followed for 
$B_s$ decays into higher charmonium resonances, which provides independent 
determinations of $\beta_s$, taking into account the different hadronic 
uncertainties.
\\
\begin{boldmath}
\subsubsection{Determination of $\beta_s$ from penguin modes}
\end{boldmath}
\label{sec:beta_sPen}

The same experimental technique can be applied also to determine
 $\beta_s$ from penguin modes. The comparison of this result to the
 value measured in $B_s \to J/\psi \phi$ allows to test the
 consistency of the SM and to constrain NP models.

A similar test is usually performed in the $B_d$ sector comparing the
value of $\sin 2\beta$ from $B_d \to J/\psi K^0_S$ to the $S$ value of
penguin dominated modes, such as $B_d \to \phi K^0_S$. 
The equality $(\sin 2\beta)_{J/ \psi K^0_S} = (\sin 2\beta)_{penguin}$ 
would be strictly 
 true only if these decays were mediated by a single
combination of CKM matrix elements, while in all the cases a CKM
suppressed amplitude is present.  At the very high precision that
is expected with the next generation of $B$ physics experiments, it is
not be possible to neglect the CKM suppressed contribution, which
introduces a theoretical error associated to the SM expectation of
$\Delta S = S(b\to s \bar q q) - \sin 2 \beta$.

The same problem is present in the case of $B_s \to K^0 \bar
K^0$, since the decay amplitude in terms of renormalization group
invariant parameters is given by the relation~\cite{burassilv}:
\newcommand{\sss}{\scriptscriptstyle} \newcommand{\nn}{\nonumber}
\begin{eqnarray}
  {\cal A}(B_s \to K^0 \bar K^0) = &-& V_{us} V_{ub}^* P_s^{\sss {\rm GIM}} - V_{ts} V_{tb}^*P_s, 
\label{eq:BsKsKs}
\end{eqnarray}
which is formally equivalent to the amplitude of the {\it golden mode}
$B_d \to \phi K^0$. The main advantage of considering this decay 
is that the theoretical error can be estimated using a data-driven 
approach~\cite{CPS_beta,Ciuchini:2007hx}, as we explain below.

As a first step, we assume that $P_s^{\sss {\rm GIM}}$ can be
neglected, which implies that $\lambda^{f}_{CP}=\eta_f e^{-2i\beta_s}$.  We use a
set of toy Monte Carlo experiments to extract the expected error on
the weak phase, applying the method described in
Sec.~\ref{sec:formalism}. We determine the error on $\beta_s$ as a
function of the integrated luminosity, as shown in the right plot of
Fig.~\ref{fig:betas_tdsign}.

Comparing the two plots of Fig.~\ref{fig:betas_tdsign}, it is clear
that $B_s \to K^0 \bar K^0$ is affected by a larger background, which
increases the error on the weak phase, the dominant contamination
coming from $q \bar q$ events ($q=u,d,s,c$). These events are
characterized by a jet-like distribution in the center of mass of the
$e^+e^-$ system, which allows to separate them from signal events. At
a new facility, the discriminating power might benefit from the
improvement of the vertexing resolution.  Using modern technology, it
is possible to improve the vertexing
precision~\cite{neri_pierini_superb}, allowing to separate the $B$ and
$D$ vertices on the tag side of a $B \bar B$ event. Since no secondary
vertex is present in $q \bar q$ events, it is not unrealistic to
imagine that such an approach will strongly suppress the background
contamination, reducing the error on $\beta_s$. A quantification of
this improvement relies on a specific detector design and goes beyond
the purpose of this paper.

The estimate of the error induced by neglecting $P_s^{\sss {\rm
GIM}}$ can be obtained considering the time-dependent study of $B_d
\to K^0 \bar K^0$.  The decay amplitude is given by
\begin{eqnarray}
  {\cal A}(B_d \to K^0 \bar K^0) = &-& V_{ud} V_{ub}^* P_d^{\sss {\rm GIM}} - V_{td} V_{tb}^*P_d, 
\label{eq:BdKsKs}
\end{eqnarray}
which is similar to Eq.~(\ref{eq:BsKsKs}), except that here
the two combinations of CKM elements have the same order of
magnitude. This difference maximizes the sensitivity to the ratio
$P^{GIM}_d/P_d$, which can be determined assuming the SM values of the
CKM parameters and using the experimental values of the BR and the CP
parameters $S$ and $C$ to constraint the hadronic parameters.  These
hadronic parameters cannot be assumed to be equal to those of
Eq.~(\ref{eq:BsKsKs}), because of SU(3) breaking
effects. Nevertheless, it is true that an SU(3) breaking larger than
$100\%$ has never been observed up to now. Thanks to this
consideration, one can determine the maximum allowed value of
$P^{GIM}_s/P_s$, taking for
the output distribution of $P^{GIM}_d/P_d$ (to take into account the
statistical error on the fit to $B_d \to K^0 \bar K^0$) a $100\%$
range centered around the mean value of $P^{GIM}_d/P_d$ (to take into
account SU(3) breaking effects). From this determination, an estimate
of the induced error on $\beta_s$ can be obtained, in analogy of what
is done for $\beta$ from $B_d \to J/\psi K^0$ using $B_d \to J/\psi
\pi^0$~\cite{CPS_beta}.

In order to give an estimate of the theoretical error induced on
$\beta_s$ by neglecting $P^{GIM}_d/P_d$, we scaled the statistical
error of the time-dependent measurement of $B_d \to K^0 \bar K^0$ by
BaBar~\cite{babarkk}, assuming an irreducible systematic error of
$0.01$ on $S$ and $C$. The output is shown in the plot on the right of
Fig.~\ref{fig:betas_tdsign} as a function of the integrated
luminosity. Here, we assume that data will be collected at the
$\Upsilon(5S)$ resonance, but it is important to recall that, for a
certain amount of integrated luminosity, the number of $B_d$ decays
available from $\Upsilon(4S)$ is larger (so that the error
corresponding to the same luminosity will be smaller, if not
systematics dominated).  \\
\begin{boldmath}
\subsection{$B_s$--$\bar B_s$ mixing phase from time-integrated measurements}\label{subsec:dgamma}
\end{boldmath}

In order to extract the weak phase $\beta_s$ of $B_s$--$\bar B_s$
mixing, one can use an independent strategy, which, as for the
measurement discussed in Sec.~\ref{sec:formalism}, does not rely on
time-dependent CP asymmetries. In fact, it is possible to use the
determination of $\Delta \Gamma_s$~\cite{Dighedgog} and the charge
asymmetry in semileptonic $B_s$ decays,
$A_{SL}^s$~\cite{Laplace:2002ik} to obtain an independent
determination of the same quantity. As we show in the following
sections, a good precision on these two quantities can be obtained at
a $B$-Factory, exploiting high statistics, high efficiency in lepton
reconstruction and a pure $B_s$ sample.  \\
\begin{boldmath}
\subsubsection{Measurement of $\Delta \Gamma_s/ \Gamma_s$}
\end{boldmath}
\label{sec:dGsOGs}
\newcommand{\vrho}  {\ensuremath{\vec{\rho}}}
\newcommand{\paral}{\ensuremath{\parallel}}

The relative decay-time difference $\Delta \Gamma_s/ \Gamma_s$ can be
determined studying the angular distribution of the $B_s \to J/\psi
\phi$ decay. Using the transversity basis, the angular distribution of
the final state can be described in terms of three complex amplitudes:
the two transverse amplitudes $A_\perp$ and $A_\paral$ (perpendicular
and parallel), and the longitudinal component $A_0$.  $A_0$ and
$A_\paral$ are CP-even, while $A_\perp$ is
CP-odd~\cite{CDFdgog}. In the SM, neglecting CP violation effects
in the $B_s$ mixing, the CP eigenstates $B_{CP}=+1$ and $B_{CP}=-1$
correspond to the mass eigenstates $B_L$ and $B_H$.  CP-even (odd)
amplitudes evolve according to the exponential factor~\footnote{Here 
$\Gamma_L$ and $\Gamma_H$ represent the lifetimes of the light and heavy 
mass eigenstates of the $B_s$--$\bar B_s$ system. Notice also that at a 
$B$-Factory the time $t$ can be substituted by $\Delta t$, the proper 
time difference between the decays of the two $B$ mesons} 
$e^{-\Gamma_L t}$ ($e^{-\Gamma_H t}$). As suggested in
ref.~\cite{Dighedgog}, if CP violation in mixing is not negligible
(which is excluded in the SM but it is possible in NP scenarios) the
time evolution of these amplitudes is modified.  In particular, an
explicit dependence on the CP violating weak phase $\beta_s$
appears.  Following these considerations, one can write the angular
distribution of $B_s$ events as:
\begin{eqnarray}
\nonumber 
   \frac{d^4\mathcal{P}(\vrho ,t)}{d\vrho \, dt} && \propto \, 
   \Biggl[ |A_0|^2 \frac{1+|cos(2\beta_s)|}{2}\cdot f_1\left(\vec\rho\right) 
   + |A_\paral|^2 \frac{1+|cos(2\beta_s)|}{2} \cdot f_2\left(\vec\rho\right) + \\ 
\nonumber && 
    + |A_\perp|^2\frac{1-|cos(2\beta_s)|}{2}\cdot f_3\left(\vec\rho\right) + 
   \frac{1}{2} |A_\paral| |A_\perp|\cos\left(\delta_1\right) \sin\left(2\beta_s\right)
   \cdot f_4\left(\vec\rho\right) + 
 \\ 
\nonumber && 
    + |A_0||A_\paral|\cos\left(\delta_2 - \delta_1\right)
   \frac{1+|cos(2\beta_s)|}{2} \cdot f_5\left(\vec\rho\right) + 
 \\
\nonumber  && 
   -  |A_0||A_\perp|\cos\left(\delta_2\right) \sin\left(2\beta_s\right) \cdot
  f_6\left(\vec\rho\right) \Biggr] e^{-\Gamma_L t} + \\ 
\nonumber && 
  +  \Biggl[
  |A_0|^2 \frac{1-|cos\left(2\beta_s\right)|}{2} \cdot f_1\left(\vec\rho\right) + 
  |A_\paral|^2 \frac{1-|cos\left(2\beta_s\right)|}{2} \cdot
  f_2\left(\vec\rho\right) + \\ 
\nonumber && 
  +  |A_\perp|^2
  \frac{1+|cos(2\beta_s)|}{2}\cdot f_3\left(\vec\rho\right) -
  \frac{1}{2} |A_\paral| |A_\perp|\cos\left(\delta_1\right) \sin\left(2\beta_s\right)
  \cdot f_4\left(\vec\rho\right) + 
\\ 
\nonumber && 
  + |A_0||A_\paral|\cos\left(\delta_2 -
  \delta_1\right)\frac{1-|cos\left(2\beta_s\right)|}{2} \cdot f_5\left(\vec\rho\right) + 
\\
  && 
  + |A_0||A_\perp|\cos\left(\delta_2\right) \sin\left(2\beta_s\right) \cdot f_6\left(\vec\rho\right)
  \Biggr] e^{-\Gamma_H t}.
\label{eq:angulardist}
\end{eqnarray}
where the functions $f_i(\vec\rho)$ and the transversity variables
$\vec\rho\equiv\{\cos\theta,\varphi,\cos\psi\}$ are defined in
~\cite{Dighedgog}, and $\delta_1$ ($\delta_2$) is the strong phase
difference between $A_\perp$ and $A_\paral$ ($A_\perp$ and $A_0$). The
arbitrary phase in the decay amplitude is removed forcing $A_0$ to be
real. In the PDF, we write $\Gamma_H$ and $\Gamma_L$ in terms of
$\Delta \Gamma = \Gamma_L - \Gamma_H$ and $\Gamma = (\Gamma_L +
\Gamma_H)/2$, and we float these two parameters in the fit, along with
the yields, the absolute values of the amplitudes, the strong phases
$\delta_1$ and $\delta_2$ and the weak phase $\beta_s$.  In the MC
generation we assume $\Delta \Gamma_s = 0.1$~ps$^{-1}$ (SM prediction
from Ref.~\cite{UTNP}), and $\Gamma_s =
1.39$~ps$^{-1}$~\cite{DGoGCDF}. It is important to stress the fact that
no assumption on the size of CP violation in $B_s$--$\bar B_s$ mixing
has been done, which is important to constrain physics beyond the SM.

We determine the shape of the $m_{miss}$ and $\Delta E$ distributions in the
$B_s \to J/\psi \phi$ sample by means of a full MC simulation, applying the same
requirements of Sec.~\ref{sec:timedepcpasymmetry}. Then, in order to determine
the expected error on $\Delta \Gamma_s/ \Gamma_s$ and $\beta_s$, we perform a
set of toy Monte Carlo experiments, generating a large number of datasets
(corresponding to the assumed luminosity, and taking the the value of
reconstruction efficiency from simulation, and $BR(B_s \to J/\psi \phi) = 9.3
\cdot 10^{-4}$) and fitting for the value of $\Delta \Gamma_s/ \Gamma_s$,
together with signal and background yields, through an extended and unbinned ML
fit, by using the likelihood function of Eq.~(\ref{eq:jpsiphilik}), with the
addition of the PDF for the angular distribution.  For the signal
component we use the expression of Eq.~(\ref{eq:angulardist}).  The
angular distribution of background events is assumed to be flat in
$\vec\rho$ variables.

\begin{figure}[!tbp]
\begin{center}
\includegraphics[width=7cm]{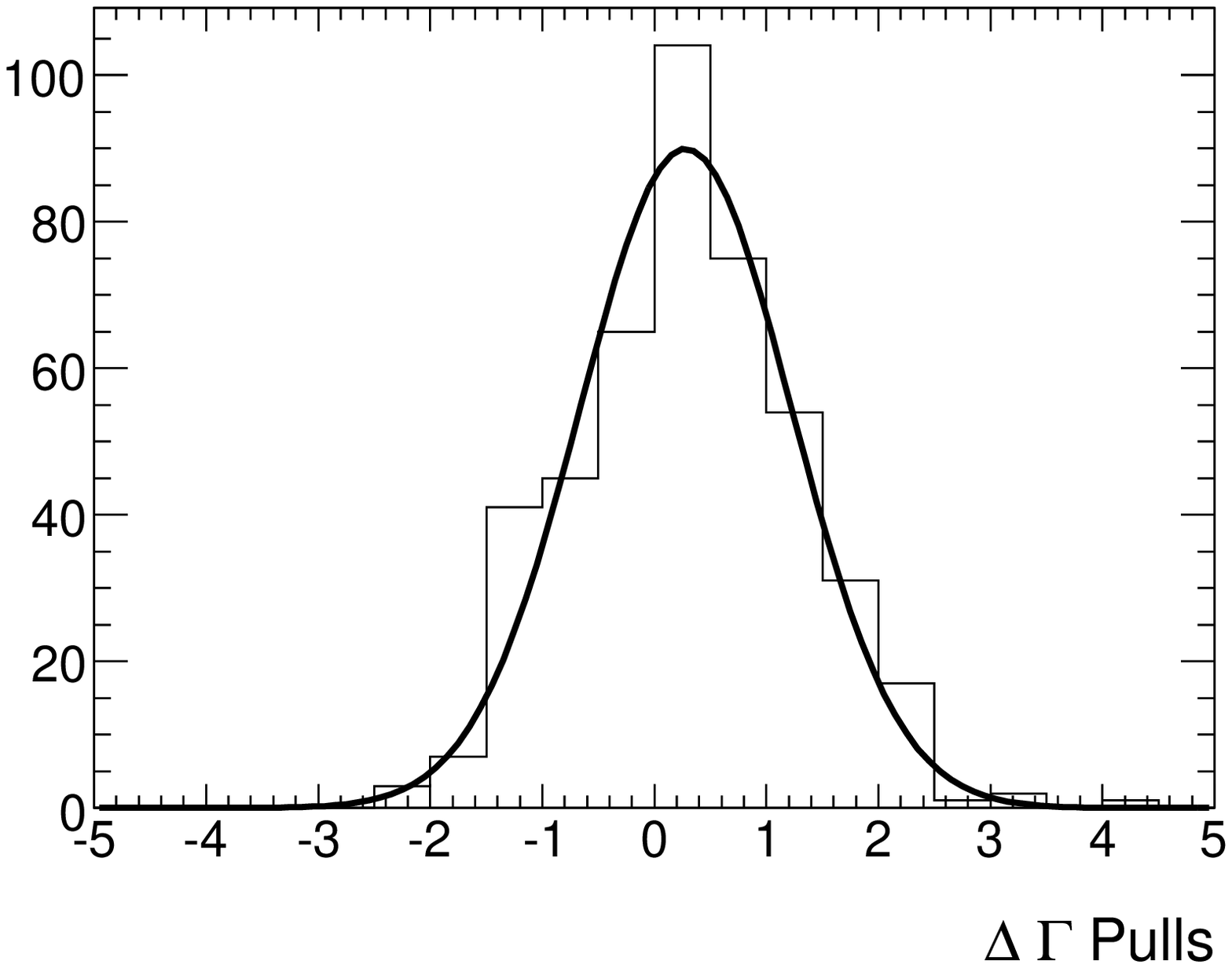}
\includegraphics[width=6cm]{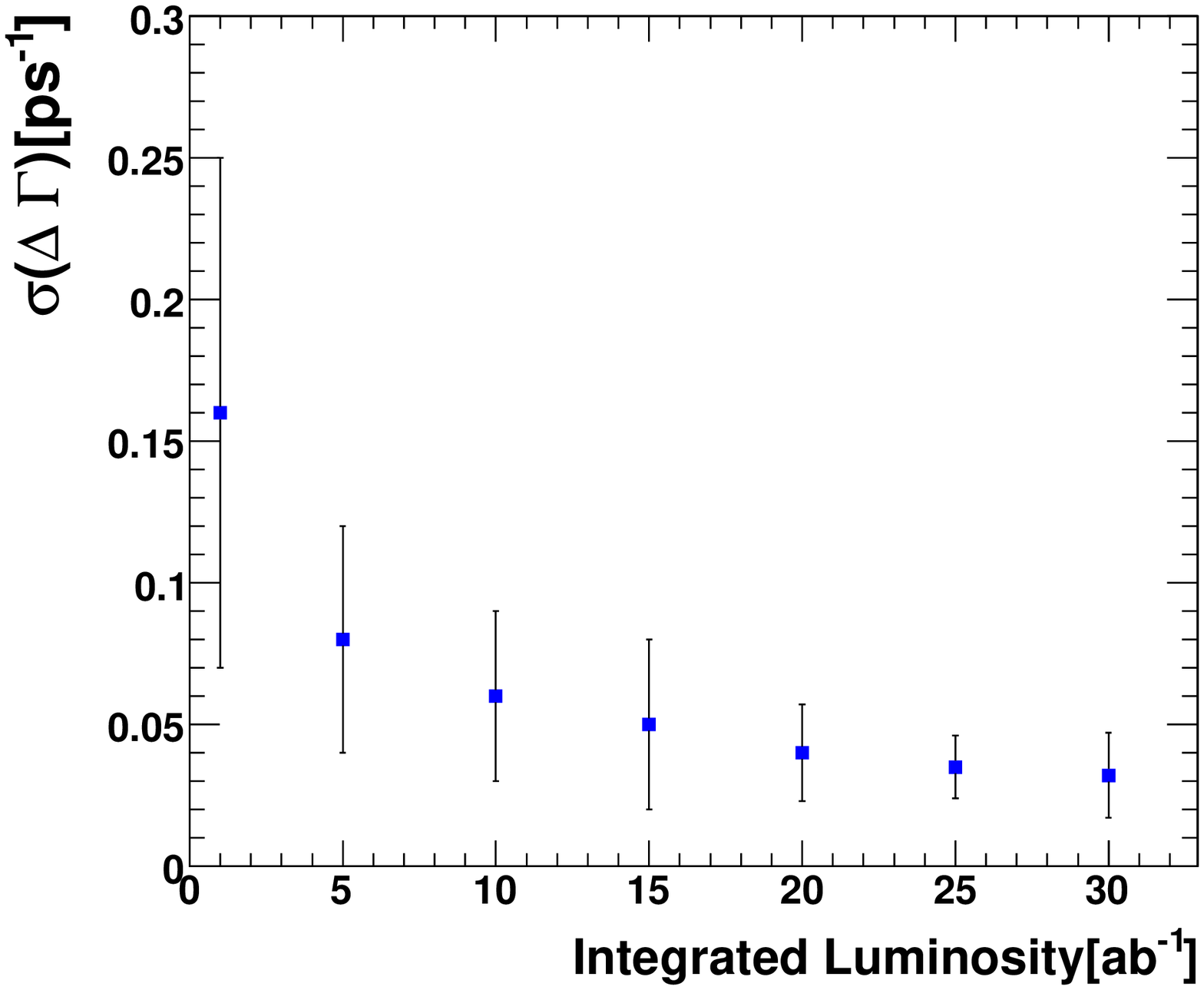}
\includegraphics[width=7cm]{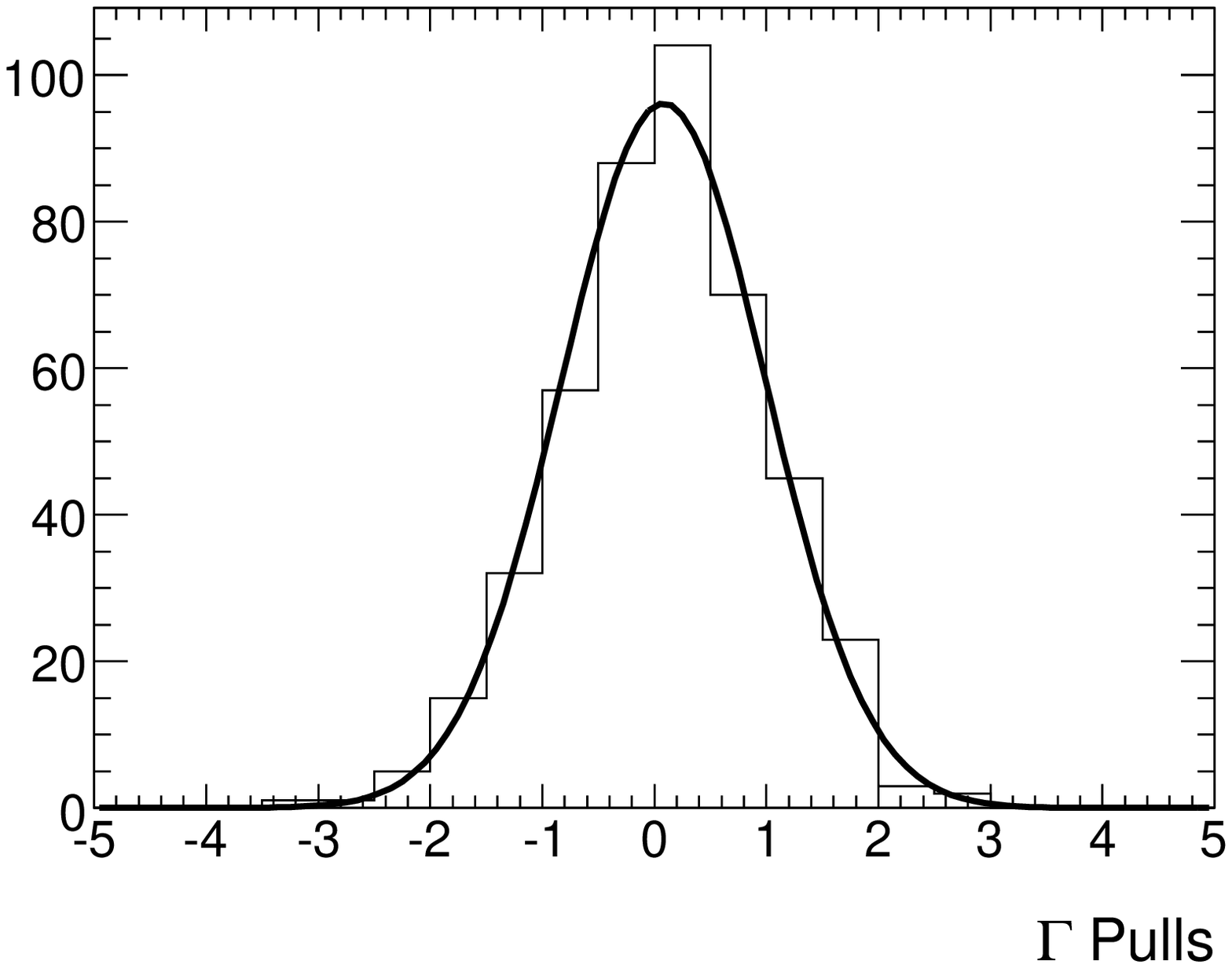}
\includegraphics[width=6cm]{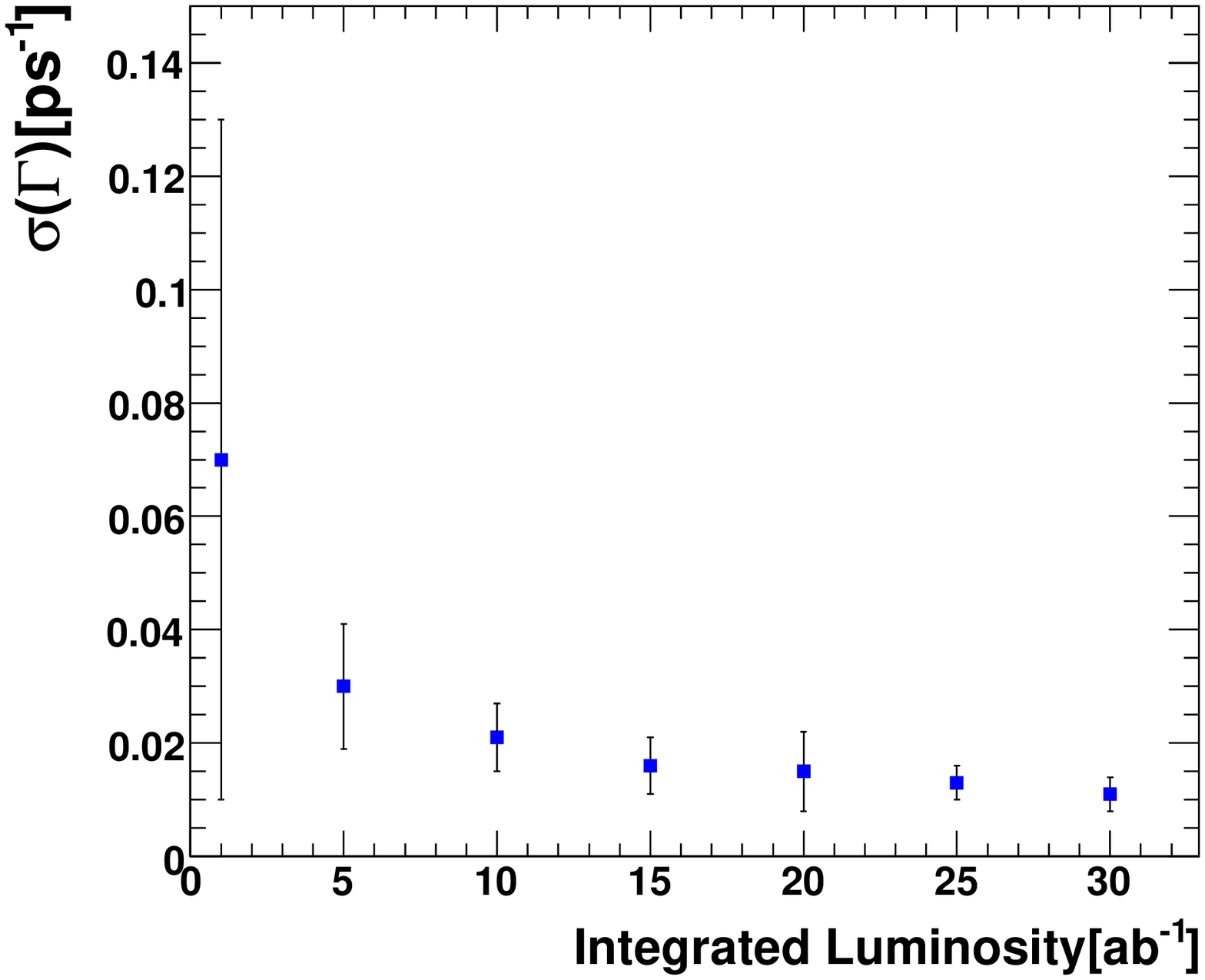}
\includegraphics[width=7cm]{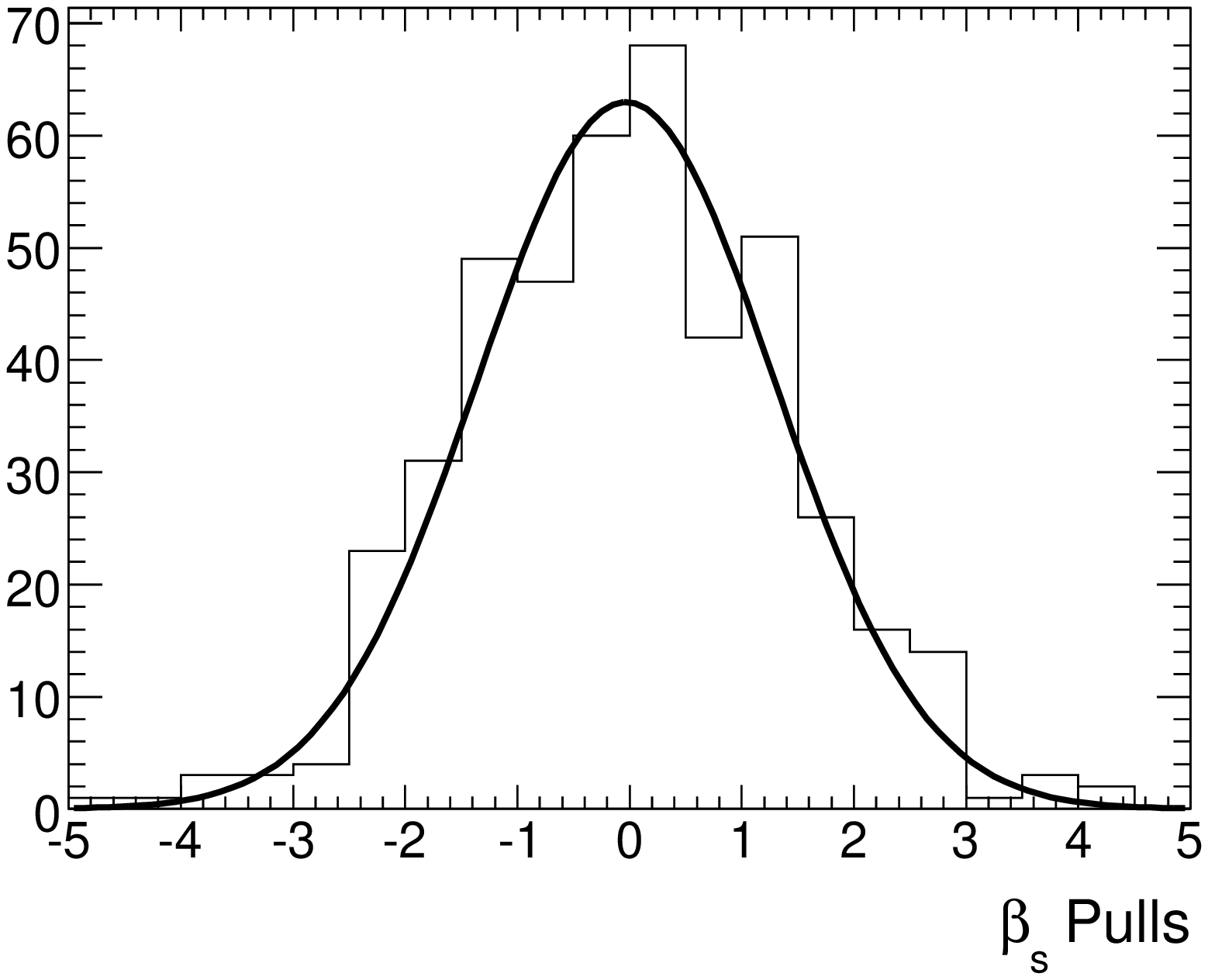}
\includegraphics[width=6cm]{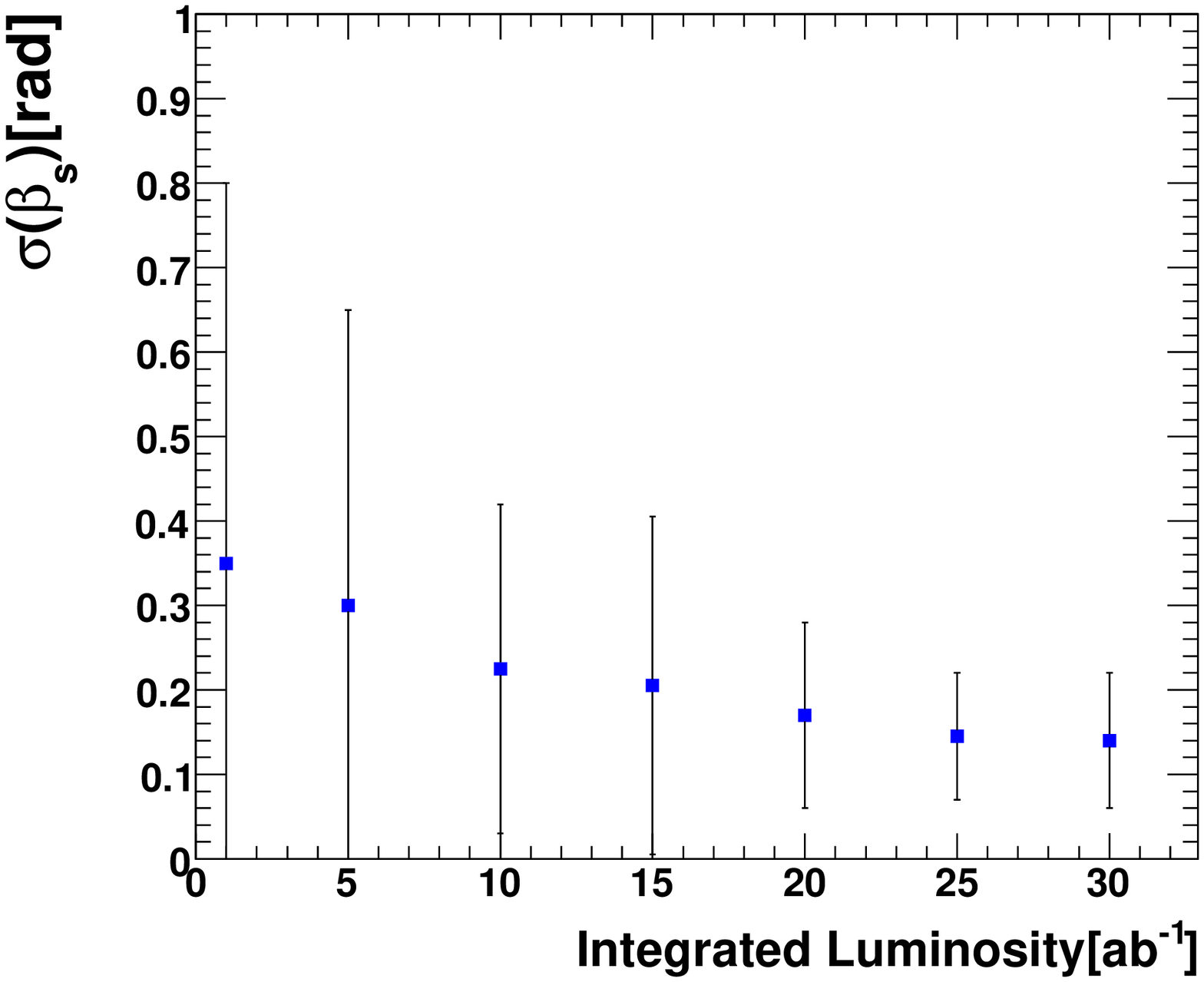}
\caption{\it Distributions of the pull for $\Delta\Gamma_s$ (top row),
$\Gamma_s$ (middle row) and $\beta_s$ (bottom row) from a set of 900
toy Monte Carlo experiments, assuming an integrated luminosity of
$30\,ab^{-1}$ (left) and the trend of the error as a function of the
integrated luminosity (right). The error bars, in the right plots,
show the RMS of the error distribution.
\label{fig:dg_30}}
\end{center}
\end{figure}

We performed different sets of toy Monte Carlo experiments,
corresponding to different integrated luminosities between
$1\,ab^{-1}$ and $30\,ab^{-1}$. Fig.~\ref{fig:dg_30} shows the pull
distributions for $\Delta \Gamma_s$, $\Gamma$ and $\beta_s$ at
$30\,ab^{-1}$ and the dependence of the statistical error on the luminosity.

We remark here that the measurement can be affected by large
correlations between the measured quantities, as experienced by the
D$\O$ Collaboration in a similar analysis~\cite{D0phase}.  Also these
correlations can be studied in the toy Monte Carlo experiments, and
they can be used to extract the corresponding constraints on the NP
parameters (see Sec.~\ref{sec:impact}) \\

\subsubsection{Charge asymmetry in semileptonic decays}

The amplitude describing $B_s$--$\bar B_s$ mixing can be
experimentally constrained looking at the difference between the $B_s$
and $\bar B_s$ semileptonic decay rates after the mixing of the $B_s$
mesons. Since the charge of the lepton tags the flavour of the $B_s$
meson at the decay point, and since the decay amplitude is
characterized by only one combination of CKM elements, any deviation
of the charge asymmetry from zero can only come from a deviation of
$|q/p|$ from one. One can define the semileptonic asymmetry as :
\begin{equation}
A_{SL}^s=\frac{BR(B_s \to \bar B_s \to D_s^{(*)-} {\it l}^+
                   \nu_{\it l}) - BR(\bar B_s \to B_s \to D_s^{(*)+}
                   {\it l}^- \nu_{\it l})} {BR(B_s \to \bar B_s \to
                   D_s^{(*)-} {\it l}^+ \nu_{\it l}) + BR(\bar B_s \to
                   B_s \to D_s^{(*)+} {\it l}^- \nu_{\it l})} =
                   \frac{1-|q/p|^4}{1+|q/p|^4}.
\end{equation}

To measure $A_{SL}^s$, we exclusively reconstruct one of the two
$B$ mesons into a self-tagging hadronic final state (such as $B_s \to
D_s^{(*)} \pi$) and look for the signature of a semileptonic decay
(high momentum lepton) in the rest of the event (ROE).  Since both the
$B$ mesons are reconstructed from self-tagging modes, it is possible
to isolate those events in which the two $B_s$ mesons have the same
flavour content.  To increase the statistics, one can also avoid to
reconstruct one of the two $B$ mesons and determine its flavour on a
probabilistic basis, using a tagging algorithm.

To evaluate the expected error associated to this strategy, we use a
toy Monte Carlo technique, generating and fitting a large set of
toy-simulated data samples.  We use two variables to discriminate
signal from background, namely the $m_{miss}^s$ of the fully
reconstructed $B_s \to D_s^{(*)} \pi$ candidate, and the missing mass
of the other $B$, $m_\nu$.  Signal events peak in the region
$m_{miss}^s\sim 5.46$ GeV and $m_\nu\sim 0$ GeV. We describe the event
distributions using the function of Eq.~(\ref{eq:Cruijff}) for
$m_{miss}^s$ and the sum of four Gaussians for $m_\nu$.  The continuum
background, coming from the combinatoric of particles in the
hadronization of $e^+e^- \to q \bar q$ events ($q=u,d,s,c$) ($\sim
10\%$ of the signal yield) is described by the phase-space threshold
function of Eq.~(\ref{eq:argus}) for $m_{miss}^s$ and a second order
polynomial for $m_\nu$.  An additional source of background comes from
other $B_s$ decays ($\sim 2.5\%$ of the signal yield), which can be
described by the $m^{s}_{miss}$ PDF used for signal and an order one
polynomial for $m_\nu$.  Background from $\Upsilon(5S) \to B B \pi$
events is strongly suppressed, since both the $B$ mesons of the event
are requested to decay into $D_s$ mesons, which occur through CKM
suppressed amplitudes for $B_d$ and $B^+$ mesons.

We calculated the expected $q \bar q$ background yield scaling the
values obtained by BaBar~\cite{Dslnu_FuLLREC}. For signal, we consider both 
$D_s l \nu$ and $D^* l \nu$ events ($l=e,\mu$), assuming
$BR(B_s\to D_s l \nu)=2\%$, $BR(B_s\to D_s^* l \nu)=4\%$, and the efficiency 
and purity of the BaBar analysis.

\begin{figure}[!tbp]
\begin{center}
\includegraphics[width=9cm]{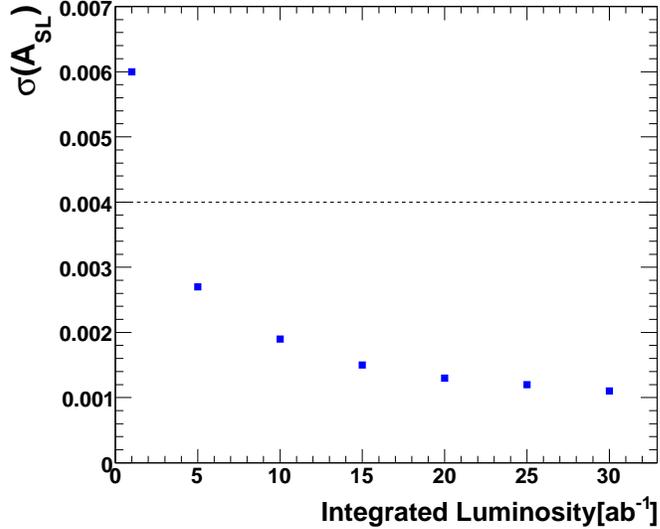}
\caption{\it Statistical error on $A_{SL}^s$, as a function of the
integrated luminosity. The dashed line represents the order of
magnitude of the current systematic error on $A_{CP}$ measurements,
shown for comparison. \label{fig:A_SLerror}}
\end{center}
\end{figure}

The result of the simulation is shown in Fig.~\ref{fig:A_SLerror},
where the statistical error on $A_{SL}$ is given as a function of the
integrated luminosity.  The dashed line on the plot represents the
current value of the systematic error typically quoted in precision
measurements of direct CP asymmetries at the $B$-Factories.  It is not
unrealistic to expect an improvement of a factor two in the evaluation
of the systematic error, considering that a sizable part of it has a
statistical nature (being related to the available statistics for
control samples).

\subsubsection{Inclusive Measurement of Dimuon Charge Asymmetry}

Running at the $\Upsilon(4S)$, the measurement of $A_{SL}^{d}$ can
also be performed in an inclusive way, identifying $B\bar B$ events
from pairs of same-sign leptons with positive ($N^{++}$) and negative
($N^{--}$) charge and calculating the charge asymmetry as
\begin{eqnarray}
A_{SL}^{d} &=& \frac{N^{--}-N^{++}}{N^{--}+N^{++}}.
\label{eq:ach4s}
\end{eqnarray}
These events correspond to $\Upsilon(4S)$ decays in which both the $B$
mesons decay into a semileptonic final state. Requiring the charge of
the leptons to be the same, is equivalent to select only events for
which the two mesons have the same $b$ flavour content, which can
happen only after a $B$--$\bar B$ mixing process.  Contrary to the
analysis presented in the previous section ($A_{SL}^{s}$), the cascade 
events component has to be controlled.  A non-null value of $A_{SL}^d$ 
is sensitive to CP violating effects in the mixing process, i.e. to the 
weak phase of the mixing.

The situation of the $\Upsilon(5S)$ is more complicated: $B_s$ and
$B_{d,u}$ mesons are produced simultaneously, but cannot be separated
during the reconstruction, since the analysis is done inclusively.  In
this case, the measured observable can be written as
\begin{eqnarray}
A_{CH} &=& \frac{N^{--}-N^{++}}{N^{--}+N^{++}} = \frac{\chi - \bar
          \chi}{\chi+\bar \chi -2 \chi \bar \chi}
\label{eq:ACH1}
\end{eqnarray}
where $\chi= (1-f_s) \chi_d + f_s \chi_s$, $\bar \chi= (1-f_s) \bar
\chi_d + f_s \bar \chi_s$, $f_s$ is the production rate of $B_s$ meson
pairs at the $\Upsilon(5S)$ and
\begin{equation}
  \label{eq:chiq}
  \stackrel{\scriptscriptstyle{(-)}}{\chi_q}=\frac{\frac{\Delta\Gamma_q}{\Gamma_q}^2+4\frac{\Delta
      m_q}{\Gamma_q}^2}{\frac{\Delta\Gamma_q}{\Gamma_q}^2(\stackrel{\scriptscriptstyle{(-)}}{z_q}-1)
      + 4(2 \stackrel{\scriptscriptstyle{(-)}}{z_q}+\frac{\Delta
      m_q}{\Gamma_q}^2(1+ \stackrel{\scriptscriptstyle{(-)}}{z_q}))}
      ~~~~~(q=d,s)
\end{equation}
with $z_q = |q/p|_q^2$ and $\bar z_q = |p/q|_q^2$.

The interesting feature of this observable is that it relates
NP effects in $B_d$ and $B_s$ sectors, providing an additional
constraint with respect to the measurement of $A^{d,s}_{SL}$. The
sample used for $A_{CH}$ can be orthogonal to the one used for
$A_{SL}^s$ ($A_{SL}^d$) measurements if one avoids the use of
semileptonic decays to reconstruct the {\it other} $B$ meson in the
latter analysis.


From the experimental point of view, the main background sources are
$q \bar q$ events ($q=u,d,s,c$) and {\it cascade-lepton} events in
which one of the two leptons is generated from the semileptonic decay
of a $D$ meson, coming from the $B$ decay.  These backgrounds can be
separated from signal looking at the distribution of the decay
time~\cite{babarASLincl}.  The charge of the muon distinguishes $B$ and
$\bar B$ mesons, while the information on the shape of the event in
the CM frame allows to improve the rejection of $q \bar q$ background.
The residual $B \bar B$ background, coming from {\it cascade-lepton}
events can be suppressed using a Neural Network algorithm, based
on:
\begin{itemize}
\item the momenta of the two leptons with the highest momentum in the
$\Upsilon(5S)$ center of mass system;
\item the total visible energy and the missing momentum of the event
in the $\Upsilon(5S)$ center of mass system;
\item the opening angle between the leptons in the $\Upsilon(5S)$
center of mass system;
\end{itemize}
and trained on simulated samples of signal and background events.

\begin{figure}[!tbp]
\begin{center}
\includegraphics[width=9cm]{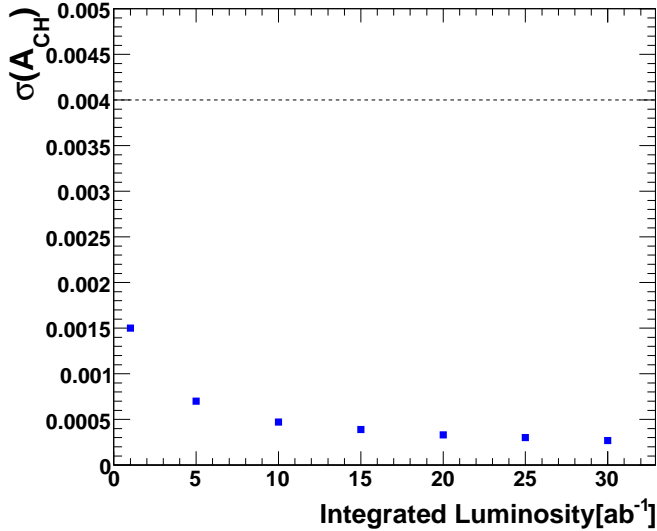}
\caption{\it Statistical error on $A_{CH}$, as a function of the
integrated luminosity. The dashed line represents the order of
magnitude of systematic error on the current measurement at
$\Upsilon(4S)$, shown for comparison. \label{fig:A_CHerror}}
\end{center}
\end{figure}

A rough estimate of the expected error achievable on the inclusive
approach can be obtained scaling the statistical error quoted in the
corresponding BaBar analysis~\cite{babarASLincl} according to the
inverse of $N_S/\sqrt{N_S+N_B}$, where $N_S$ and $N_B$ are the
expected yields for signal and background respectively.  The result,
as a function of the luminosity, is given in Fig.~\ref{fig:A_CHerror}.
The dashed line represents the order of magnitude of the current
systematic error on $A_{CP}$ measurement, shown for comparison.  
The measurement will be systematics dominated after a
relatively small period of data taking, even though also in this case an
improvement of the systematic error with an increased size of the
control samples is possible.  Notice that the information about
$A_{SL}$ comes only from same sign events (i.e. with 2 leptons of the
same sign), so only these events are taken into account in the error
estimate.

\begin{boldmath}
\section{Benchmark Measurements of $B_s$ rare decays}\label{sec:bsmeasurements}
\end{boldmath}

The production of $B_s$ mesons at the $\Upsilon(5S)$ allows to study
the decay rates of the $B_s$ sector with the same completeness and
accuracy that is currently available for the $B_d$ and charged $B$
mesons, improving our understanding of $B$ physics and helping to
reduce the theoretical uncertainties related to NP-sensitive
quantities.  For instance, using the measurements of BR's of $B_s$
mesons to open charm and to charmless decays it is possible to close
the SU(3) multiplets of $b$ decays~\cite{SU3Gristain} and to control
SU(3) breaking effects in channels like $B_d \to \phi
K^0_S$~\cite{phiksGrossman}. In addition, even without accessing
time-dependent CP asymmetries, it is possible to access weak phases of
$B_s$ decays. This is the case of the determination of $\gamma$ from
the tree-level amplitudes of $B_s \to K \pi \pi^0$ decays, for which
it is possible to disentangle penguin and tree contributions using the
distribution of the events on the Dalitz plot~\cite{CPSBs}.

On the other hand, $B_s$ physics provides a direct way to probe NP
effects in the $b \rightarrow s$ transitions.  Effects generated from
NP particles are expected to be more evident in $b \to s$ than in $b
\to d$ transitions, not only because, thanks to the CKM hierarchy
among the amplitudes, the effect of hadronic uncertainties on the
calculation of the SM is smaller than in the case of $b \to d$
processes, but also because this scenario is still allowed by the
generalized UT analysis beyond the SM~\cite{UTNP} and is expected in
several extensions of the SM~\cite{Pomarol}.

In the following we concentrate on this point. We show what could be
learned from a high statistics study of $B_s$ mesons by giving some
benchmark examples. In some cases, only at very high statistics and
with particular detector performances, one can reach the same precision 
as at LHCb. Anyhow, one should take into account that
the set of measurements we consider is not an exhaustive
representation of the potential of a $B$-Factory running at the
$\Upsilon(5S)$ and that a quantitative comparison between the
performance of LHCb and a next-generation $B$-Factory, based on a
common set of measurements, goes beyond the purpose of this paper.
What we rather want to prove is that, contrary to what is usually
claimed, all the interesting measurements (rates of rare decays,
angular analyses, measurement of direct CP asymmetry and weak phases
in $B_s$--$\bar B_s$ oscillations) can be performed even if
$B_s$--$\bar B_s$ oscillation cannot be accessed.  To demonstrate this
point, we already documented in Sec.~\ref{sec:BsBsbarphase} how the
$B_s$--$\bar B_s$ mixing phase $\beta_s$ can be measured in tree-level
and in penguin modes, testing the SM in $B_s$--$\bar B_s$ mixing as
well as in $b \to s$ transitions. In this section, we discuss two other
milestones of a $B_s$ physics program, namely the measurement of
the BR of rare decays of the $B_s$, and the test of the SM through the
determination of $|V_{td}/V_{ts}|$ in penguin dominated modes.

The main result of this study is that a $B$-Factory running at the
$\Upsilon(5S)$ will not be limited by detector performances. Using the
analyses techniques we briefly describe here, it will be possible to
study several decays which might go beyond the capability of LHCb,
such as channels without primary tracks in the decay and/or channels
with only photons in the final state. Furthermore, the possibility of
studying channels with open kinematic (such as $b \to u l \nu$ and $b
\to s \nu \bar \nu$), even with a factor three reduction in statistics
with respect to the $\Upsilon(4S)$ represents something that only at a
$B$-Factory can be determined. In our opinion, the sum of all these
arguments establishes the complementarity between the physics program of
LHCb and the possibility of studying $B$ mesons at the $\Upsilon(5S)$.
So, while LHCb will provide high precision measurements for a limited set of
observables, a B factory running at the $\Upsilon(5S)$ will be able to 
acccess (not nececssarely with a comparable precision) a much larger 
ensamble of quantities. If this is done before the LHCb will start collecting 
data, already with two ab$^{-1}$ it will be possible to exclude NP 
contributions comparable to the SM amplitude, as well as to provide a set 
of measurements of rates and asymmetries to be used by LHCb as a calibration 
of the physics measurements in the very first run period.
\\

\begin{boldmath}
\subsection{Probing New Physics in penguins with $|V_{td}/V_{ts}|$}
\end{boldmath}\label{subsec:vtdvts}

\begin{figure}[!tbp]
\begin{center}
\includegraphics[width=7cm]{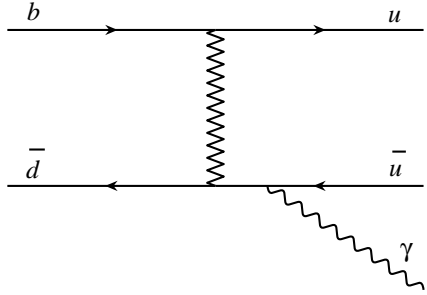}
\caption{\it Annihilation contribution to the $b \to d$ amplitude of $B
\to \rho \gamma$ decays.}
\label{fig:bdannihilation}
\end{center}
\end{figure}

The measurement of $\Delta m_s$ at the Tevatron~\cite{dmsexp} has
provided a further evidence of the consistency between experimental
data and the predictions of the SM. The agreement between the
prediction by {\bf UT}{\it fit}~\cite{UTdmspred} and the measured
value of $|V_{td}/V_{ts}|$ has reduced the possibility of observing
large NP effects in $B_s$--$\bar B_s$
oscillations~\cite{UTNP}. Nevertheless, it is still interesting to
probe NP in $b \to s$ channels using an independent determination of
$|V_{td}/V_{ts}|$, which might be sensitive to NP in a different way
than $\Delta m_s$.  It has been proposed in literature that such a
test can be provided by the ratio $R=BR(B^0 \to \rho^0 \gamma)/BR(B
\to K^{*0} \gamma)$~\cite{refbtovg}. This quantity can be expressed as
\begin{equation}
 R \equiv c_\rho^2\frac{r_m}{\xi^2} 
     \frac{|a_7^c(\rho \gamma)|^2}{|a_7^c(K^{*0} \gamma)|^2}
     \frac{|V_{td}|^2}{|V_{ts}|^2}(1+\Delta R)
\label{eq:btovg}
\end{equation}
where $\frac{|V_{td}|}{|V_{ts}|}$ is the combination of CKM elements
one wants to determine; $\frac{|a_7^c(\rho \gamma)|^2}{|a_7^c(K^{*0}
\gamma)|''}$ is the difference in the penguin factorized decay
amplitude, due to the different mesons in the final state; $r_{m}$ is
defined as $r_m = (\frac{m_{B}^2 - m_{\rho}^2}{m_{B}^2 - m_{K^{*0}}^2})^3$; 
$\xi$ is the ratio of form
factors $F(B \to K^{*0})/F(B \to \rho)$, and $\Delta R$ is a correction
generated by the annihilation contribution to $B \to \rho \gamma$ (see
Fig.~\ref{fig:bdannihilation}), which is not present in $B \to K^{*0}
\gamma$.

This ratio provides an independent test of the SM prediction of
$|V_{td}/V_{ts}|$, since radiative decays are particularly sensitive
to NP contributions. As these processes violate
chirality, the decay amplitude in the SM is proportional to $m_q$, the
mass of the quark ($q=b,s$) generating the helicity flip.  In presence
of NP, the helicity flip is obtained by the mass insertion of a new
heavy state, determining the enhancement of the NP amplitude with
respect to the SM by a factor $m_{HS}/m_q$, where $m_{HS}$ is the mass
of the new heavy state.  This implies that it is possible to observe
NP effects in these channels even when the contribution is too small
to be determined in the $B$--$\bar B$ mixing.

A limitation to the effectiveness of this test comes from the presence
of the $\Delta R$ term in Eq.~(\ref{eq:btovg}). In the SM this
contribution is not only expected to be
$\mathcal{O}(\Lambda_{QCD}/m_b)$ but it is also CKM suppressed, being
proportional to $\cos(\alpha)$~\cite{refbtovg}. These two suppressions
reduce the sensitivity to non perturbative QCD effects that determines
the value of $\Delta R$. The main limitation is related to the fact
that, beyond the SM, the CKM factor is not necessarily small.  While
in MFV models the $\cos(\alpha)$ suppression still reduces the
sensitivity to the hadronic corrections, this is not the case for
models with a generic flavour structure.  This is why it is
particularly interesting to look for a similar observable which is not
affected by the presence of the annihilation term.

Such observable can be provided by the ratio $R=BR(B_s^0 \to K^{*0}
\gamma)/BR(B_d \to K^{*0} \gamma)$. In this case, in fact, the two
decays are not affected by annihilation contributions and the ratio is
governed by Eq.~(\ref{eq:btovg}), with $\frac{|a_7^c(\rho
\gamma)|}{|a_7^c(K^{*0} \gamma)|}$ and $\xi$ replaced by the similar
factors, and where the $\Delta R$ term vanishes.

The BR measurement for $B_d \to K^{*0} \gamma$ has been performed at the
current $B$-Factories~\cite{Aubert:2004te} and the present error is
(almost) dominated by the systematic error. We assume $BR(B_d \to K^{*0}
\gamma) = (40.1 \pm 2.4) \cdot 10^{-6}$, corresponding to the current
world average, with the error given by the current systematic error.
 
An estimate of the error that can be obtained for $BR(B_s^0 \to K^{*0}
\gamma)$ has been extracted performing toy Monte Carlo simulations as
a function of the integrated luminosity. The SM expectation value for
$BR(B_s^0 \to K^{*0} \gamma)$ is about two times its $B_d$
counterparts, $B_d^0 \to \rho^0 \gamma$. As a reference, we use
$BR(B_s^0 \to K^{*0} \gamma) = 1.8 \cdot 10^{-6}$, obtained using the
central value of the current world average for $B_d^0 \to \rho^0
\gamma$~\cite{expBRrho0gamma} and the relative factor two, coming from
the Clebsch-Gordon coefficient of the $\rho^0$.

\begin{figure}[!tbp]
\begin{center}
\includegraphics[width=7cm]{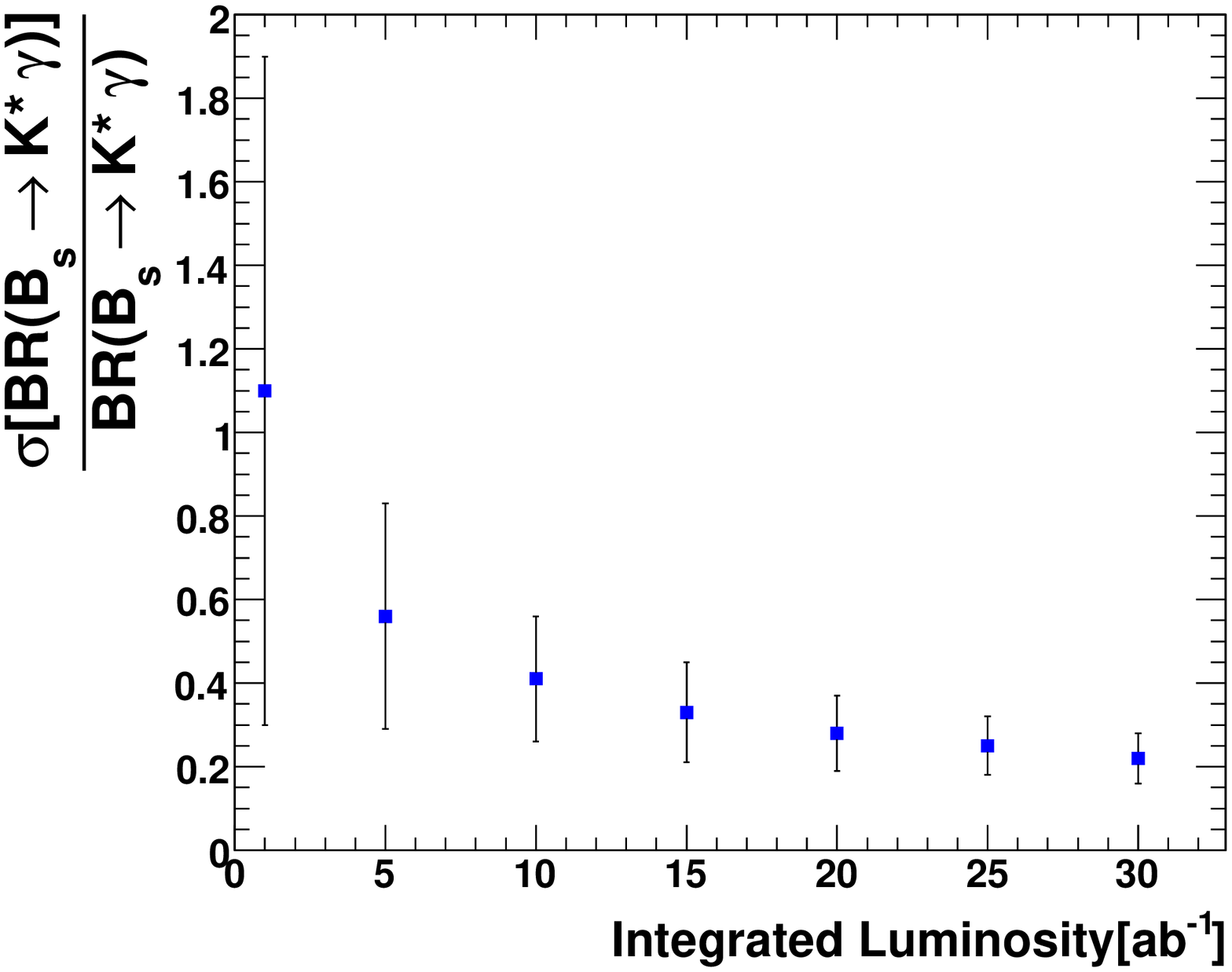}
\includegraphics[width=7cm]{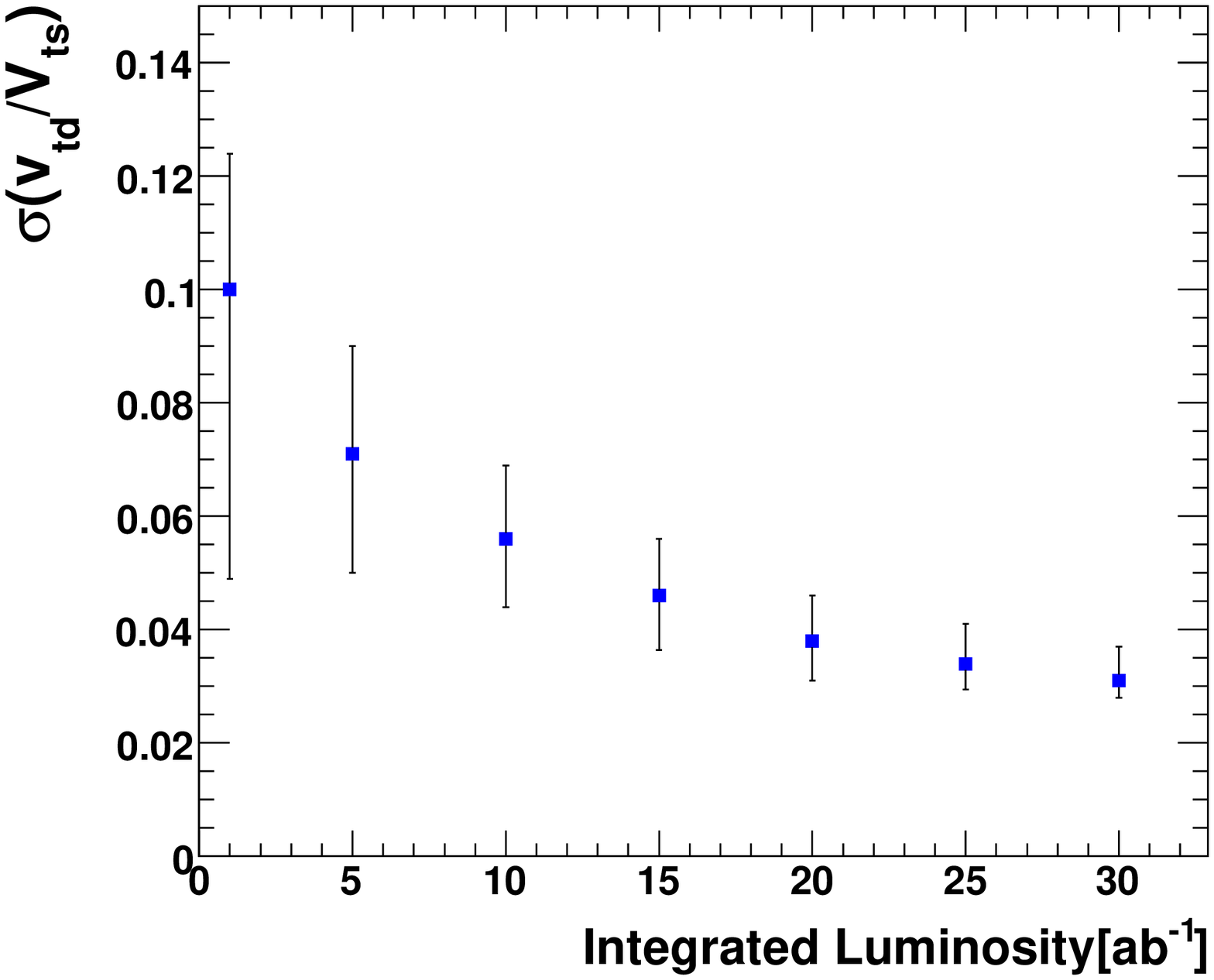}
\caption{\it Distribution of the statistical error on $BR(B_s^0 \to K^{*0}
\gamma)$ from toy Monte Carlo experiments (left) and the expected
error of the $|V_{td}/V_{ts}|$ measurement (right), as a
function of the integrated luminosity. \label{fig:radiative_err}}
\end{center}
\end{figure}

The distribution of the relative errors on the BR from the toy
experiments as a function of the integrated luminosity is shown in the
left plot of Fig.~\ref{fig:radiative_err}. The information coming from
the toy experiments, combined to the current predictions on $\xi$
($\xi = 1.17 \pm 0.09$~\cite{Ball:2006nr}), allows to determined the
value of $\frac{|V_{td}|}{|V_{ts}|}$. The result is shown in the right
plot of Fig.~\ref{fig:radiative_err}.  By repeating the exercise after
reducing the error on $\xi$, we verified that the determination of
$\frac{|V_{td}|}{|V_{ts}|}$ is dominated by the statistical error even
for large values of integrated luminosity. \\

\begin{boldmath}
\subsection{$B_s \to \mu \mu$ and the flavour structure of New Physics}
\end{boldmath}

This decay plays a crucial role in study the structure of
the Higgs sector and assessing the value of $\tan\beta$.
In the SM, the BR of this process is written as~\cite{bsmumuformula}
\begin{equation}\label{bbll}
Br(B_s\to l^+l^-)=\tau(B_s)\frac{G^2_{\rm F}}{\pi}
\left(\frac{\alpha}{4\pi\sin^2\theta_{w}}\right)^2 f^2_{B_s}m^2_l
m_{B_s} \sqrt{1-4\frac{m^2_l}{m^2_{B_s}}} |V^\ast_{tb}V_{ts}|^2
Y^2(x_t),
\end{equation}
where $x_t = m_t^2/m_W^2$, $f_{B_s}$ is the $B_s$ meson decay constant
and, within the SM, $Y(x_t)$ is the Inami-Lim function~\cite{inamilim},
given by the expression
\begin{equation}\label{Y0}
Y(x_t)={{x_t}\over8}\; \left[{{x_t -4}\over{x_t-1}} + {{3
x_t}\over{(x_t -1)^2}} \ln x_t\right].
\end{equation}

The SM expectation for this decay rate is $BR(B_s \to \mu\mu) = (3.35
\pm 0.32)\cdot 10^{-9}$~\cite{buras_bs_MFV}. The decay occurs in the
SM through loop diagrams, which make this process particularly
sensitive to NP contributions. For instance, the observation of this
decay above the SM prediction will provide a clear evidence of new
heavy states contributing to virtual processes. Recently, a combined
analysis of $B$ and $K$ rare decays~\cite{MFVfit} has addressed this
point in a quantitative way in the context of MFV models at small
$\tan \beta$, obtaining $BR(B_s \to \mu\mu)<7.4\cdot 10^{-9}$ at
$95\%$ probability. This result indicates that in MFV at small $\tan
\beta$ we do not expect a large enhancement of the decay rate.
Conversely, in very large $\tan \beta$ scenarios~\cite{isidori_tanb}, $\tan
\beta$--enhanced terms can still give sizable contributions resulting
in a enhancement of the decay rate well beyond the above limit.

In order to estimate the expected experimental sensitivity in the SM,
we assume the expectation value quoted above. We only consider
$B_s$--$\bar B_s$ pairs from VV events. Since $BR(B_d \to \mu\mu)$ is
suppressed by more than a factor of ten with respect to $BR(B_s \to
\mu\mu)$, both in SM and in NP scenarios, we can neglect $\bdpi$
events.  The contamination coming from $B_s \to \pi \pi$ events is
lower than in the case of $B_d$, since the decay is CKM suppressed. As
for the search of $B_d \to \mu \mu$ performed by the
$B$-Factories~\cite{refmumuexp}, the hadronic background can be
suppressed imposing a set of requirements on PID, typically combined
into a PID neural net. The cost is about a factor $60\%$ for the
reconstruction efficiency.  With respect to the $B$-Factories, we
can also imagine a detector configuration with larger acceptance,
thanks to a reduction of the Lorentz boost . Assuming a typical value of the
reconstruction efficiency of $60\%$ we expect $6$ signal events and
$960$ background events in $30$ ab$^{-1}$ of integrated data.

We use a set of toy Monte Carlo to evaluate the sensitivity of this
measurement using $m_{miss}$, $\Delta E$ and the Fisher discriminant
$\cal F$ as discriminating variables. We first generate signal and
background events according to the shape of kinematic ($\mmiss$ and
$\de$) and topological ($\cal F$) variables, as determined from fully
simulated Monte Carlo samples.  Signal events peak at a value of
$\mmiss \sim 5.46$ GeV and $\de \sim - 47$ MeV.  Both 
distributions are described by Eq.~(\ref{eq:Cruijff}). For background
events we use the threshold function of Eq.~(\ref{eq:argus}) for
$\mmiss$ and a second order polynomial for $\de$.  For $\cal F$, we
use a bifurcated Gaussian to parameterize the signal distribution and
the sum of two Gaussians for the background.

\begin{figure}[!tbp]
\begin{center}
\includegraphics[width=7cm]{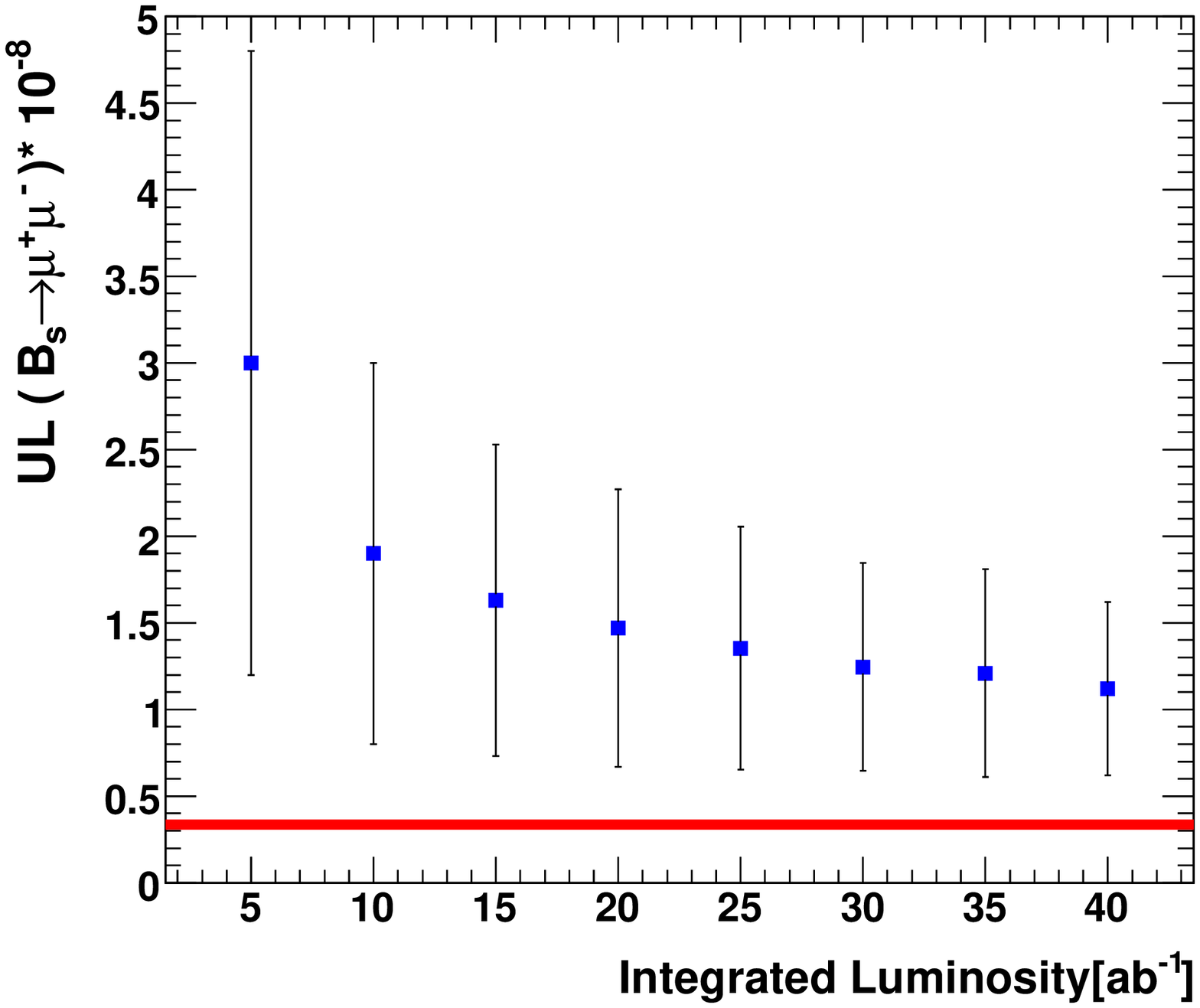}
\includegraphics[width=7cm]{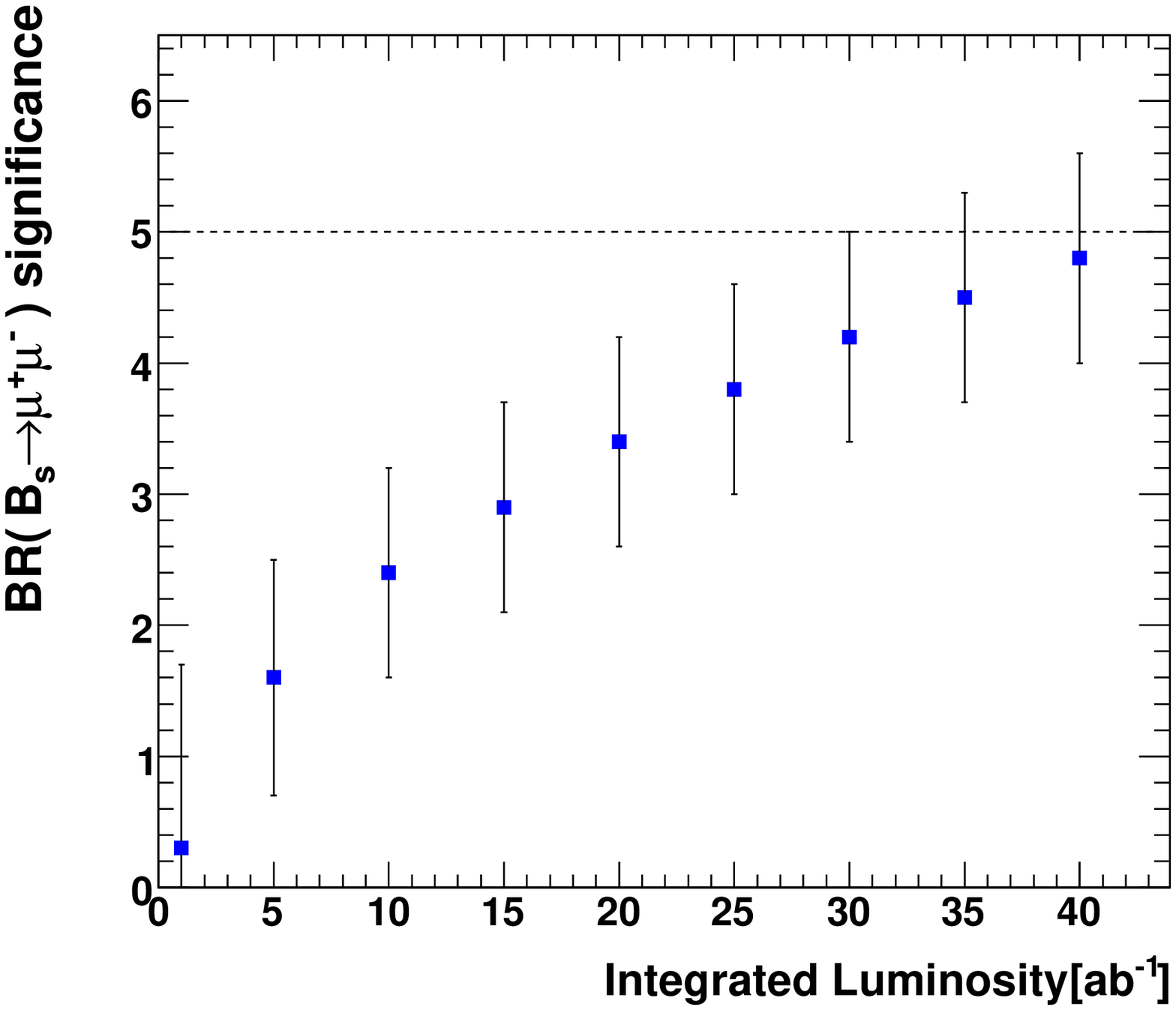}
\caption{\it Distribution of the upper limit at $90 \%$ C.L. 
(left) and statistical significance on the BR (right) for
$BR(B_{s}\to \mu^+\mu^-)$, as a function of the
integrated luminosity. Error bars represent the RMS of the UL
distribution for each set of toys. For the left (right) plot,
the SM expectation for the BR (ten times the SM value) was 
assumed. On the right plot, the dashed line
represent the $5 \sigma$ evidence.\label{fig:BSmumu_UL}}
\end{center}
\end{figure}

For each generated datasets, we perform a ML fit as a function of the
signal yield and obtain a shape of the likelihood for $N_{\mu\mu}$.
We set the $90\%$ upper limit to the value $N_{UL}$ such that
\begin{equation}
\frac{\int_{0}^{N_{UL}} {\cal L}(N_{\mu\mu})
dN_{\mu\mu}}{\int_{0}^{\infty} {\cal L}(N_{\mu\mu}) dN_{\mu\mu}} =
90\%.
\end{equation}
We repeat this procedure for a large set of toys, as a function of the
integrated luminosity. The result is shown in
Fig.~\ref{fig:BSmumu_UL}, where the full line represents the SM
expectation and the errors bars the RMS of the upper limit
distribution for each set of toy experiments.  For comparison, the
current $90\%$ C.L. experimental UL ($BR(B_{s}\to \mu^+\mu^-)<8.0
\cdot 10^{-8}$\cite{CDF_Bsmumu}) falls outside the range of the plot.

Since in several NP scenarios we expect a sizable contribution to the
decay amplitude from new heavy states, we repeated the exercise
assuming a decay rate one order of magnitude larger than the SM
prediction. In this case, it should be possible to measure the BR with
a meaningful statistical significance, rather than setting an upper
limit. This is shown in Fig.~\ref{fig:BSmumu_UL}, where the
statistical significance is given as a function of the integrated
luminosity.

\begin{figure}[!tbp]
\begin{center}
\includegraphics[width=7cm]{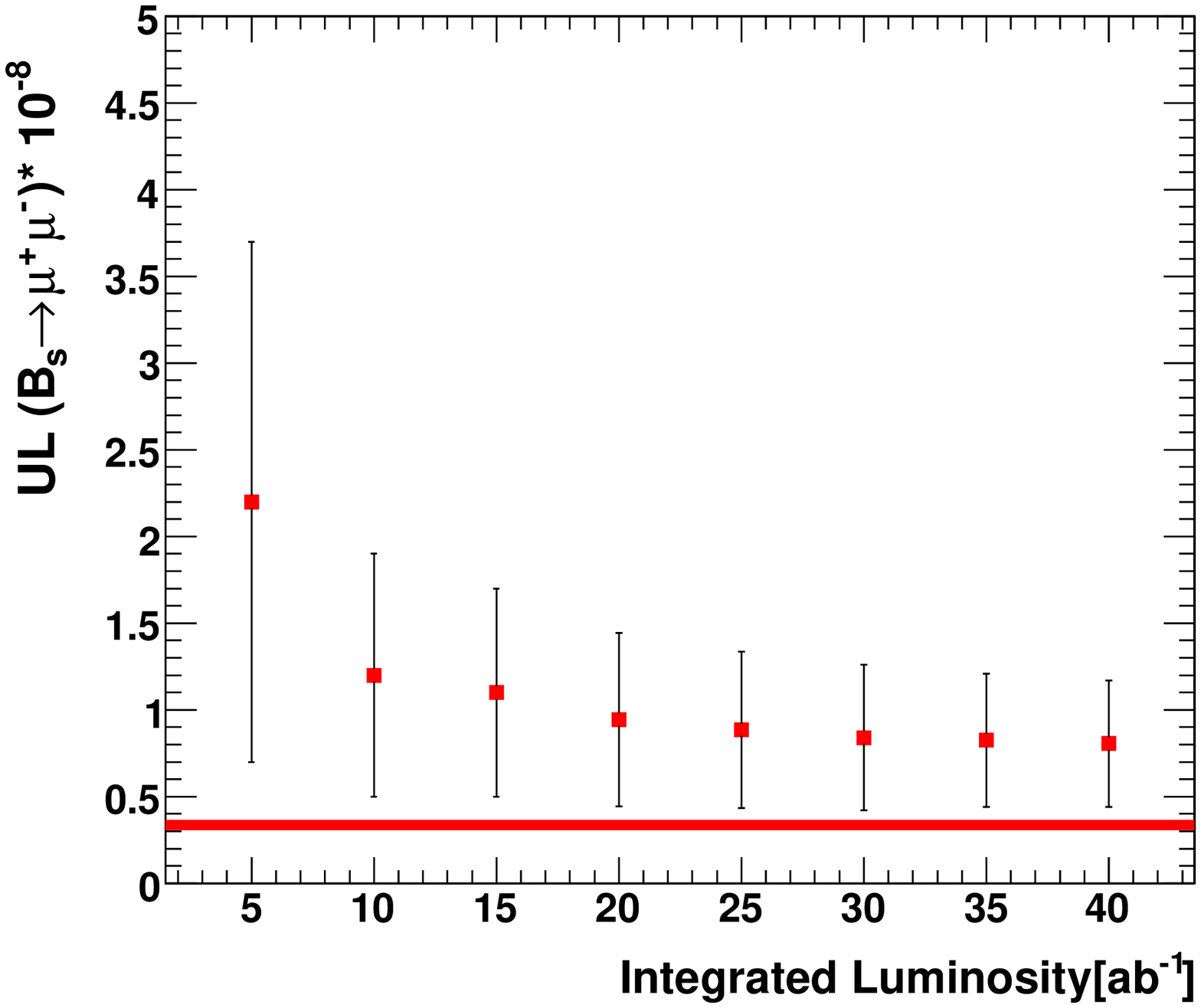}
\includegraphics[width=7cm]{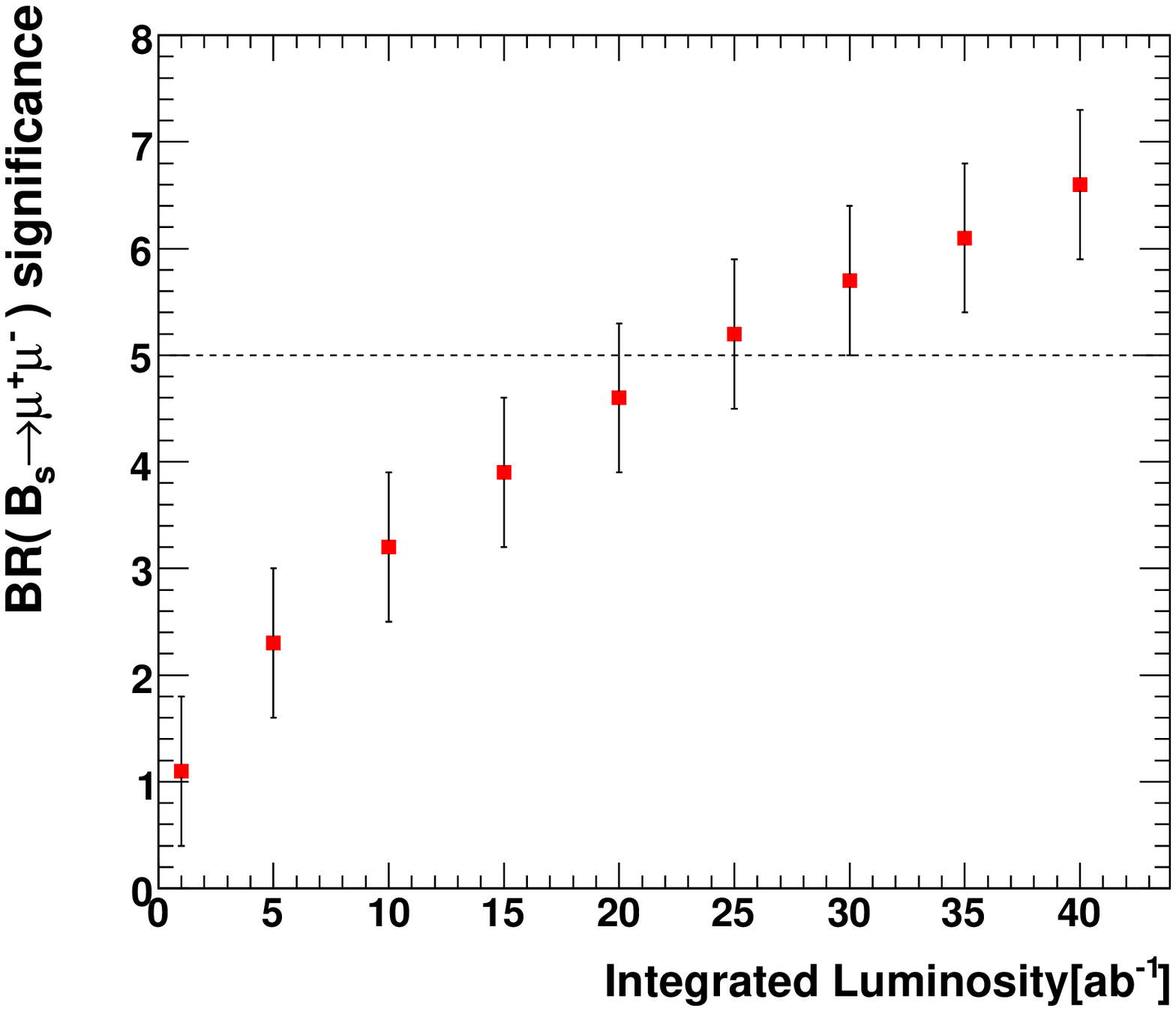}
\caption{\it On the left upper limit distribution at $90 \%$ C.L. for
$BR(B_{s}\to \mu^+\mu^-)$ in a Standard Model scenario, assuming a
reduction of the background of a factor five as a function of the
integrated luminosity.  Error bars represent the RMS of the UL
distribution for each set of toys. The horizontal band represents the
Standard Model prediction\label{fig:BSmumu_1_5_UL}. On the right
significance of the measurement for $BR(B_{s}\to \mu^+\mu^-)$ in a New
Physics scenario with an order of magnitude enhancement with respect
to the Standard Model assuming a reduction of the background of a
factor five, as a function of the integrated luminosity. Error bars
represent the RMS of the significance distribution for each set of
toys. The dashed line represent the $5 \sigma$ evidence.}
\end{center}
\end{figure}

As for the case of $B_s \to K^0 \bar K^0$, discussed in
Sec.~\ref{sec:beta_sPen}, this channel is affected by a large
background contamination from $q \bar q$ continuum events
($q=u,d,s,j$), which are characterized by a jet-like structure, with
all the particles coming from the primary vertex. Considering the
possibility of an upgrade of the vertexing detector with modern
technologies, it is possible to use the improved vertexing
resolution~\cite{neri_pierini_superb} to separate the $B$ and $D$
vertexes on the tag side for signal events and reject $q \bar q$
events, for which no secondary vertex is present.  Using this
additional requirement, it could be possible to strongly reduce the
continuum background, improving the experimental precision. To give an
idea of the impact of this improvement on the search of rare
decays, we repeated the previous exercise assuming a reduction of
continuum background by a factor of five, which might not represent an
optimistic expectation.\footnote{A better estimation of the impact of
the vertex improvements on the data analysis can be obtained with a
more accurate simulation of event reconstruction, which implies a
specific design of the detector. This goes beyond the purpose of this
paper, but it represents an interesting exercise to be considered by
present and future experimental collaborations.} We show in
Fig.~\ref{fig:BSmumu_1_5_UL} the impact of this improvement on the
result of our toy studies, for both the SM and the enhanced values of
the BR considered above. In this case, one can obtain an observation
of the BR in the enhanced scenario with a relatively small amount of
statistics for a $10^{36}$cm$^{-2}$sec$^{-1}$ super $B$-Factory.  \\
It is clear that in this case LHCb can perform much better, given the 
larger amount of $B_s$ produced in its environment and the clear 
signature of the muon tracks. In fact, LHCb is expected to be able 
to exclude the SM branching ratio already with 0.5 $fb^{-1}$
of collected data~\cite{schune_paris} and to reach $3\sigma$
and $5\sigma$ evidence of the SM signal respectively with 
2 $fb^{-1}$ (one year) and of integrated luminosity.
Moreover, also ATLAS and CMS experiments are expected to give 
$4\sigma$ signal after one year of nominal data taking
at the luminosity of $\mathcal{L} = 10^{34}$cm$^{-2}$sec$^{-1}$.\\

\begin{boldmath}
\subsection{Measurement of $B_s \to \gamma \gamma$}
\end{boldmath}\label{sec:bsgammagamma}

For several years, $b \to s \gamma$ has been considered 
as the {\it golden mode} to probe NP in the flavour sector.
From an experimental point of view, any attempt to study this
transition has to face the fact that the present error on $BR(b \to s
\gamma)$ is dominated by systematic effects.  It is then interesting
to look for other channels that can play a similar r\^ole in
constraining NP, but do not imply the experimental complexities that are
introduced in $b \to s \gamma$ by the knowledge of the photon
spectrum.  In this context, an interesting candidate is provided by
$B_s \to \gamma \gamma$ decays. The final state contains
CP-odd and CP-even states, allowing to study CP violating effects with
the tagging algorithms usually adopted at $B$-Factories.  The SM
expectation for the BR is $BR(B_s \to \gamma \gamma)= 2-8 \cdot 10^
{-7}$\cite{Reina:1997my}. NP effects are expected to give sizable
contributions to the decay rate in some particular scenarios.  For
instance, in R-parity violating SUSY models, neutralino exchange can
enhance the decay rate up to $BR(B_s \to \gamma \gamma)\simeq 5 \cdot
10^ {-6}$\cite{Gemintern:2004bw}.  On the other hand, in R-parity
conserving SUSY models
$BR(B_s \to \gamma \gamma)$ is found to be highly correlated with
$BR(b \to s \gamma)$\cite{Bertolini:1998hp}. In fact, thanks to an
extension of the Low theorem\cite{Lin:1989vj}, the $b \rightarrow s
\gamma \gamma$ operator can be expanded at $O(G_{f})$ on the standard
operator basis needed for $b \rightarrow s \gamma$. Doing so, one can
relate deviations from SM in $b \to s \gamma$ and $B_s \to \gamma
\gamma$ processes, provided the fact that NP effects appear only in
the matching of Wilson coefficients of the standard basis of
operators of the $b \to s \gamma$ effective Hamiltonian.

\begin{figure}[!tbp]
\begin{center}
\includegraphics[width=7cm]{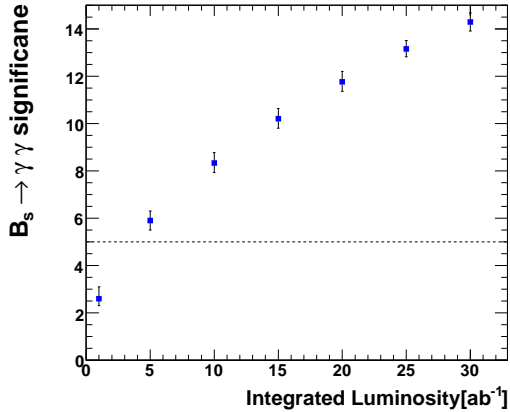}
\caption{\it Statistical significance of the measurement of $BR(B_s \to
\gamma \gamma)$ as a function of the integrated luminosity. Error bars
represent the RMS of the significance distribution for each set of toys.
\label{fig:Bsgammagamma}}
\end{center} 
\end{figure}

As already mentioned, the knowledge of the photon spectrum will always
be a limiting factor for the measurement of $b \to s \gamma$ decay
rate, whose error is presently systematically dominated and it is not
expected to improve in the near future. On the other hand, the
exclusive measurement of $B_s \to \gamma \gamma$ decay is very similar
to other measurements already performed at $B$-Factories (such as $B^0
\to \pi^0 \pi^0$) and it is not affected by a theoretical error
associated to the measurement (unlike the case of $b \to s \gamma$
analyses).
From an experimental point of view, the limiting systematic factor is
represented by the knowledge of the efficiency for photon
reconstruction, which can be reduced with dedicated studies on control
samples with similar energy range for the photons. On the other hand,
the presence of two hard photons represents a clear signature
for signal events, in particular if studied with recoil techniques.

In order to estimate the expected experimental sensitivity, we assume
the same technique and the same performance in terms of efficiency and
background rejection of the current BaBar search of $B_d \to \gamma
\gamma$~\cite{BaBarBdgg}. 
The main source of background comes from hard photons from
$\pi^0$ or $\eta$ decays, which are rejected requiring the invariant
mass to be not within $\pm 3 \sigma$ of the nominal $\pi^0$ or $\eta$
mass. Other sources of background, as initial state radiation, are
removed with cuts on topological variables. The efficiency on signal
event is $10\%$ and we expect $14$ signal events and $20$ background
events in a sample of $1$ab$^{-1}$ of integrated data.  We perform a
set of toy Monte Carlo experiments, using $m_{ES}$, $\Delta E$, and
$L_{2}/L_{0}$ in a ML fit. The result in term of the
integrated luminosity is shown in Fig.~\ref{fig:Bsgammagamma}.
Considering that the measurement at $1$ab$^{-1}$ has already a $5
\sigma$ significance, we conclude that a super $B$-Factory will be
able to precisely measure the BR even at low statistic.  Moreover, one
will achieve a $7\%$ statistical precision on the measurement at
$30$ab$^{-1}$, with a systematic error that will be smaller
~\footnote{Current analyses of $B^0 \to \pi^0 \pi^0$ have a
systematic error of $10\%$, generated by the presence of four photons
in the final state and other sources of systematic, which have a
statistical nature and can be strongly reduced at high integrated 
luminosity.} than $5\%$.
The large amount of events available even with few ab$^{-1}$ of integrated
data will allow to measure also $B_s \to \gamma \gamma$ CP asymmetry.
This will be possible with good accuracy exploiting a tagging
algorithm which determines whether the other $B_s$ meson in the event
decayed as a $B_s$ or a $\bar B_s$ (flavour
tag)\cite{{Aubert:2003sg}}.  \\

\section{The impact on flavour physics: A possible future scenario}\label{sec:impact}
In the previous sections we tried to give an idea of how rich is the
set of physics measurements a $B$-Factory can perform while running at
the $\Upsilon(5S)$. The processes described represent a
minimum set of measurements demonstrating the possibility of accessing
rates and weak phases of rare decays, i.e. the necessary ingredients to
test the SM.

Even with the limited amount of measurements we discussed, a facility
like the one we are considering will have an impact on the knowledge
of the flavour sector of NP. This might be clear enough from the
previous chapters. In this section we make it more
evident, showing how the unitarity triangle (UT) analysis will
benefit from it. In order to do that, we consider two different
scenarios: 
\begin{itemize}
\item a {\it low-statistics} case, assuming that after collecting
  $2~{\rm ab}^{-1}$ of data at the $\Upsilon(4S)$, (one of) the two
  existing $B$-Factories will move to the $\Upsilon(5S)$, collecting
  $1~{\rm ab}^{-1}$ before LHCb will produce the first physics
  results.
\item a {\it high-statistics} case, in which a high luminosity
  $B$-Factory will collect $75~{\rm ab}^{-1}$ at the $\Upsilon(4S)$
  and then $30~{\rm ab}^{-1}$ at the $\Upsilon(5S)$~\cite{superBcdr}.
\end{itemize}
The {\it low-statistics} ({\it high-statistics}) scenario corresponds
approximatively to fall 2009 (2015-2020). Taking into account the
different time scales of the two projections, we consider two
different sets of theoretical inputs: while for the {\it
low-statistics} case we use the current values of the determinations
of $f_{B_s}\sqrt{B_{B_s}}$, $\xi$, and $\hat B_K$, in the {\it
high-statistics} case we assume that the increase of computation power
will bring down the error of lattice QCD (LQCD) calculations to the
percent level, while no improvement on the theory is
considered. Indeed, since a constant progress in
the calculation techniques has characterized the recent history of
LQCD calculations, we can consider the estimates of the future errors
relatively conservative.

\begin{table}[tb!]
  \begin{center}
    \begin{tabular}{ccc} 
      \hline \hline
      Observable                      & {\it low-statistics}  &   {\it high-statistics} \\ 
                                      & Error at $2 \ {\rm ab}^{-1}$    &    Error at $75 \ {\rm ab}^{-1}$  \\
      \hline\hline
      $\left| V_{cb} \right|$ (exclusive)       &      $ 4\%$       &       $1.0\%$ \\
      $\left| V_{cb} \right|$ (inclusive)       &      $ 1\%$       &       $0.5\%$ \\
      $\left| V_{ub} \right|$ (exclusive)       &      $ 8\%$       &       $2.0\%$ \\
      $\left| V_{ub} \right|$ (inclusive)       &      $ 8\%$       &       $2.0\%$ \\
      $BR(B \to \tau \nu)$                      &      $20\%$               &       $ 4\%$ \\
      \hline
      $f_{B_s}\sqrt{B_{B_s}}$                   &      $13\%$               & $1\%$ \\
      $\xi-1$                                   &      $26\%$               & $5\%$ \\
      $\hat B_K$                                &      $5\%$ $10\%$         & $1\%$ \\
      \hline
      $\sin(2\beta)$ ($J/\psi\,K^0$)            &       0.018               &        0.005 \\
      $\gamma$ ($B \to DK$, combined)           &   $\sim  6^\circ$         &        $1^\circ$ \\
      $\alpha$ (combined)                       &   $\sim  6^\circ$         &        $1^\circ$ \\
      $2\beta+\gamma$ ($D^{(*)\pm}\pi^\mp$, $D^\pm K_S \pi^\mp$) 
                                                & $20^\circ$                &        $5^\circ$        \\
      \hline \hline
    \end{tabular}
  \end{center}
  \caption{\it Summary of the expected precision on the inputs of the UT
    analysis, in the {\it low-statistics} and {\it high-statistics}
    scenarios. The central values are taken in order to provide a
    perfect agreement among the constraints in the total fit.}
  \label{tab:inputs1} 
\end{table}

In Tab.~\ref{tab:inputs1}, we quote the errors on the experimental and
theoretical inputs currently used in the UT analysis, based on
measurements performed at the $\Upsilon(4S)$, together with
$\epsilon_K$ from $K$--$\bar K$ mixing and $\Delta m_s$ from
$B_s$--$\bar B_s$ mixing. In Tab.~\ref{tab:bs-input} we give the errors
on the measurements performed at the $\Upsilon(5S)$, taken from the
study presented in this paper. As mentioned in Sec.~\ref{sec:betas_dtsign}, 
we took into account an increase of the error of $\beta_s$ from 
$J/\psi \phi$ with tagged rates at positive and negative $\Delta t$, due to 
the presence of a two-fold ambiguity around $\beta_s \sim 0$.

\subsection{The Unitarity Triangle in the Standard Model}

\begin{figure}[tb!]
\begin{center}
\includegraphics[width=7cm]{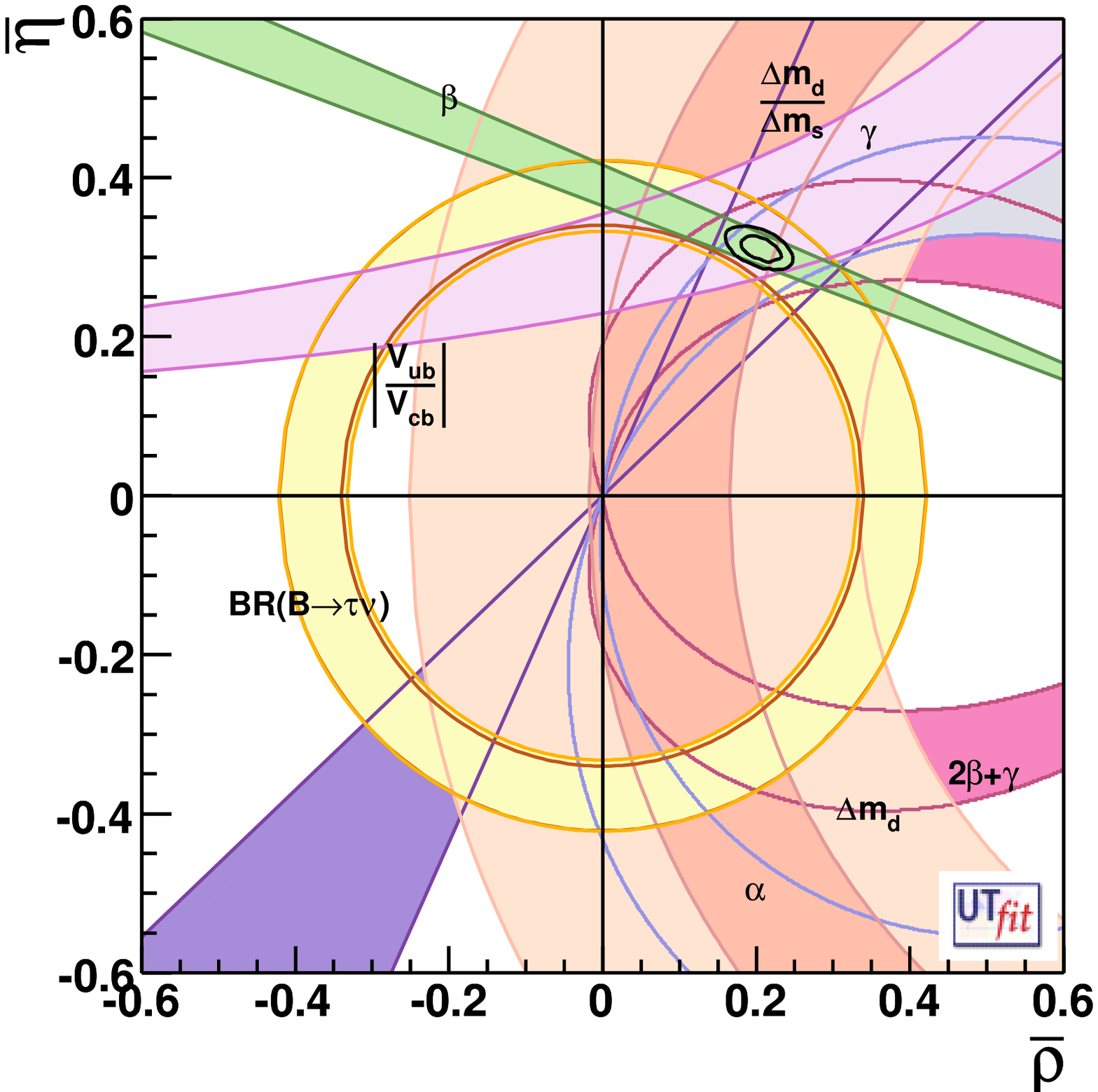}
\includegraphics[width=7cm]{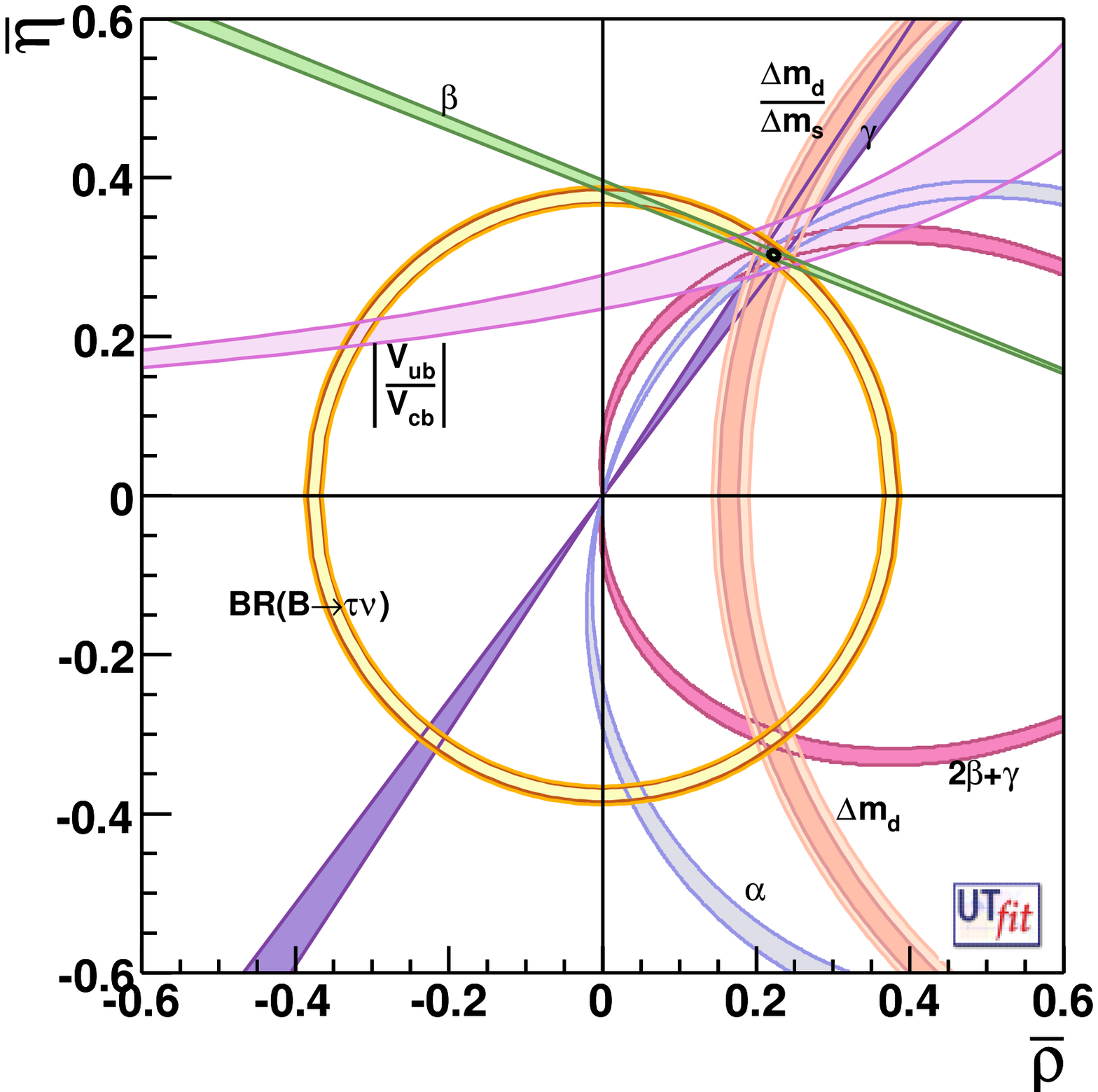}
\caption{\it Determination of $\bar \rho$ and $\bar \eta$ in the
Standard Model UT analysis for $2~{\rm ab}^{-1}$ (left) and $75~{\rm
ab}^{-1}$ (right) projections of the errors as given in
Tab.~\ref{tab:inputs1}\label{fig:UTSM}}
\end{center}
\end{figure}

We show in Fig.~\ref{fig:UTSM} how the SM analysis of the UT will look like in
the two scenarios described above. The two plots correspond to an
overall error of $\sigma_{\bar\rho} = 9\%$ ($\sigma_{\bar\eta} = 4\%$)
on $\bar\rho$ ($\bar\eta$) for the {\it low-statistics} scenario and
to an overall error $\sigma_{\bar\rho} = 1.4\%$ ($\sigma_{\bar\eta} =
0.8\%$) on $\bar\rho$ ($\bar\eta$) for the {\it high-statistics} one.
For comparison, the current knowledge of the UT triangle gives
$\sigma_{\bar\rho} = 17\%$ and $\sigma_{\bar\eta} = 5\%$.

\begin{table}[tbp!]
\begin{center}
\begin{tabular}{ccc}
\hline
\hline
      Observable                      & {\it low-statistics}  &   {\it high-statistics} \\ 
                                      & Error at $1 \ {\rm ab}^{-1}$    &    Error at $30 \ {\rm ab}^{-1}$   \\
\hline\hline
$\Delta \Gamma$                                  & 0.16 ps$^{-1}$       &  0.03 ps$^{-1}$          \\  
$\Gamma$                                         & 0.07 ps$^{-1}$       &  0.01 ps$^{-1}$	   \\
$\beta_s$ from angular analysis                  & $20^\circ$           &  $8^\circ$		   \\
$A^s_{SL}$                                       & 0.006                &  0.004		   \\
$A_{CH}$                                         & 0.004                &  0.004                   \\
$\beta_s$ from $J/\psi \phi$ $\Delta t$ sign     & 10$^\circ$           &  3$^\circ$               \\
$V_{td}/V_{ts}$                                  & 0.10                 &  0.031                   \\
$BR(B_s \rightarrow \mu^+ \mu^-) \cdot 10^{-8   }$      &$<$10 at 90$\%$ prob. & $<$ 1.30 at 90$\%$ prob. \\
$BR(B_s \rightarrow \gamma \gamma)$                  &  38$\%$              & 7$\%$\\
\hline\hline
\end{tabular}
\caption{\it Expected errors for different observables considering an integrated
luminosity of 1ab$^{-1}$ and 30ab$^{-1}$.}
\label{tab:bs-input}
\end{center}
\end{table}

\subsection{The Unitarity Triangle beyond the Standard Model}

In a generic NP scenario the effect of $\Delta F=2$ NP contributions in
$B_q$--$\bar B_q$ mixing can be parameterized in terms of two
quantities, $C_{B_q}$ and $\phi_{B_q}$, which relate the experimental
observables to the SM quantities. They are defined
as~\cite{formuleNPmodelindep}:
\begin{eqnarray} 
 &&C_{B_q} \, e^{2 i \phi_{B_q}} =\frac{\langle
  B_q|H_\mathrm{eff}^\mathrm{full}|\bar{B}_q\rangle} {\langle
  B_q|H_\mathrm{eff}^\mathrm{SM}|\bar{B}_q\rangle}= \nonumber \\ &&=
  \frac{A_q^\mathrm{SM} e^{2 i \phi_q^\mathrm{SM}} + A_q^\mathrm{NP}
  e^{2 i (\phi_q^\mathrm{SM} + \phi_q^\mathrm{NP})}}{A_q^\mathrm{SM}
  e^{2 i \phi_q^\mathrm{SM}}}\,,
  \label{eq:paranp}
\end{eqnarray}
where $H_\mathrm{eff}^\mathrm{SM}$ is the effective Hamiltonian,
$H_\mathrm{eff}^\mathrm{full}$ is the Hamiltonian also including NP
contributions, and $A_q^\mathrm{SM} e^{2 i \phi_q^\mathrm{SM}}$
($A_q^\mathrm{NP} e^{2 i (\phi_q^\mathrm{SM} + \phi_q^\mathrm{NP})}$)
is the SM (NP) amplitude.  The two parameters $C_{B_q}$ and
$\phi_{B_q}$ introduced in Eq.~(\ref{eq:paranp}) allow to relate the
SM value of a certain observable to its measured value.  For instance,
the measured values of the size and phase of $B_d$--$\bar B_d$ mixing
are given by $\Delta m_d^{EXP} = C_{B_d} \Delta m_d^{SM}$ and
$\beta^{EXP} = \beta+\phi_{B_d}$. Similar relations hold for the other
observables of Tab.~\ref{tab:inputs1}.  Neglecting the case of NP
contributions entering at tree-level processes, there are only two
observables which do not depend on the presence of NP, namely
$|V_{ub}/V_{cb}|$ and $\gamma$.~\footnote{In the case of $\gamma$,
this is true up to a contribution in $D$--$\bar D$ mixing, which is
estimated to be at the level of a few percent~\cite{DDmixing1} and
which can be taken into account following the approach suggested in
ref.~\cite{DDmixing2}.}

\begin{figure}[tb!]
\begin{center}
\includegraphics[width=7cm]{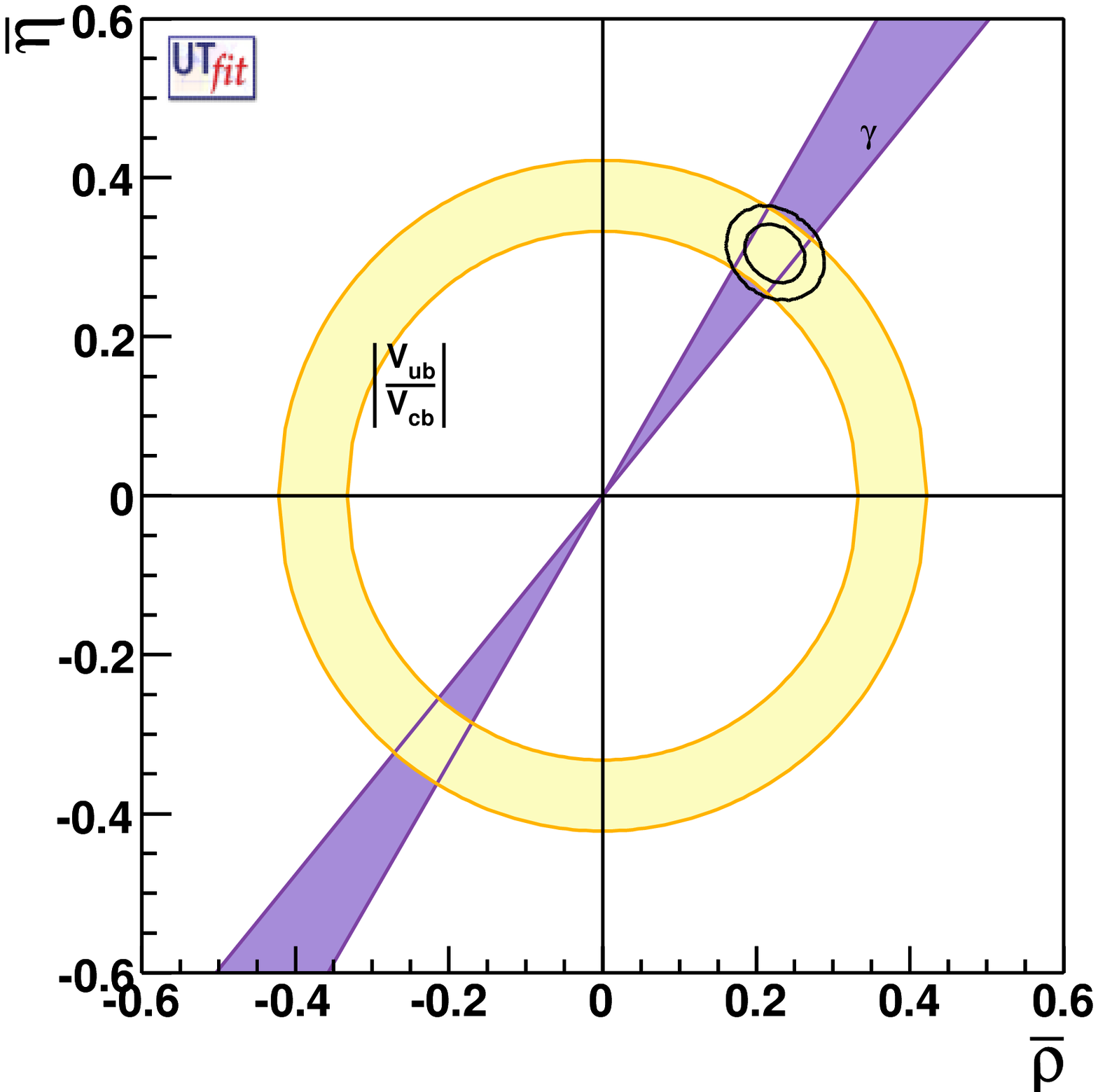}
\includegraphics[width=7cm]{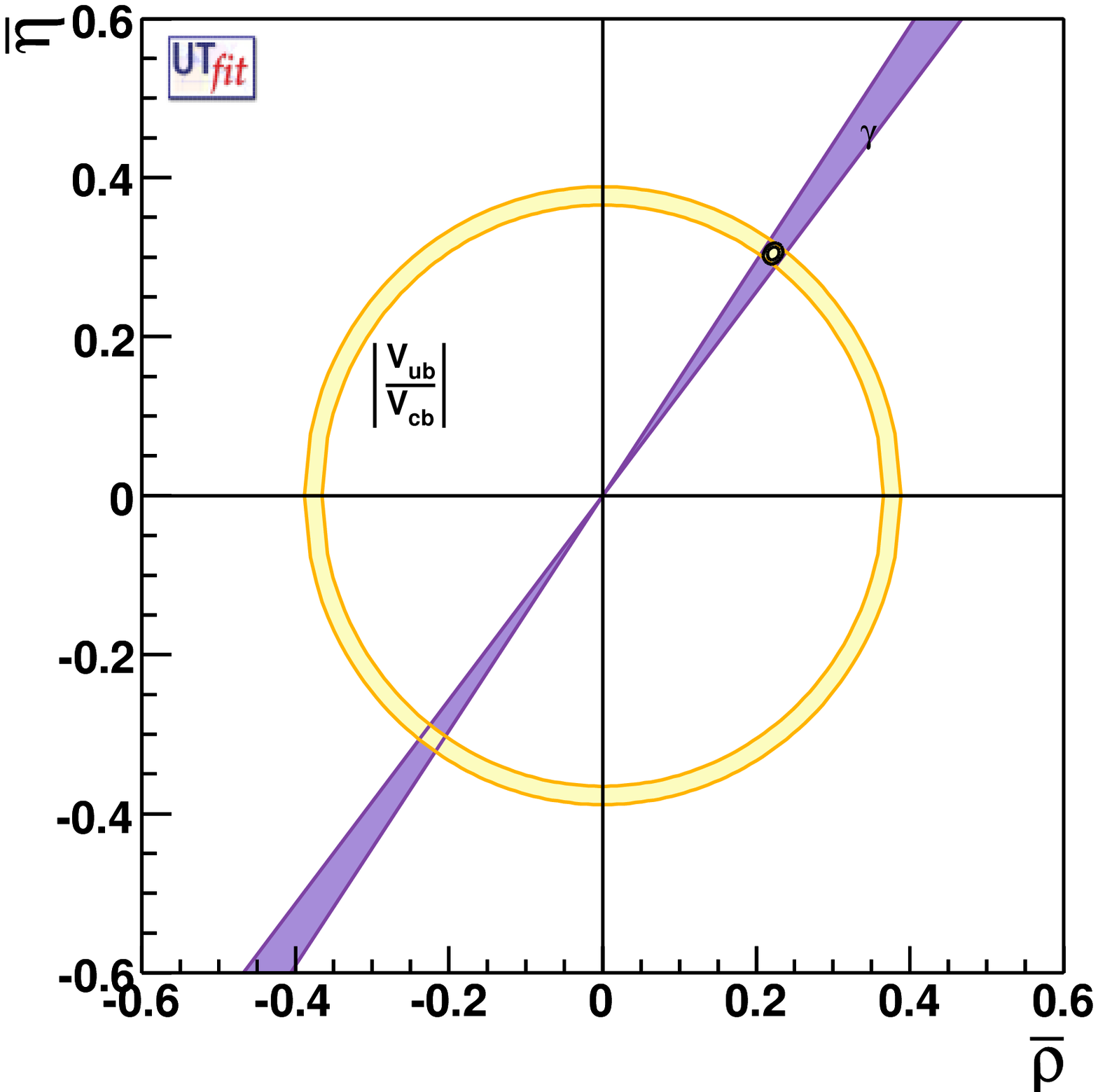}
\caption{\it Determination of $\bar \rho$ and $\bar \eta$ in the
generalized UT analysis, obtained using all the available constraints
for $2~{\rm ab}^{-1}$ (left) and $75~{\rm ab}^{-1}$ (right)
projections of the errors as given in
Tab.~\ref{tab:inputs1}\label{fig:nprhoeta}}
\end{center}
\end{figure}

\begin{figure}[tb!]
\begin{center}
\includegraphics[width=7cm]{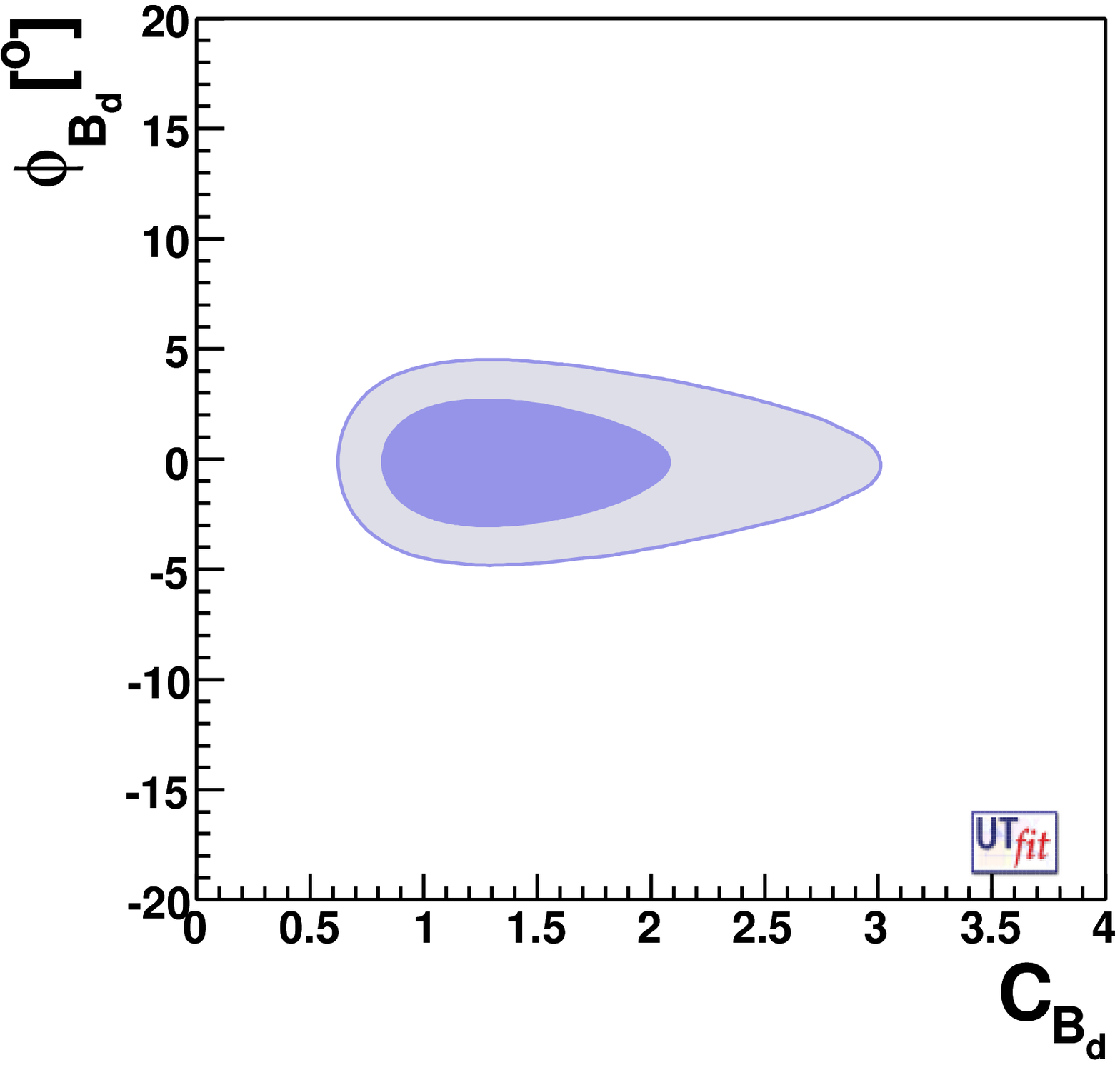}
\includegraphics[width=7cm]{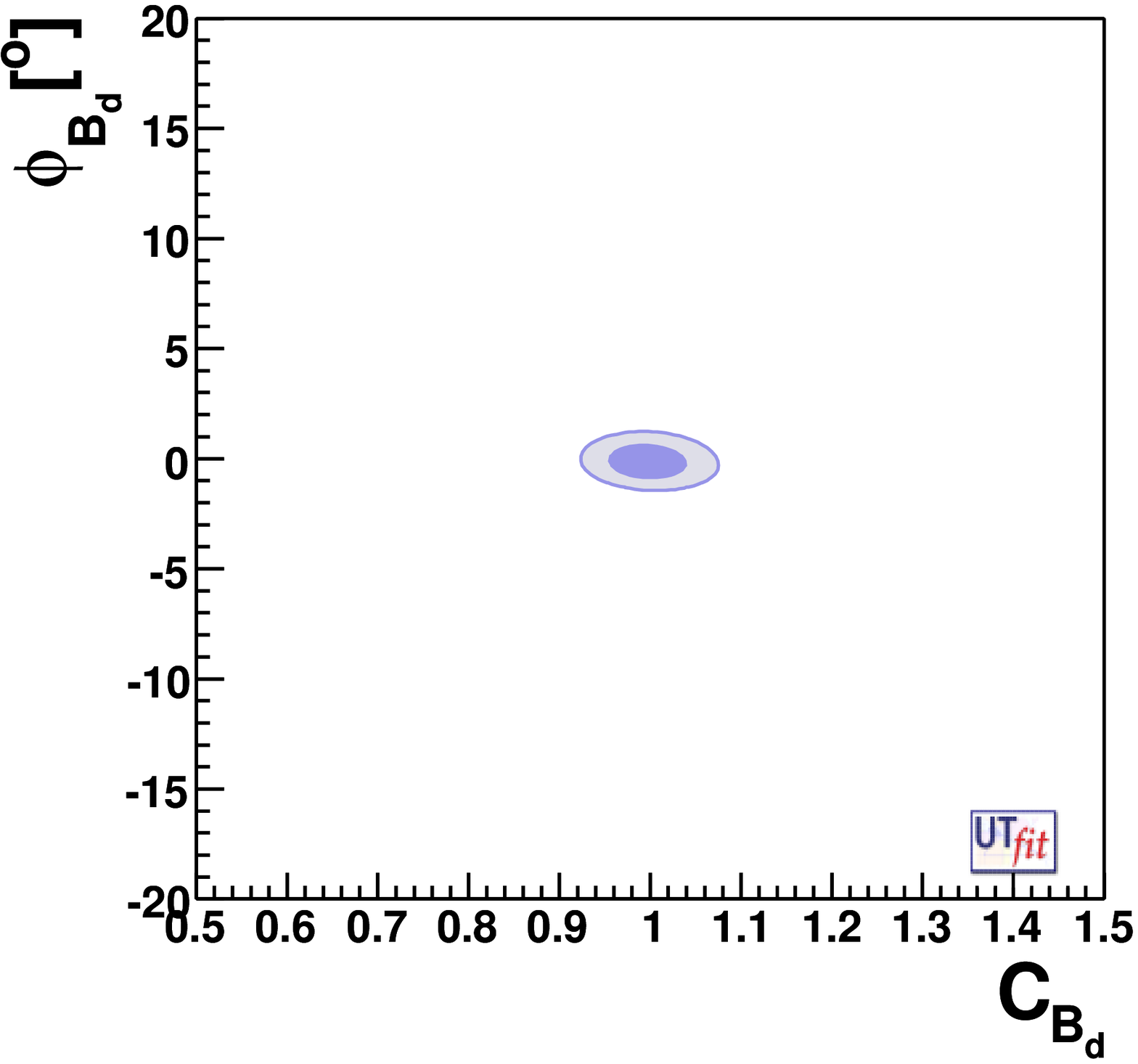}
\caption{\it Determination of $C_{B_d}$ and $\phi_{B_d}$ in the
generalized UT analysis for $2~{\rm ab}^{-1}$ (left) and $75~{\rm
ab}^{-1}$ (right) projections of the errors given in
Tab.~\ref{tab:inputs1}, combined to the low-statistics (left) and
high-statistics (right) scenarios of
Tab.~\ref{tab:bs-input}. \label{fig:cbdphibd}}
\end{center}
\end{figure}

Since under these assumptions we can parameterize NP effects just with
two real parameters, without assuming any specific model, it is
possible to determine the allowed region for NP, fitting simultaneously
$C_{B_d}$, $\phi_{B_d}$, $\bar \rho$, and $\bar \eta$. Assuming the
two scenarios of Tab.~\ref{tab:inputs1}, we obtain the two plots given
in Fig.~\ref{fig:nprhoeta} and the errors quoted in the first two raws
of Tab.~\ref{tab:NPresults}. The determination of $C_{B_d}$ and 
$\phi_{B_d}$ in the two cases is given in Fig.~\ref{fig:cbdphibd}.

\begin{table}[h]
\begin{center}
\begin{tabular}{@{}cccc}
\hline\hline
    Parameter           & {\it low-statistics} & {\it high-statistics}  \\   
                        &      Error           &        Error           \\
\hline\hline
$\overline {\rho}$      &  $12\%$  & $2.3\%$  \\ 
$\overline {\eta}$      &  $8\%$   & $1.8\%$  \\ 
\hline
$C_{B_d}$               &  $0.43$  & $0.032$  \\ 
$\phi_{B_d} [^{\circ}]$ &  $1.7^o$ & $0.44^o$ \\ 
\hline
$C_{B_s}$               &  $0.33$  & $0.026$  \\ 
$\phi_{B_s} [^{\circ}]$ &  $7^o$   & $1.9^o$  \\ 
\hline
\hline
\end{tabular}
\end{center}
\caption{\it Determination of UT and NP parameters
from the generalized UT analysis, in the two scenarios of
Tab.~\ref{tab:inputs1}. For the $B_s$ sector, values from
Tab.~\ref{tab:bs-input} are used.}
\label{tab:NPresults}
\end{table}

On top of this information, one can use the experimental information
provided by the $\Upsilon(5S)$ run to constrain the values of $C_{B_s}$ and
$\phi_{B_s}$.  $C_{B_s}$ is directly related to the value of $\Delta
m_s$, in analogy to the $B_d$ case, while $\phi_{B_s}$ is defined in
such a way that $\beta_s^{EXP} = \beta_s-\phi_{B_s}$. The relation
between the other constraints of Tab.~\ref{tab:bs-input} and the NP
parameters is more complicated.  It has been shown that
the semileptonic CP asymmetry~\cite{Laplace:2002ik} and the value of
$\Delta \Gamma_s/\Gamma_s$~\cite{Dighedgog} are sensitive to NP
contributions to $\Delta F=2$ effective Hamiltonian. We use the NLO
expression of these observables in terms of the parameters $C_{B_s}$
and $\phi_{B_s}$~\cite{UTNP}.

\begin{figure}[tb!]
\begin{center}
\includegraphics[width=7cm]{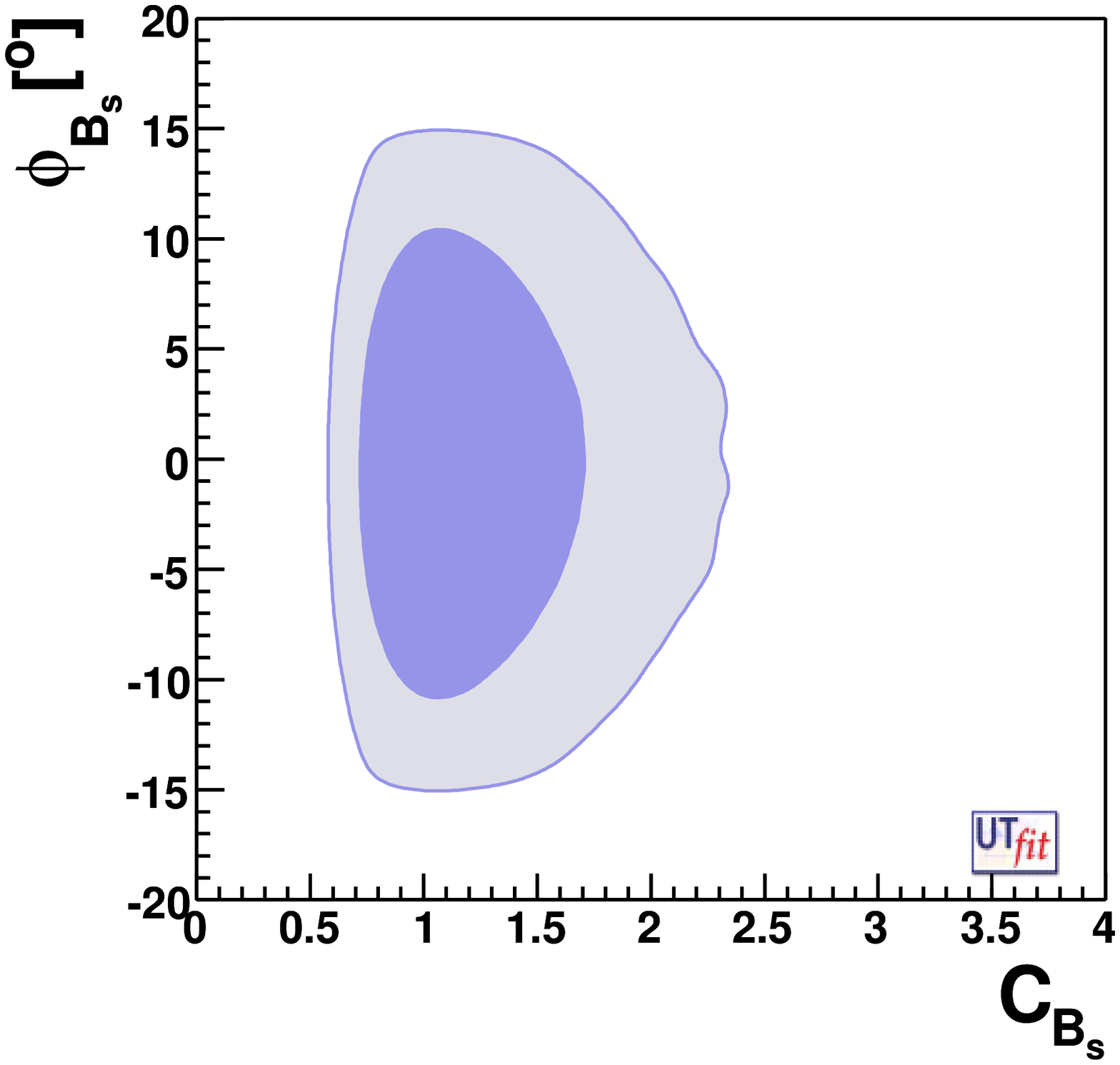}
\includegraphics[width=7cm]{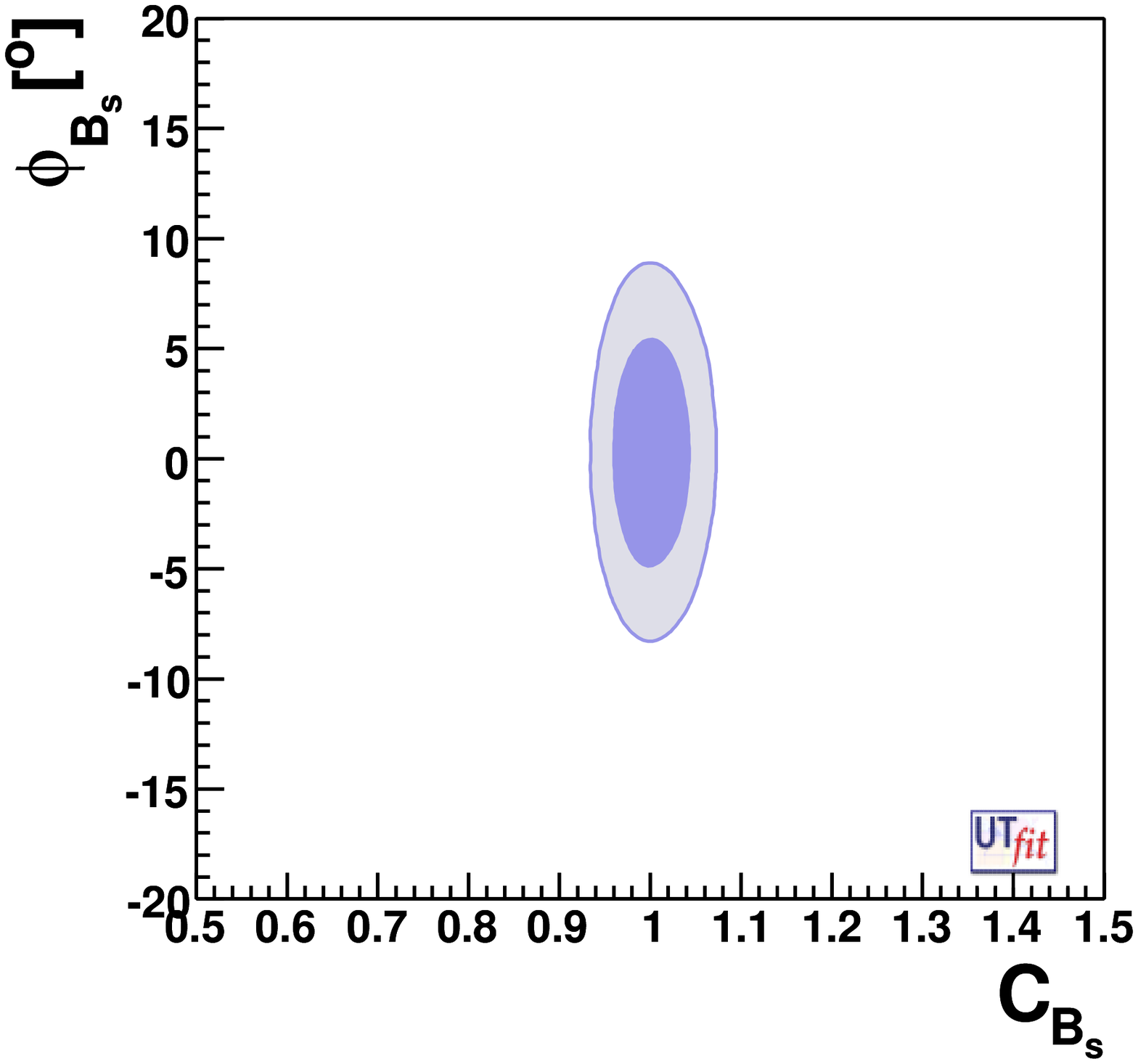}
\caption{\it Determination of $C_{B_s}$ and $\phi_{B_s}$ in the
generalized UT analysis for $1~{\rm ab}^{-1}$ (left) and $30~{\rm ab}^{-1}$
(right) projections of the errors for the input of
Tab.~\ref{tab:bs-input}. In the left (right) plot, $2~{\rm ab}^{-1}$
($75~{\rm ab}^{-1}$) inputs of Tab.~\ref{tab:inputs1} are used for the other
determinations.\label{fig:cbsphibs}}
\end{center}
\end{figure}

The result of the combination of all these measurements is shown in
the two plots of Fig.~\ref{fig:cbsphibs} and quantified in the bottom
part of Tab.~\ref{tab:NPresults}. On top of this analysis, one can
further test the presence of NP using $\Delta F=1$ processes, as
already discussed in Sec.~\ref{sec:beta_sPen}, relating  $\Delta F=1$
and  $\Delta F=2$  NP effects by using a specific NP model
(see for example~\cite{silvestrini_review}).

\subsection{A comparison with the reach of the LHCb experiment}

Tab.~\ref{tab:y5s_vs_lhcb} shows a comparison between the results presented
in this paper and the expected sensitivity of LHCb for different 
benchmark measurement. The LHCb estimates are shown for an integrated
luminosity of 2 $fb^{-1}$ (one year of nominal data taking) and
10 $fb^{-1}$ (full data taking), using the errors on lattice predictions 
expected on 2009 and 2015 respectively. The two scenarios described in the 
previous sections are used for a $B$-Factory running at the $\Upsilon(5S)$ resonance.
The measurement of $\Delta m_s$ at LHCb is supposed to be dominated by a 
systematic error of 0.09 $ps^{-1}$~\cite{schune_paris}. The relative error 
on $B_d \to K^{*0} \gamma$ is estimated to be $\sim$ 5\% (dominated by
the systematics on the branching ratio's normalization). The relative error 
on $B_s \to K^{*0} \gamma$ is estimated to be $\sim$ 20\% in the 2 $fb^{-1}$
scenario and $\sim$ 10\% in the 10 $fb^{-1}$ scenario~\cite{dickens,vagnoni}.
In the $A^s_{SL}$ and $A_{CH}$ measurements both statistical and systematic
uncertainties are shown.

In doing this comparison, the reader should remember that a reliable estimate of 
important systematic effects is not available for several LHCb measurements,
while these uncertainties are typically well known and under control at 
the $B$-Factories. Moreover, we want to stress here that we presented only a 
minimal subsample of the possible measurements that can be performed at a 
$B$-Factory running at the $\Upsilon(5S)$, aiming to demonstrate the experimental 
feasibility of the corresponding analysis techniques. In the channels we chose,
our estimates are mostly not competitive with respect to LHCb, but through them 
we proved the possibility to widely apply these techniques, in order to investigate
several channels that cannot be studied in the LHC environment.

\begin{table}
\begin{center}
\begin{tabular}{c|cc|cc}
\hline
\hline
           & \multicolumn{2}{c}{LHCb} & \multicolumn{2}{c}{$\Upsilon(5S)$} \\
\hline
Observable & $2fb^{-1}$ & $10fb^{-1}$ & $1ab^{-1}$ & $30ab^{-1}$ \\
\hline
$|V_{td}/V_{ts}|$ from $\Delta m_s$ & $0.010$ & $0.002$ &
- & - \\
$\Delta \Gamma / \Gamma$      & 0.0092 & 0.004 & 0.12 & 0.02\\
$\beta_s$ from angular analysis  & $0.66^\circ$ & $0.29^\circ$ & $20^\circ$ &
$8^\circ$ \\
$A^s_{SL}$                    & - & - &
$\pm 0.006 \pm 0.004$ & $\pm 0.001 \pm 0.004$\\
$A_{CH}$                      & - & - &
$\pm 0.0015 \pm 0.004$ & $\pm 0.0003 \pm 0.004$\\
$\beta_s$ from $J/\psi \phi$ $\Delta t$ sign & - & - &$20^\circ$ &$8^\circ$
\\
$BR(B_s \to \mu \mu)$ & $1.2 \cdot 10^{-9}$ & $0.7 \cdot 10^{-9}$ & $<10^{-7}$ &
$<1.30 \cdot 10^{-8}$ \\
$|V_{td}/V_{ts}|$  from radiative decays & 0.03 & 0.015 & 0.10 & 0.031 \\
$BR(B_s \to \gamma \gamma)$   & - & - & 38\% & 7\% \\
\hline
\end{tabular}
\caption{\it Expected errors for different observables at 
LHCb~\cite{schune_paris,dickens,vagnoni} and at a $B$-Factory running at the 
$\Upsilon(5S)$ resonance.}
\label{tab:y5s_vs_lhcb}
\end{center}
\end{table}

\section{Conclusions}\label{sec:conclusions}

In this paper we discussed the physics potential of a $B$-Factory
running at the $\Upsilon(5S)$. Assuming the performance of the
detectors currently taking data at the two $B$-Factories, we proved
that it is possible to obtain a pure sample of $B_s$ mesons and at the
same time to cover the physics program of a traditional $B$-Factory.
We discussed the detector limitations in studying $B_s$--$\bar B_s$
oscillations and we showed that, even if time-dependent
measurements of $B_s$ are not accessible with the current vertex
resolution, the same physics information can be obtained using
time-integrated measurements.  We also illustrated some benchmark
analyses of rare $B_s$ decays, interesting to test the SM
and obtain information on the flavour structure of the NP.
With this limited set of examples, we proved that this
facility can perform similar measurements using alternative
techniques with respect to hadron collider experiments. On the other
hand, thanks to the clean environment and the progress made by BaBar
and Belle in the recent past, this facility allows to extend the LHCb
physics program to a larger set of interesting channels which, not
having prompt charged tracks and/or being characterized by the
presence of photons in the final state, are difficult to study at a
hadron collider. Examples of this kind are $B_s \to
\gamma \gamma$ and $B_s \to K^0 \bar K^0$. To conclude, we
extrapolated the impact of the $B_s$ physics program on the unitarity
triangle analysis beyond the standard model, considering the
possibility of $1{\rm ab}^{-1}$ collected by the existing
$B$-Factories after accumulating $2{\rm ab}^{-1}$ at the $\Upsilon(4S)$,
or $30{\rm ab}^{-1}$ collected by a high luminosity $B$-Factory after
collecting $75{\rm ab}^{-1}$ at the $\Upsilon(4S)$.  We stress that
the experimental precision on these observables is not necessarily
better than what LHCb can achieve~\cite{Schneider}. Nevertheless, this
limited set of examples demonstrates that, even using different
experimental techniques, the proposed machine can measure absolute
values and (strong and weak) phases of decay amplitudes, even without
directly accessing the time-dependent structure of $B_s$--$\bar B_s$
oscillation.  Furthermore, thanks to the very clean environment, it is
possible to apply the discussed experimental techniques to all $B$
decays, without any restriction related to the need of prompt tracks
in the event.  We do not cover all the examples of such additional
potentiality with respect to hadron collider experiments, but only
outline the cases of the measurements of the rate of $B_s \to \gamma
\gamma$ decays and of the weak phase in $B_s \to K^0 \bar K^0$ decays.

Concluding, we would like to stress that, unlike the case of
$B_d$ and charged $B$ decays, the $B_s$ sector is presently largely
unstudied.  Based on what we discussed in this paper, we think that it is
particularly interesting to consider the possibility of studying $B_s$
decays with BaBar and Belle before LHCb starts. This would
maximize the physics impact of the luminosity that can be collected
(less than $2$ab$^{-1}$), while a super $B$-Factory would be needed to
obtain the required luminosity during and after LHC data taking.

\section{Acknowledgments}

We would like to thank G.~Isidori and G.~Martinelli for their help
in writing Sec.~\ref{sec:timeevolution},
V.~Lubicz for providing us the extrapolation of lattice QCD errors,
R.~Faccini for useful discussions and suggestions and for the
feed-back provided during all the development of this work and E.~Emili
and D.~Granata for their contribution in our early studies for this paper.
We would
also like to thank T.~Browder and M.~Hazumi for the interest shown for
this work and for inviting us to take part at the BNMS2006 workshop
series, which allowed us to compare our studies to the activity of the
Belle community.  This work has been supported in part by the
France-Japon Particle Physics Laboratory FJ-PPL, the EU networks ``The
quest for unification'' under the contract MRTN-CT-2004-503369,
``FLAVIAnet'' under the contract MRTN-CT-2006-035482, and ``HEPTOOLS''
under the contract MRTN-CT-2006-035505.

\end{document}